\documentclass[11pt,a4paper]{article}
\pdfoutput=1
\usepackage{jcappub}
\usepackage{rotating}
\usepackage{array}
\usepackage{amsmath}
\usepackage[normalem]{ulem}
\usepackage{slashed}
\usepackage{booktabs}
\usepackage[pdftex,table]{xcolor}
\usepackage{units}
\usepackage[title]{appendix}
\usepackage{xspace}
\usepackage{physics}
\usepackage{listings}

\usepackage{multirow}

\usepackage{url}
\usepackage[normalem]{ulem}

\newcommand{\darkbit}{\textsf{DarkBit}\xspace}
\newcommand{\cosmobit}{\textsf{CosmoBit}\xspace}
\newcommand{\colliderbit}{\textsf{ColliderBit}\xspace}
\newcommand{\specbit}{\textsf{SpecBit}\xspace}

\newcommand{\gambit}{\textsc{Gambit}\xspace}
\newcommand{\obscura}{\textsc{obscura}\xspace}
\newcommand{\darkcast}{\textsc{DarkCast}\xspace}
\newcommand{\darksusy}{\textsc{DarkSUSY}\xspace}
\newcommand{\ddcalc}{\textsc{DDCalc}\xspace}
\newcommand{\alterbbn}{\textsc{AlterBBN}\xspace}
\newcommand{\darkages}{\textsc{DarkAges}\xspace}
\newcommand{\hazma}{\textsc{Hazma}\xspace}
\newcommand{\BdNMC}{\textsc{BdNMC}\xspace}
\newcommand{\CalcHEP}{\textsc{CalcHEP}\xspace}
\newcommand{\FeynRules}{\textsc{FeynRules}\xspace}

\newcommand{\doublecrosssf}[2]{\hyperref[#2]{\textbf{\textsf{#1}}}}
\newcommand{\term}[1]{\texttt{#1}\xspace}
\definecolor{solarizedblue}{HTML}{268BD2}
\newcommand\cpp[1]{{\lstinline!#1!}}
\lstdefinestyle{cpp}
{
  language=C++,
  basicstyle=\footnotesize\ttfamily,
  basewidth={0.53em,0.44em},
  numbers=none,
  tabsize=2,
  breaklines=true,
  escapeinside={@}{@},
  showstringspaces=false,
  identifierstyle=\color{solarizedblue},
  literate={~} {\customtilde}1,
  moredelim=*[directive]\ \ \#, 
  moredelim=*[directive]\ \ \ \ \#
}
\arxivnumber{TTP24-015, P3H-24-033 }

\title{Resonant or asymmetric: \\ The status of sub-GeV dark matter}

\author[a]{Sowmiya Balan,}
\author[b]{Csaba Bal\'{a}zs,}
\author[c]{Torsten Bringmann,}
\author[d,e,f]{Christopher~Cappiello,}
\author[g]{Riccardo Catena,}
\author[h]{Timon Emken,}
\author[a]{Tom\'as~E.~Gonzalo,}
\author[g]{Taylor R. Gray,}
\author[i,j]{Will Handley,}
\author[b]{Quan Huynh,}
\author[a]{Felix Kahlhoefer}
\author[d,e,f]{and Aaron C. Vincent}

\affiliation[a]{\small Institute for Theoretical Particle Physics (TTP), Karlsruhe Institute of Technology (KIT), 76128 Karlsruhe, Germany}
\affiliation[b]{\small School of Physics and Astronomy, Monash University, Melbourne VIC 3800, Australia}
\affiliation[c]{\small Department of Physics, University of Oslo, N-0316 Oslo, Norway}
\affiliation[d]{\small Arthur B. McDonald Canadian Astroparticle Physics Research Institute, Kingston ON K7L 3N6, Canada}
\affiliation[e]{\small Department of Physics, Engineering Physics and Astronomy, Queen’s University, Kingston ON K7L 3N6, Canada}
\affiliation[f]{\small Perimeter Institute for Theoretical Physics, Waterloo ON N2L 2Y5, Canada}
\affiliation[g]{\small Chalmers University of Technology, Department of Physics, SE-412 96 Göteborg, Sweden}
\affiliation[h]{\small The Oskar Klein Centre, Department of Physics, Stockholm University,
AlbaNova, SE-10691 Stockholm, Sweden}
\affiliation[i]{\small Kavli Institute for Cosmology, University of Cambridge, Madingley Road, Cambridge, CB3 0HA, UK}
\affiliation[j]{\small Cavendish Laboratory, University of Cambridge, JJ Thomson Avenue, Cambridge, CB3 0HE, UK}

\emailAdd{sowmiya.balan@kit.edu}
\emailAdd{taylor.gray@chalmers.se}
\emailAdd{kahlhoefer@kit.edu}

\abstract{Sub-GeV dark matter (DM) particles produced via thermal freeze-out evade many of the strong constraints on heavier DM candidates but at the same time face a multitude of new constraints from laboratory experiments, astrophysical observations and cosmological data. In this work we combine all of these constraints in order to perform frequentist and Bayesian global analyses of fermionic and scalar sub-GeV DM coupled to a dark photon with kinetic mixing. For fermionic DM, we find viable parameter regions close to the dark photon resonance, which expand significantly when including a particle-antiparticle asymmetry. For scalar DM, the velocity-dependent annihilation cross section evades the strongest constraints even in the symmetric case. Using Bayesian model comparison, we show that both asymmetric fermionic DM and symmetric scalar DM are preferred over symmetric fermionic DM due to the reduced fine-tuning penalty. Finally, we explore the discovery prospects of near-future experiments both in the full parameter space and for specific benchmark points.
We find that the most commonly used benchmark scenarios are already in tension with existing constraints and propose a new benchmark point that can be targeted with future searches.}


\begin{document}
\maketitle
\flushbottom

\section{Introduction}

The most plausible explanation for the wealth of astrophysical and cosmological evidence for dark matter (DM) is the existence of 
new elementary particles. Without additional assumptions, however, the allowed mass range for such particles spans almost 50 
orders of magnitude, with the lower bound of around $10^{-30} \, \mathrm{GeV}$ set by quantum mechanics (the De Broglie 
wavelength must be smaller than both the size of known astrophysical objects~\cite{Zimmermann:2024xvd} and the correlation length of structures in the Lyman alpha forest~\cite{Irsic:2017yje,Rogers:2020ltq}) and the upper 
bound of around $10^{19} \, \mathrm{GeV}$ set by gravity (the mass of a fundamental particle must be smaller than the Planck 
mass). Even larger DM masses are possible when considering composite states.
 This enormous window can be narrowed down significantly, if we consider DM particles that obtain their relic abundance in 
the early universe by freezing-out from the SM thermal bath. In this case, a lower bound on the DM mass of around 10 MeV is imposed by 
cosmology, specifically constraints on the number of relativistic degrees of freedom from Big Bang Nucleosynthesis (BBN) and 
the Cosmic Microwave Background (CMB)~\cite{Depta:2019lbe,Sabti:2019mhn}, and an upper bound of around 100 TeV is 
imposed by the unitarity requirement that the annihilation cross section of a particle cannot be arbitrarily 
large~\cite{Griest:1989wd}. Over the past few decades, the central part of this range, from a few GeV to a few TeV, has 
been the target of a large number of direct and indirect detection experiments and collider searches, and their null results have 
strongly constrained the corresponding DM models~\cite{Arcadi:2017kky,Arcadi:2024ukq}. {For many models, the leading constraints come from direct detection experiments searching for nuclear recoils, which however rapidly lose sensitivity for DM particles below the GeV scale, because the typical kinetic energy of such particles lies below the detector threshold.}

In the present work we therefore focus on sub-GeV thermal DM particles. This mass range has traditionally received less 
attention because of the Lee-Weinberg bound~\cite{Lee:1977ua}, which states that for sub-GeV DM particles the known 
interactions of the Standard Model (SM) are insufficient to reproduce the observed DM relic abundance through thermal 
freeze-out.
In recent years, however, many new DM models have been developed, which -- in addition to the DM particle -- feature a new 
interaction, mediated for example by the gauge boson of a new $U(1)'$ gauge group, called dark photon (for a review see 
for example~ref.~\cite{Knapen:2017xzo}). At the same time, direct detection experiments have substantially improved their sensitivity to 
sub-GeV DM particles by lowering their thresholds for nuclear recoils and searching for electron recoil 
signatures~\cite{Lin:2019uvt}, while accelerator experiments have performed various dedicated searches for dark photon signals 
in visible and invisible final states~\cite{Fabbrichesi:2020wbt}. In combination, these developments make sub-GeV thermal DM 
one of the most exciting frontiers of particle physics~\cite{Battaglieri:2017aum}.

Nevertheless, models of sub-GeV DM face a key challenge, which is that they need to satisfy strong constraints on the DM 
annihilation cross section from the CMB~\cite{Slatyer:2009yq,Planck:2018vyg} and, as recently pointed out in 
ref.~\cite{Cirelli:2023tnx}, from searches for X-ray emission. There are essentially three ways to evade these constraints: by 
considering a strongly velocity-dependent annihilation cross section arising for example from $p$-wave 
suppression~\cite{Boehm:2003hm} or resonant enhancement~\cite{Bernreuther:2020koj,Brahma:2023psr,Belanger:2024bro}, by allowing for a particle-antiparticle 
asymmetry suppressing annihilation rates in the present universe~\cite{Lin:2011gj}, or by considering DM particles that only 
constitute a fraction $f$ of the total DM density, such that annihilation signals are suppressed proportional to $f^2$ 
even in the absence of an asymmetry. In the present work we consider all three possibilities simultaneously and determine which 
ones are favoured or disfavoured by data.\footnote{Another possibility would be to consider alternative ways for DM particles to obtain their relic abundance, such as secluded or forbidden annihilations, or non-thermal production via the freeze-in mechanism. These possibilities typically imply considerably smaller couplings that are much harder to probe experimentally and therefore outside the scope of the present work.}

A key finding of our analysis is that these solutions only work in special regions of parameter space, which may face other 
constraints or require substantial tuning. To study this interplay of different constraints and parameters it becomes necessary to 
explore the full parameter space of sub-GeV DM models. This approach differs from the one commonly adopted in the literature~\cite{Izaguirre:2015yja,Alexander:2016aln,Berlin:2018bsc,Beacham:2019nyx}, 
where certain parameter combinations (such as the ratio of DM to dark photon mass or the dark fine-structure constant) 
are fixed to specific benchmark values. In many cases the most interesting parameter regions can only be found when departing 
from these benchmark points~\cite{Berlin:2020uwy}. 

The goal of the present work is to  perform global fits of scalar and fermionic sub-GeV DM particles coupled to dark photons in 
order to compare different models and identify the viable regions of parameter space. For this purpose we calculate likelihoods 
for a wide variety of results from laboratory experiments, astrophysical observations and cosmological data and then perform 
parameter scans over all model parameters simultaneously. The results can then be analysed in the framework of frequentist or 
Bayesian statistics, in order to determine which models are disfavoured or preferred by data. This process can be greatly 
simplified using the \gambit global fitting 
framework v2.5.0~\cite{GAMBIT:2017yxo,GAMBITDarkMatterWorkgroup:2017fax,GAMBITCosmologyWorkgroup:2020htv,Bloor:2021gtp}, 
which has previously been applied to various DM 
models~\cite{GAMBIT:2017gge,GAMBIT:2018eea,Hoof:2018ieb,GAMBIT:2021rlp,Chang:2022jgo,Chang:2023cki} and other BSM 
scenarios~\cite{Athron:2020maw,Balazs:2022tjl}.

\begin{figure}[t]
    \centering
    \includegraphics[width=0.49\textwidth]{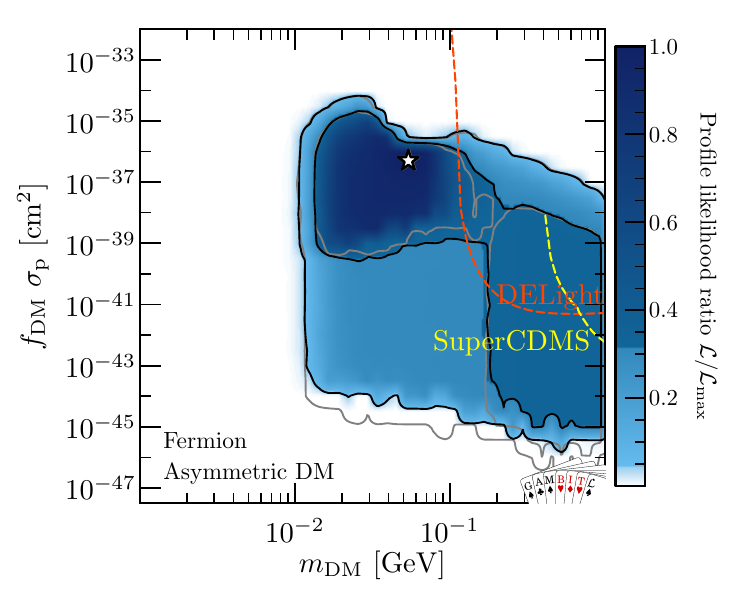}
    \includegraphics[width=0.49\textwidth]{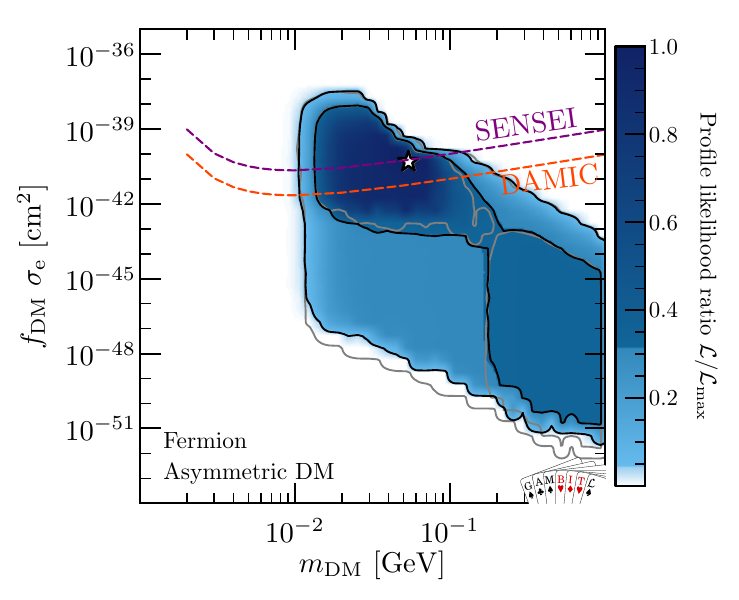}
    \includegraphics[width=0.49\textwidth]{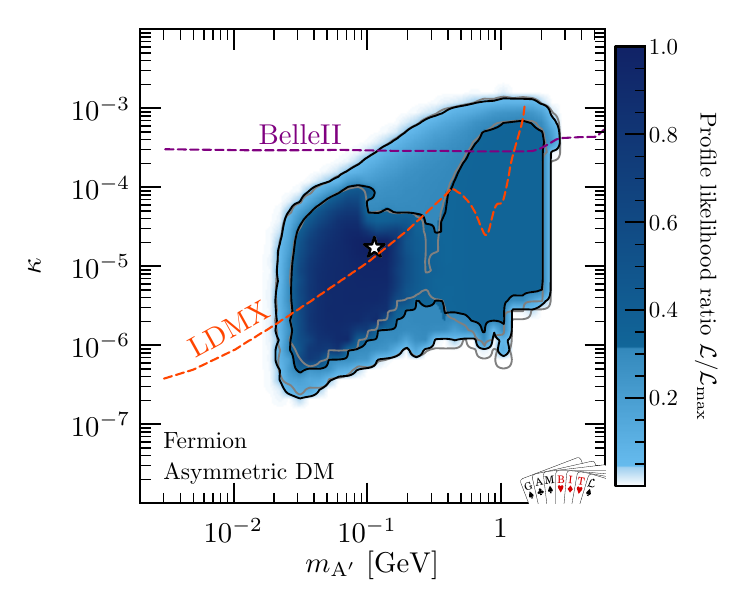}
    \includegraphics[width=0.49\textwidth]{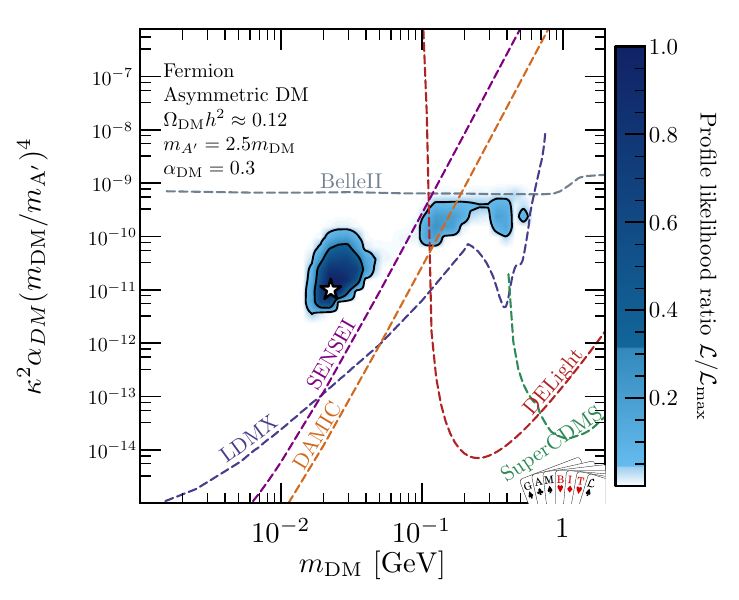}
    \caption{Allowed parameter regions for asymmetric fermionic DM, with the star indicating the best-fit point, compared to the projected sensitivities of various experiments.  The different panels show the rescaled DM-nucleus scattering cross section versus the DM mass (top-left), the rescaled DM-electron scattering cross section versus the DM mass (top-right) and the kinetic mixing parameter versus the dark photon mass (bottom-left). In the bottom-right panel we have fixed   $m_{A'} = 2.5 m_\text{DM}$ and $\alpha_\mathrm{DM} = 0.3$, and we show constraints in terms of the effective coupling $\kappa^2 \alpha_\mathrm{DM} (m_\mathrm{DM} / m_{A'}^4)$ versus the DM mass.}
    \label{fig:exec_summary}
\end{figure}

\paragraph{Executive summary.} For the case of fermionic DM, we find that some tuning of the parameters, in particular the mass ratio of dark photon and DM, is 
needed to satisfy all constraints. This tuning can be significantly relaxed when including an asymmetry, making it possible in 
particular to achieve large self-interaction cross sections that can be probed with astrophysical observations. From the Bayesian 
viewpoint, we find a substantial preference for the model with asymmetry over the symmetric model, which does however 
depend on the choice of priors. The scalar DM model naturally evades indirect detection constraints, leading to larger allowed 
parameter regions (and correspondingly larger Bayes factors) in the symmetric case.
For both asymmetric fermionic DM and symmetric scalar DM we find that the benchmark points most commonly used in the 
literature are already in considerable tension with data. We therefore propose a new benchmark point that is consistent will all 
current constraints and provides an attractive target for future experiments.

Our main findings are summarized in figure~\ref{fig:exec_summary}, which shows the allowed parameter regions for 
asymmetric fermionic DM in terms of different model parameters and observables compared to the projected sensitivities of 
various experiments {(the corresponding results for scalar DM will be presented in figure~\ref{fig:projections} and discussed in section~\ref{sec:discussion})}. In the bottom-right panel we have fixed two of the model parameters to the proposed benchmark values. 
Doing so leads to a more restricted parameter space, which can be fully probed by next-generation experiments. 

\paragraph{Outline.} The remainder of the work is structured as follows. In section~\ref{sec:models} we introduce the models of sub-GeV DM that we 
will study and derive the relevant cross sections and decay widths in terms of the model parameters. Section~\ref{sec:likelihoods} 
is dedicated to a detailed discussion of all likelihoods that constrain the parameter spaces under consideration. In 
section~\ref{sec:results}, we then describe the various parameter scans that we perform and present the corresponding results. We discuss the implications for future DM searches in section~\ref{sec:discussion}.
Our conclusions are summarized in section~\ref{sec:conclusions}. Additional technical details are provided in appendices~\ref{app:darksusy}--\ref{app:relcs}, while additional figures with results from our scans are provided in appendix~\ref{app:bayesian}. Details regarding the implementation in \gambit can be found in appendix~\ref{app:gambit}.

\section{Models of sub-GeV dark matter}
\label{sec:models}

We consider the gauge boson of a new $U(1)'$ gauge group, called the dark photon $A'$, which obtains a mass $m_{A'}$ via the 
Stueckelberg mechanism~\cite{Stueckelberg:1900zz}.\footnote{
It is of course also possible to generate the dark photon mass through a dark Higgs mechanism. This does not affect the 
phenomenology of dark photons as long as the dark Higgs boson is sufficiently heavy and sufficiently weakly coupled to the SM 
Higgs boson. The DM phenomenology, on the other hand, can become more complicated, if the dark Higgs field also couples to 
the DM particle, such that spontaneous symmetry breaking contributes to its mass. We do not consider this possibility further in 
the present work.}
Couplings of the dark photon to SM particles arise 
from kinetic mixing with the SM hypercharge field $B$~\cite{Holdom:1985ag,Babu:1997st}:
\begin{equation}
    \mathcal{L} \supset - \frac{\kappa}{2 \cos \theta_\mathrm{w}} \hat{A}^{\prime \mu \nu} \hat{B}_{\mu\nu} \,,
\end{equation}
where $X^{\mu\nu} = \partial^\mu X^\nu - \partial^\nu X^\mu$ with  $X = \hat{A}', \hat{B}$ is the field-strength tensor before 
diagonalisation, $\theta_\mathrm{w}$ denotes the weak mixing angle and $\kappa$ the kinetic mixing parameter.
After electroweak symmetry breaking, we can transform the fields to mass eigenstates with canonical kinetic terms. 
For $m_{A'} \ll m_Z$ the Lagrangian of the $A'$ is then given by
\begin{align}
\mathcal{L}_\text{int} = -\frac{1}{2} m_{A'}^2 A'^\mu A'_\mu - \frac{1}{4} A'^{\mu\nu}A'_{\mu\nu} -\kappa e A'^\mu \sum_{f} q_f \overline{f} \gamma_\mu f \,,
\end{align}
where $f$ denote the SM fermions and $q_f$ denotes their electric charges.

We furthermore consider a DM candidate with mass $m_\text{DM}<m_{A'}/ 2$ in the sub-GeV range, 
which can either be a complex scalar $\Phi$ or a Dirac 
fermion $\psi$, that couples to the dark photon. The corresponding additions to the Lagrangian are given by
\begin{align}
 \mathcal{L}_\Phi & = |\partial_\mu \Phi|^2 - m_\text{DM}^2 |\Phi|^2 + i g_\text{DM} A'^\mu \left[\Phi^\ast (\partial_\mu \Phi) - (\partial_\mu \Phi^\ast) \Phi\right] - g_\text{DM}^2 A'_\mu A'^\mu |\Phi|^2 \, ,\\
\mathcal{L}_\psi & = \bar{\psi}(i \slashed{\partial} - m_\text{DM}) \psi + g_\text{DM} A'^\mu \bar{\psi} \gamma_\mu \psi \,.
\end{align}
We note that in the absence of a dark Higgs mechanism, the couplings of a massive fermion $\psi$ must be vector-like, i.e.\ there is no axial-vector 
coupling. 

Since in both cases the DM particle is different from its anti-particle, there may be an asymmetry between their respective number densities, which we denote by~\cite{Lin:2011gj}
\begin{equation}
\label{eq:etadef}
 \eta_\text{DM} \equiv \frac{n_\chi - n_{\overline{\chi}}}{s} \,,
\end{equation}
where $\chi=\Phi,\psi$ and $s$ denotes the entropy density of the universe. In the absence of processes that violate entropy conservation or DM 
number conservation, $\eta_\text{DM}$ is constant throughout the evolution of the universe. The two models that we consider 
therefore have 5 parameters each: $m_{A'}$, $m_\text{DM}$, $\kappa$, $g_\text{DM}$ and $\eta_\text{DM}$.

The main difference between the two models lies in their annihilation cross section. For scalar DM, $s$-wave annihilation 
via an $s$-channel dark photon is forbidden because of angular momentum conservation,
which means that the leading contribution to the annihilation cross 
section in the non-relativistic limit scales with the DM velocity $v$ as $v^2$, such that indirect detection constraints are largely
absent~\cite{Boehm:2003hm,Boehm:2020wbt}. For Dirac DM, on the other hand, $s$-wave annihilation is allowed, such that the 
annihilation cross section scales as $v^0$, and indirect detection constraints exclude large parts of the parameter space where 
the observed DM relic abundance can be reproduced.

Nevertheless, there are three ways in which the fermionic model can evade indirect detection constraints. First, the annihilation 
cross section relevant for DM freeze-out in the early universe may be resonantly enhanced relative to the annihilation cross 
section relevant in the present universe if the resonance parameter
\begin{equation}
    \epsilon_R \equiv \frac{m_{A'}^2 - 4 m_\text{DM}^2}{4 m_\text{DM}^2}
    \label{eq:resparam}
\end{equation}
is much smaller than unity: $\epsilon_R \ll 1$~\cite{Feng:2017drg,Bernreuther:2020koj}. Second, in the presence of a particle-
antiparticle asymmetry, only the symmetric component can annihilate, which leads to a strong suppression of indirect detection 
signals if the asymmetric component dominates. Finally, we may consider the case where the DM particle under consideration 
constitutes only a fraction $f_\text{DM} < 1$ of the total DM abundance. 
In our scans we will comprehensively 
explore all three possibilities.

\subsection{Decays, annihilations and self-interactions}
\label{sec:decays}

In our model there are three competing decay modes of the dark photon: leptonic, hadronic and invisible decays. The leptonic 
decay width is given by $\Gamma_\text{lep} = \sum_\ell \Gamma_{\ell\ell}$ with $\ell = e, \mu, \tau$ and
\begin{align}
\label{eq:widthSM}
\Gamma_\mathrm{\ell\ell} & =  \frac{\kappa^2 e^2 m_{A^\prime}}{12\pi}\sqrt{1-\left(\frac{2m_\ell}{m_{A^\prime}}\right)^2}\left(1 + \frac{2m_\ell^2}{m_{A^\prime}^2}\right)\,.
\end{align}

The hadronic decay width can be expressed in terms of the ratio 
$R(\sqrt{s}) = \sigma(e^+ e^- \to \text{hadrons}) / \sigma(e^+ e^- \to \mu^+ \mu^-)$ via off-shell SM photons with 
centre-of-mass energy $\sqrt{s}=m_{A'}$~\cite{Ezhela:2003pp,Ilten:2018crw,Zyla:2020zbs}:
\begin{equation}
    \Gamma_\text{had} = R(m_{A'}) \Gamma_{\mu\mu} \,.
\end{equation}
For certain experimental constraints it may be interesting to split $\Gamma_\text{had}$ into different exclusive final states, the 
most relevant of which are $\pi^+ \pi^-$ and $K^+ K^-$. We obtain the corresponding branching ratios from 
\darkcast~\cite{Ilten:2018crw}, which provides a data base of different models of light gauge bosons that has been interfaced with \gambit.

Finally, the invisible decay width (to DM particles) is given by~\cite{Feng:2017drg}
\begin{align}
\Gamma_\mathrm{inv} &=  \begin{cases}
    \frac{g_\text{DM}^2 m_{A^\prime}}{48 \pi} \left(1 - \frac{4 m_\text{DM}^2}{m_{A'}^2}\right)^{3/2} & \text{scalar DM}\,, \\
    \frac{g_\text{DM}^2 m_{A^\prime}}{12\pi}\left(1-\frac{4m_\text{DM}^2}{m_{A^\prime}^2}\right)^{1/2}\left(1+\frac{2m_\text{DM}^2}{m_{A^\prime}^2}\right)  & \text{fermionic DM}\,.
    \end{cases}
\end{align}
From these contributions we can calculate the total decay width of the dark photon as
\begin{equation}
 \Gamma_{A'} = \Gamma_\text{lep} + \Gamma_\text{had} + \Gamma_\text{inv} \,.
\end{equation}

As we will discuss in more detail below, the kinetic mixing parameter $\kappa$ is experimentally constrained to be well below 
$10^{-3}$. The DM coupling $g_\text{DM}$, on the other hand, may be close to the perturbative bound 
$\alpha_\text{DM} = g_\text{DM}^2 / (4\pi) < 1$, i.e.\ $g_\text{DM} < \sqrt{4 \pi}$. If invisible decays are kinematically allowed, 
i.e.~if $m_{A'} > 2 m_\text{DM}$, we expect this decay mode to dominate the annihilation cross section with a branching ratio 
close to 100\%. In the opposite regime, where $m_{A'} < 2 m_\text{DM}$, the dark photon can only decay into SM final states. 
This latter case is constrained by a wide variety of fixed-target experiments and searches at low-energy colliders, which are 
difficult to reinterpret due to a complicated dependence of the constraints on the dark photon decay length. In the present work 
we therefore restrict ourselves to the former case. In terms of the resonance parameter introduced in eq.~\eqref{eq:resparam}, 
this implies $\epsilon_R > 0$.\footnote{%
For $\epsilon_R \ll 1$ it is conceivable that the visible branching ratio becomes non-negligible due to a strong phase-space 
suppression for the invisible decay. In our scans we will check explicitly for all parameter points that it is a good approximation to 
neglect laboratory searches for visible dark photon decays.
}

The DM annihilation cross section into charged lepton pairs is given by~\cite{Feng:2017drg}
\begin{equation}
\label{eq:sigmall}
(\sigma v)_{\ell\ell} =  \frac{g_\text{DM}^2 \kappa^2 e^2}{(s-m_{A'}^2)^2+m_{A'}^2  \Gamma_{A'}^2}  \frac{(2 m_\ell^2 + s)(y - 4)}{12 \pi (y - 2)} \sqrt{1- \frac{4 m_\ell^2}{s}}
\end{equation}
for scalar DM and 
\begin{equation}
\label{eq:sigmall}
(\sigma v)_{\ell\ell} = \frac{g_\text{DM}^2 \kappa^2 e^2}{(s-m_{A'}^2)^2+m_{A'}^2 \Gamma_{A'}^2} \frac{(2 m_\ell^2 + s) (y + 2)}{12 \pi (y - 2)} \sqrt{1- \frac{4 m_\ell^2}{s}} \,,
\end{equation}
for fermionic DM, where 
$y \equiv s/m_\text{DM}^2$. 
For the annihilation into hadrons, we proceed in analogy to the decay width discussed above and write
\begin{equation}
    \label{eq:sigmahad}
    (\sigma v)_\text{had} = (\sigma v)_{\mu\mu} \, R(\sqrt{s}) \,.
\end{equation}
In principle, DM particles could also annihilate into dark photon pairs, but for the spectrum that we consider 
($m_{A'} > 2 m_\text{DM}$) this decay mode is strongly suppressed and plays a negligible role in the calculation of the DM 
relic abundance. We also note that ref.~\cite{Mohlabeng:2024itu} recently showed that higher-order corrections to the annihilation cross section may be non-negligible for large values of $g_\mathrm{DM}$. Given other large theory uncertainties that enter the relic density calculation (see below), these corrections are not included in the present work.

To understand the importance of the resonance parameter $\epsilon_R$ introduced in eq.~\eqref{eq:resparam}, we define
\begin{equation}
    \epsilon = \frac{s - 4 m_\text{DM}^2}{4 m_\text{DM}^2}
\end{equation}
as a dimensionless measure of the kinetic energy available in an annihilation process. In the non-relativistic limit, $\epsilon = v_\text{DM}^2$, where $v_\text{DM}$ denotes the velocity of each DM particle in the CM frame. With this definition, the dark photon propagator becomes
\begin{equation}
    \frac{1}{(s - m_{A'})^2 + m_{A'}^2 \Gamma_{A'}^2} = \frac{1}{16 m_\text{DM}^4 (\epsilon - \epsilon_R)^2 + m_{A'}^2 \Gamma_{A'}^2} \; .
\end{equation}
Hence, we see that for $\epsilon \approx \epsilon_R$ the annihilation cross section receives a resonant enhancement, 
provided that $m_{A'} \Gamma_{A'}\ll m_\text{DM}^2$,
whereas the propagator becomes independent of $\epsilon$ for $\epsilon \ll \epsilon_R$.

\begin{figure}[t]
    \centering
    \includegraphics[width=0.6 \textwidth]{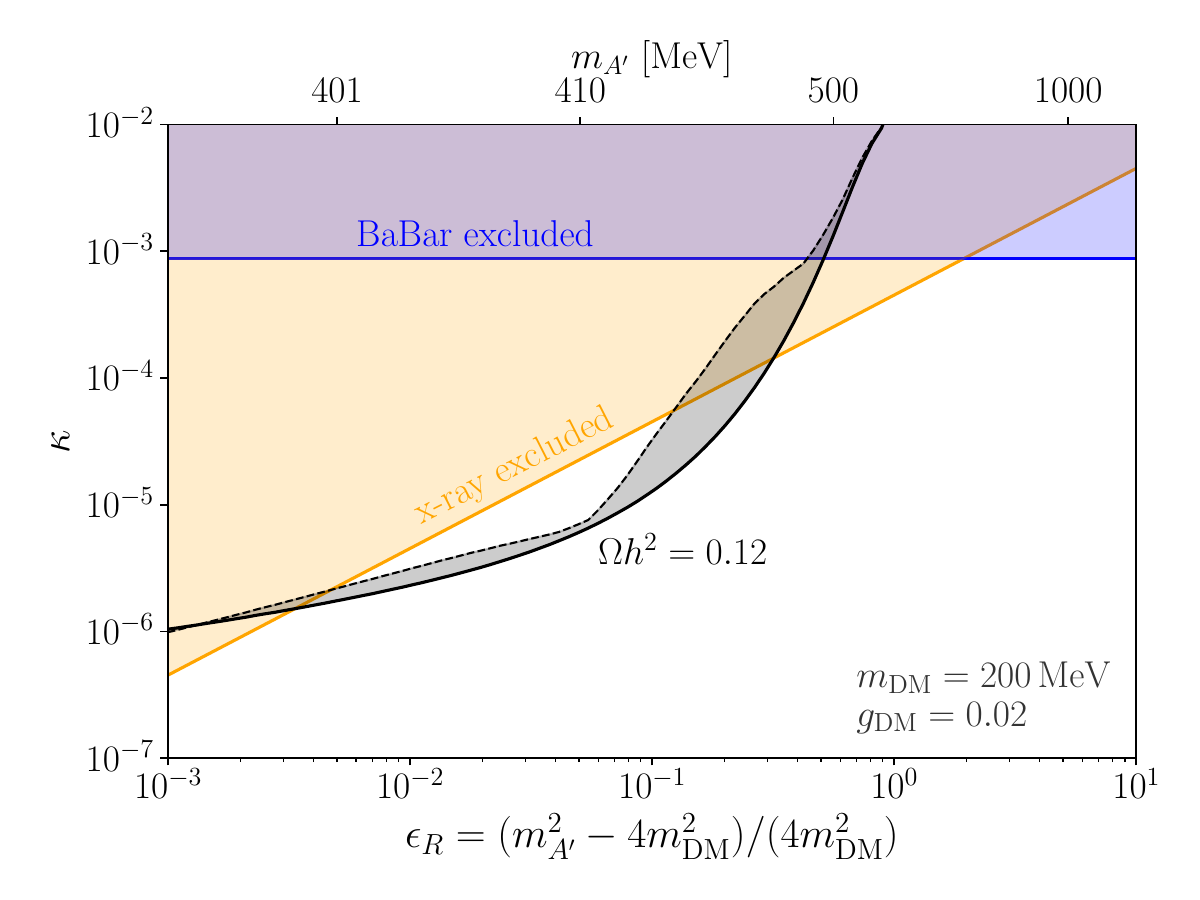}

    \caption{The black line shows the value of the kinetic mixing parameter $\kappa$ needed to reproduce the observed DM relic 
    abundance $\Omega_\text{DM} h^2 = 0.12$ as a function of $\epsilon_R$, for $m_\text{DM} = 200\,\mathrm{MeV}$ and 
    $g_\text{DM} = 0.02$, when adopting the standard way of calculating the relic density. The shaded region above this line 
    indicates our conservative estimate of the uncertainty on the standard relic density calculation close to very narrow 
    resonances, following ref.~\cite{Binder:2021bmg} (see also appendix~\ref{app:RD_cBE}). 
    For comparison we show the parameter regions excluded by 
    X-ray observations and by the missing energy search from BaBar. See section~\ref{sec:likelihoods} for details.
    }
    \label{fig:epsR}
\end{figure}

In the temperature range relevant for DM freeze-out, $\epsilon$ is typically of order $0.1$, 
whereas for indirect detection one can to very good approximation set $\epsilon \to 0$ 
(for DM particles bound to the Milky Way halo, e.g., we have $\epsilon\sim10^{-6}$). Thus, we require $\epsilon_R \sim 0.1$ in 
order for annihilations to maximally benefit from resonant enhancement during freeze-out without 
enhancing indirect detection constraints. 
Choosing $\epsilon_R \ll 0.1$, on the other hand, has the opposite effect and has been used as a means to {\it boost} 
indirect detection signals from thermally produced DM~\cite{Ibe:2008ye,Guo:2009aj}.
This is illustrated in figure~\ref{fig:epsR}, which shows the value of $\kappa$ required to reproduce the 
observed DM relic abundance as well as the bounds on $\kappa$ from indirect detection and missing energy searches as a 
function of $\epsilon_R$, 
for $m_\text{DM} = 200\,\mathrm{MeV}$ and $g_\text{DM} = 0.02$ (see section~\ref{sec:likelihoods} for details on both the 
relic density calculation and the constraints shown in the figure). We observe that for this parameter choice indirect detection 
constraints imply $3 \cdot 10^{-3} < \epsilon_R < 0.3$. If these constraints are absent (for example due to a non-zero 
asymmetry), the lower bound on $\epsilon_R$ disappears and the upper bound on $\epsilon_R$ is somewhat relaxed. Because 
of its central role in the phenomenology of sub-GeV DM, we will in the following use $\epsilon_R$ as an independent model 
parameter instead of $m_{A'}$.

\bigskip

The self-interactions that affect dynamics in astrophysical systems include those between $\chi \chi$, $\bar{\chi} \bar{\chi}$ and $\chi \bar{\chi}$. 
The differential elastic scattering cross sections for the fermionic DM model are calculated in the non-relativistic limit and are given by
\begin{align}
    \left(\dv{\sigma}{\Omega}\right)_{\psi\psi} = \ \sigma_0, \quad
    \left(\dv{\sigma}{\Omega}\right)_{
    \psi\bar{\psi}} = \ \sigma_0 &\left[1 + \frac{12w^2}{(1-4w^2)^2 + \Gamma_{A'}^2/m_{A'}^2} \right] .
\end{align}
where $\sigma_0 \equiv (\alpha_{\text{DM}}^2 m_{\text{DM}}^2)/m_{A'}^4$ and $w \equiv m_{\text{DM}}/m_{A'}$. Note that we have made use of the Born approximation, which is accurate for sufficiently large dark photon masses and perturbative couplings.
The differential cross-sections for the scalar DM model in the non-relativistic limit are given by,
\begin{align}
    \left(\dv{\sigma}{\Omega}\right)_{\Phi \Phi} = \ 4\sigma_0 , \quad
    \left(\dv{\sigma}{\Omega}\right)_{
    \Phi \Phi^\ast} = \ \sigma_0\,. 
\end{align}
Note that for scalar DM and in the non-relativistic limit, the $t$-channel dominates over the $s$-channel even in the resonance 
region.

\subsection{Parameter ranges and priors}
\label{sec:priors}
   
To conclude this section, let us briefly state the parameter regions that we will explore in our scans, summarized in table~\ref{tab:parameters}. By choice we restrict 
ourselves to sub-GeV DM particles, i.e.\ $m_\text{DM} < 1 \, \mathrm{GeV}$. A practical reason for this specific upper bound is that there is currently no tool or method to accurately predict the injection spectra from DM 
annihilations for $1 \, \mathrm{GeV} < m_\text{DM} < 5 \, \mathrm{GeV}$. For the lower bound we take $m_\text{DM} > 1 \, \mathrm{MeV}$, well below the bound imposed by cosmological constraints. For 
the kinetic mixing parameter, a lower bound of $\kappa > 10^{-8}$ is chosen to ensure that dark and visible sector remain in 
kinetic equilibrium during freeze-out~\cite{Evans:2017kti}, whereas the upper bound is set to $\kappa < 10^{-2}$, well 
above the experimental constraints. For the DM coupling, on the other hand, we take a lower bound of $10^{-2}$, such that 
$g_\text{DM} > \kappa$ everywhere in our parameter space and the dark photon invisible width dominates over the visible one,
and impose the perturbativity bound 
$g_\text{DM} < \sqrt{4\pi}$ (see above). Finally, we vary the resonance parameter $\epsilon_R$ in the range $[10^{-3}, 8]$, which 
corresponds to $2 m_\text{DM} < m_{A'} < 6 m_\text{DM}$, where the upper bound is chosen to lie well above the parameter range allowed by the relic density requirement (see below).

\begin{table}[tp]
\centering
  \caption{List of model parameters and their ranges. For frequentist scans, the prior is only used to determine the sampling strategy. {Our scans also include several nuisance parameters as discussed in the text. The likelihoods that we consider are presented in section~\ref{sec:likelihoods}  and summarized in appendix~\ref{app:gambit}}.\label{tab:parameters}}
  \vspace{2mm}
   \begin{tabular}{lllll}
   \hline
   \hline\\[-4mm]
   \bf Parameter name & \bf Symbol & \bf Unit & \bf Range & \bf Prior \\[1pt]
   \hline\\[-4mm]
   Kinetic mixing & $\kappa$ & -- & $[10^{-8},10^{-2}]$ & logarithmic \\[5pt]
   Dark sector coupling & $g_\text{DM}$ & -- & $[10^{-2},\sqrt{4\pi}]$ & logarithmic \\[5pt]
   Asymmetry parameter & $\eta_\text{DM}$ & -- & $[0,10^{-9}\,\mathrm{GeV}/m_\mathrm{DM}]$& linear \\[5pt]
      Dark matter mass & $m_\text{DM}$ & MeV & [1,1000] & logarithmic \\[2pt]
      \hline\\[-4mm]
   Dark photon mass & $m_{A'}$ & MeV & [2,6000] with $m_{A'} \geq 2 m_\text{DM}$ & logarithmic \\
    \qquad \qquad \emph{or} \\
   Resonance parameter & $\epsilon_R$ & -- & $[10^{-3}$,8] & logarithmic \\[2pt]
   \hline
   \hline
   \end{tabular}
\end{table}
For the asymmetry parameter, we consider the range 
\begin{equation}
    \label{eq:m_eta}
    0 \leq \frac{m_\text{DM}}{1 \, \mathrm{GeV}} \eta_\text{DM} < 10^{-9} \, ,
\end{equation}
where the lower bound corresponds to the fully symmetric case and the upper bound is obtained from the observation that the 
asymmetric component alone should not overclose the universe. The chosen parameter range for $\eta_\text{DM} m_\text{DM}$ implies $\Omega_\text{DM} h^2 < 0.275$ for the total 
cosmological DM density, which extends well beyond the observational bound $\Omega_\text{DM,obs} h^2 \leq 0.12$, see~section \ref{sec:rd}.

For our Bayesian scans, we need to specify the prior probabilities in addition to the prior ranges. Since the couplings and DM 
mass span several orders of magnitude with no preferred scale, we choose logarithmic priors. The parameter $\epsilon_R$ 
however is not a fundamental parameter and therefore not suitable for defining the prior. In our Bayesian scans, we therefore 
instead take a logarithmic prior on $m_{A'}$ between $2 \, \mathrm{MeV}$ and $6 \, \mathrm{GeV}$, removing all points with 
$m_{A'} < 2 m_\text{DM}$ at the prior level. We note that this introduces a bias towards smaller DM masses in the scans, for which there is a larger prior volume in $m_{A'}$.
Finally, when including the asymmetry parameter, we choose a flat prior on $\eta_\text{DM} m_\text{DM}$. The reason for 
choosing this particular combination of parameters is that it directly enters the relic density calculation, such that the prior range 
can be restricted by the observed DM relic abundance, see section~\ref{sec:rd}. A flat prior is chosen such that the symmetric 
case $\eta_\text{DM} = 0$ is included in the scan but does not dominate the prior volume.

Finally, we include three nuisance parameter in our scans, namely the local DM density $\rho_0$ and the velocity dispersion $v_0$ and escape velocity $v_\text{esc}$ of the local DM velocity distribution in the Standard Halo Model. Following previous \gambit studies, we allow for $\rho_0$ to take a rather broad range of values to reflect the spread of different results from the 
literature, rather than the quoted uncertainty of any individual measurement. Specifically, we take a log-normal distribution with 
mean $\mu = 0.4 \, \mathrm{GeV \, cm^{-3}}$ and spread $\sigma = 0.15  \, \mathrm{GeV \, cm^{-3}}$. To avoid extreme values, 
the parameter range is restricted to $0.2 \, \mathrm{GeV \, cm^{-3}} \leq \rho_0 \leq 0.8 \, \mathrm{GeV \, cm^{-3}}$. The velocity parameters are constrained through measurements to $v_0 = 240 \pm 8 \, \mathrm{km \, s^{-1}}$~\cite{Reid:2014boa} and $v_\text{esc} = 528 \pm 25 \, \mathrm{km \, s^{-1}}$~\cite{Deason:2019tpl}. All other astrophysical parameters (such as the peculiar motion of the Sun) are fixed to the default parameters of the respective likelihood codes as described in ref.~\cite{GAMBITDarkMatterWorkgroup:2017fax}.

\section{Constraints and likelihoods}
\label{sec:likelihoods}

In this section we discuss the relevant constraints on the models under consideration, and how we obtain and implement the 
corresponding likelihoods. The constraints can be divided into four categories: cosmological constraints 
(section~\ref{sec:cosmo}) from the CMB and from 
BBN, astrophysical constraints  (section~\ref{sec:astro})  from X-rays and observations of the Bullet cluster, 
accelerator constraints  (section~\ref{sec:acc}) from beam-dumps and 
electron-positron colliders and direct detection constraints  (section~\ref{sec:dd})  from searches for electron and nuclear recoils.

\subsection{Cosmological constraints}
\label{sec:cosmo}

\subsubsection{Relic density}
\label{sec:rd}

To calculate the DM relic density from thermal freeze-out, we pass the annihilation cross sections from eqs.~\eqref{eq:sigmall} 
and \eqref{eq:sigmahad} to \darksusy v6.4~\cite{Bringmann:2018lay}, which performs the thermal averaging, taking special care 
of centre-of-mass energies close to the dark photon mass and hadronic resonances. DarkSUSY then solves the Boltzmann 
equation describing the number density of DM particles (including a potential asymmetry $\eta_\text{DM}$, see 
appendix~\ref{app:darksusy} for details about this newly added feature) in order to return the 
present-day abundance of DM particles and anti-particles $\Omega_\chi h^2$ and $\Omega_{\overline{\chi}} h^2$,
where $\chi=\phi,\psi$.
The sum of these numbers can then be compared to the Planck measurement~\cite{Planck:2018vyg}
\begin{equation}
    \Omega_\text{DM,obs} h^2 = 0.120 \pm 0.001
\end{equation}
in order to define the relative cosmological abundance
\begin{equation}
    f_\text{DM} \equiv \frac{\Omega_\text{DM} h^2}{\Omega_\text{DM,obs} h^2} \equiv \frac{\Omega_\chi h^2 + \Omega_{\overline{\chi}} h^2}{\Omega_\text{DM,obs} h^2} \,.
\end{equation}

In our analysis we will consider two possible interpretations of this result. The first is that the predicted abundance must match 
observations, i.e.\ $f_\text{DM} = 1$ within observational and theoretical uncertainties.\footnote{
For the theoretical uncertainty we assume $10\%$. This is significantly larger than the precision achieved by DarkSUSY,
for the annihilation cross sections of our model,
when solving the standard Boltzmann equation~\cite{Gondolo:1990dk} 
adapted to the case with $\eta_{\rm DM}>0$. In the presence of very narrow resonances as in our models, however, 
the standard Boltzmann equation can underestimate the
relic density by a much larger factor~\cite{Binder:2017rgn,Binder:2021bmg}. 
As already indicated in figure~\ref{fig:epsR}, this is particularly relevant for $\epsilon_R\sim\mathcal{O}(0.1)$.
We refer to appendix~\ref{app:RD_cBE} for a more detailed
discussion, demonstrating that these complications do not affect our results and conclusions based on {\it scans} over the 
parameter space (which we perform by using the numerically much faster 
standard approach, deliberately choosing a `too large' overall theoretical uncertainty in view of this discussion).}
The second option is to allow for the possibility that the DM particles under consideration constitute only a 
DM sub-component, i.e.~we implement a one-sided likelihood that penalises $f_\text{DM} > 1$ according to the observational 
and theoretical uncertainty, but does not penalise $f_\text{DM} < 1$. If the observed DM abundance is not saturated, we rescale 
the local DM density 
$\rho_\text{DM}$ under the assumption that $f_\text{DM}$ is constant everywhere, i.e.
\begin{equation}
 \rho_\text{DM} \to f_\text{DM} \rho_\text{DM} \,.
\end{equation}
Constraints on the DM-nucleon and DM-electron scattering cross sections from direct detection experiments are therefore relaxed by a factor $f_\text{DM}^{-1}$.

For observations that probe the DM annihilation cross sections, we furthermore need to account for the particle-antiparticle 
asymmetry. For this purpose we define the symmetric DM fraction as\footnote{%
Here we assume that $\eta_\text{DM}$ is positive, i.e.\ that we define particles to be the more abundant component and 
anti-particles to be the less abundant component.
}
\begin{equation}
    f_\text{sym} \equiv \frac{2 \, \Omega_{\overline{\chi}} h^2}{\Omega_\text{DM,obs} h^2} 
    = f_\text{DM}-\eta_\text{DM} m_\text{DM} \frac{s_0 h^2}{\rho_\text{DM,obs} h^2} 
    \leq f_\text{DM}\,,
\end{equation}
where $s_0/\rho_\text{DM,obs}\approx 2.755\times10^{8}\, (\Omega_\text{DM,obs} h^2)^{-1}\,\text{GeV}^{-1}$. 
We note that the asymmetry parameter {introduced in eq.~\eqref{eq:etadef}} must satisfy the inequality
\begin{equation}
\eta_\text{DM} \leq \frac{4.33 \times 10^{-10} \, \mathrm{GeV}}{m_\text{DM}} \equiv \eta_\text{asym}(m_\text{DM}) \, ,
\label{eq:eta_asym}
\end{equation}
where $\eta_\text{asym}$ denotes the value of $\eta_\text{DM}$ for which the asymmetric component alone saturates the relic density.
Compared to the fully symmetric case, indirect detection constraints are relaxed by a factor
\begin{equation}
    \xi_\text{sym} \equiv  \frac{n_\chi  n_{\overline{\chi}}}{\frac{1}{4}(n_\chi + n_{\overline{\chi}})^2} = 
    \frac{f_\text{sym}}{f_\text{DM}}\left(2-\frac{f_\text{sym}}{f_\text{DM}}\right),
\label{eq:xi_sym}
\end{equation}
which may be tiny if the symmetric component efficiently annihilates away~\cite{Graesser:2011wi,Bell:2014xta}.
Of course, indirect detection rates still scale with the {\it local} DM density as $\rho_{\rm DM}^2$ 
(as long as both $f_\text{DM}$ and $\xi_\text{sym}$ are constant everywhere).

\begin{figure}[t]
    \centering
    \includegraphics[width=0.49\textwidth]{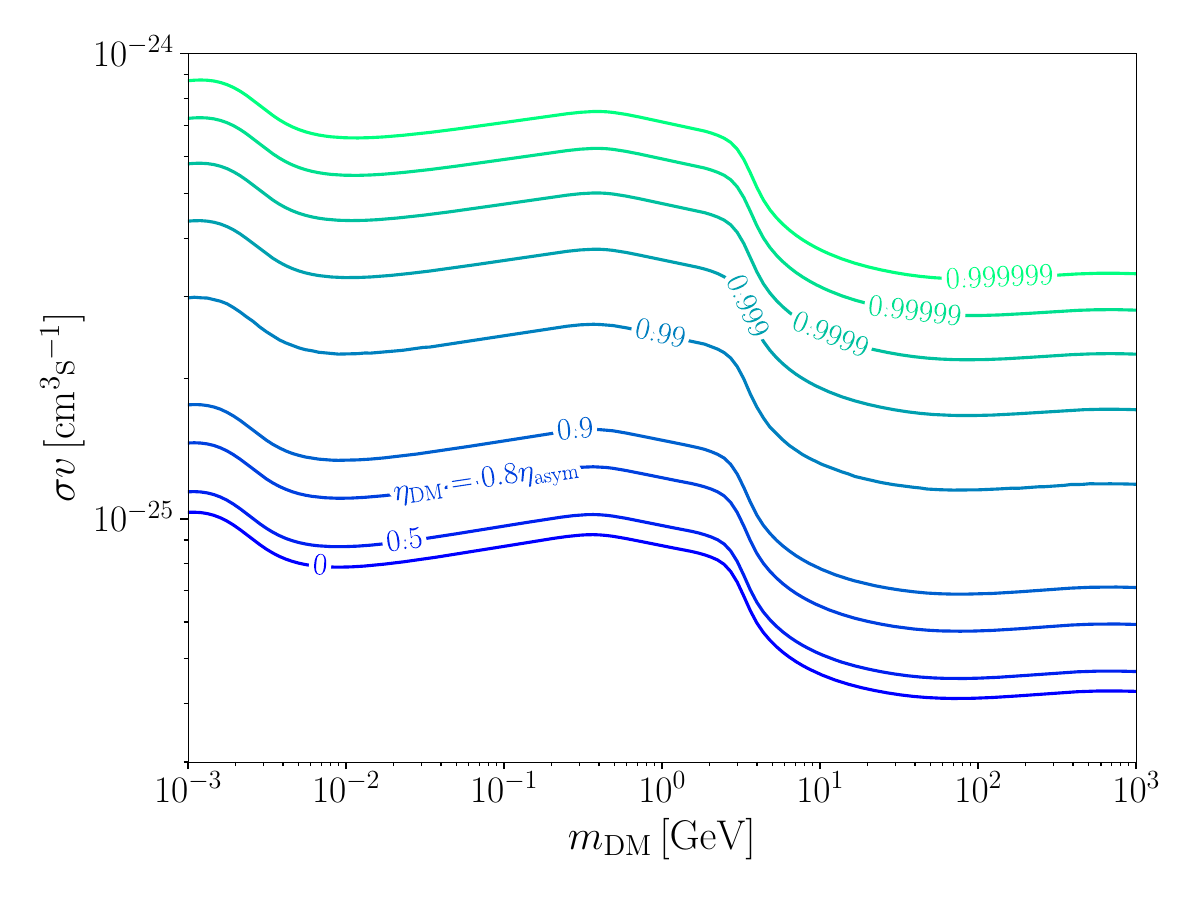}    \includegraphics[width=0.49\textwidth]{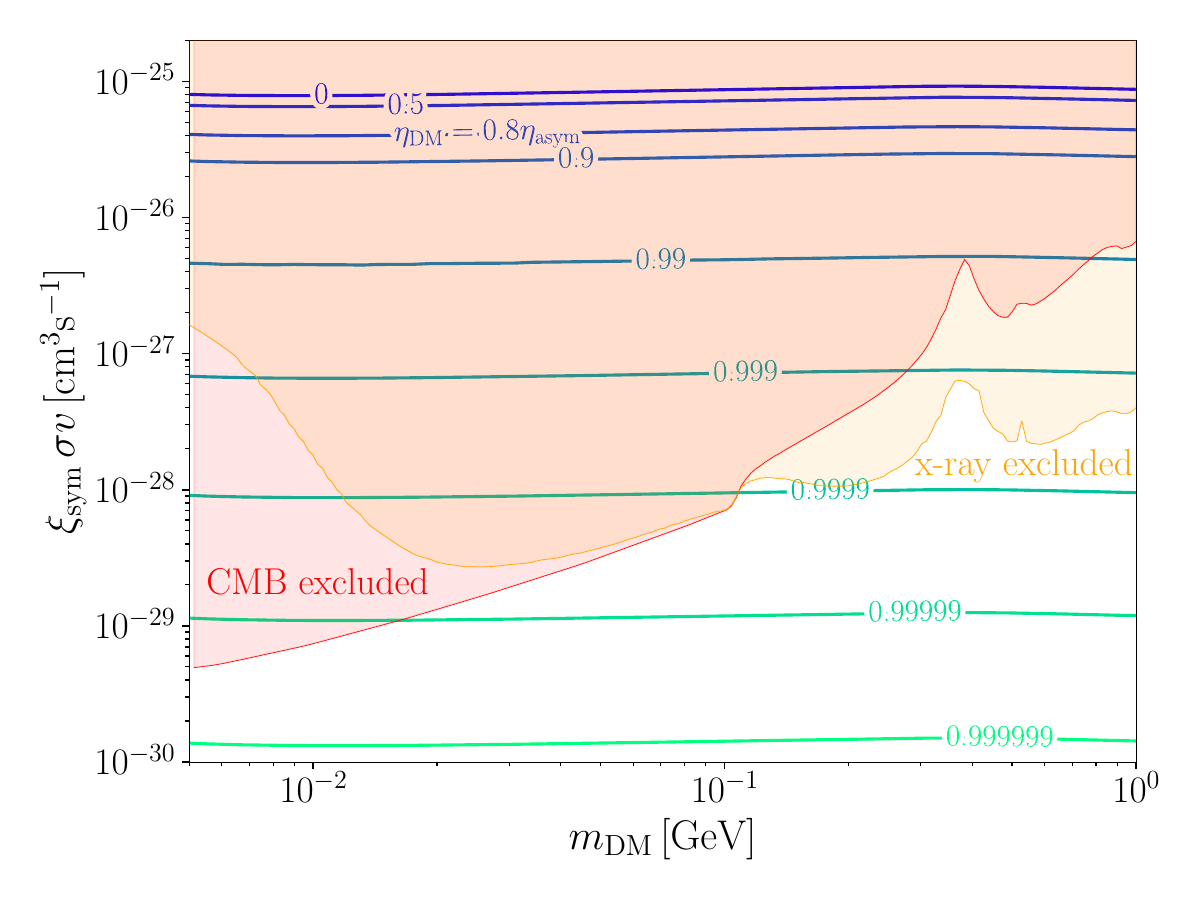}
\caption{Left: constant annihilation cross section $\sigma v$ required to reproduce the observed DM abundance 
$\Omega_\text{DM} h^2 = 0.12$ as a function of the DM mass for different values of the asymmetry parameter $\eta_\text{DM}$, 
with $\eta_\text{asym}$ defined in eq.~\eqref{eq:eta_asym}. {Relic density calculations have been performed with \darksusy v6.4~\cite{Bringmann:2018lay}}. Right: corresponding rescaled annihilation cross section 
$\xi_\text{sym}\times \sigma v$ relevant for indirect detection, see eq.~\eqref{eq:xi_sym}, compared to the constraints on the fermionic 
DM model from X-rays and the CMB.}
    \label{fig:relic}
\end{figure}

In figure~\ref{fig:relic} (left panel) we show the (velocity-independent) annihilation cross section needed to reproduce the 
observed DM relic abundance $\Omega h^2 = 0.12$ for different values of $\eta_\text{DM}$. We find that $\eta_\text{DM}$ 
needs to be very close to $\eta_\text{asym}$ in order to allow for annihilation cross sections well above the standard 
thermal cross section (corresponding to $\eta_\text{DM} = 0$).
{For a given value of $\eta_\text{DM}$, the mass dependence of these curves reflects as usual the different number of 
relativistic degrees of freedom in the SM heat bath during DM freeze-out.}
 In the right panel we show the corresponding effective annihilation cross section 
$\xi_\text{sym} \times \sigma v$ that determines the annihilation rate relevant for indirect detection constraints. 
For comparison, we also indicate the parameter regions excluded by CMB constraints and X-ray constraints, see below, for the 
fermionic DM  model.

\subsubsection{Relativistic degrees of freedom}

DM particles can either increase or decrease the effective number of relativistic degrees of freedom $N_\text{eff}$, depending on 
their properties: if their mass is close to the MeV scale, they may still be semi-relativistic at the time when neutrinos freeze out, 
giving a non-negligible contribution to the (radiation) energy density and thus increasing $N_\text{eff}$. On the other hand, 
residual DM annihilations into electrons after neutrino decoupling would raise the photon temperature relative to the one of 
neutrinos, and as a result reduce $N_\text{eff}$.

To calculate the impact of these effects on the abundances of light elements, we make use of the \gambit interface with 
\alterbbn v2.2~\cite{Arbey:2018zfh}, which provides routines to study the effects of DM annihilations. More concretely, \alterbbn returns the abundances of $^4$He and 
D, which are then compared to observations in order to calculate the likelihood 
{as described in the \cosmobit documentation}~\cite{GAMBITCosmologyWorkgroup:2020htv} (see also ref.~\cite{Balazs:2022tjl} for a discussion of the most recent data). 
AlterBBN also returns the neutrino temperature at the end of BBN, which can be translated into the value of $N_\text{eff}$ 
relevant for recombination.

{For the CMB constraint on $N_\text{eff}$, we adopt the
likelihood} provided by the Planck collaboration~\cite{Planck:2018vyg}:
\begin{equation}
    N_\text{eff} = 2.99 \pm 0.17\,,
\end{equation}
using the combination  TT,TE,EE+lowE+lensing
+BAO. We find that in practice the constraints from BBN and CMB are very similar and exclude DM masses below about 10 MeV, {in agreement with ref.~\cite{Sabti:2019mhn}}.

\subsubsection{Exotic energy injection}
\label{sec:energy_injection}

Residual DM annihilations can lead to an injection of exotic energy into the photon-baryon plasma that may spoil the successful 
predictions of recombination. The rate of injected energy per unit volume depends on the effective parameter~\cite{Slatyer:2015kla}
\begin{equation}
p_\text{ann} = \frac{f_\text{DM}^2 \xi_\text{sym}}{2}f_\text{eff} \frac{(\sigma_\text{ann} v)_0}{m_\text{DM}} \,.
\end{equation}
Here the subscript $0$ indicates the $v \to 0$ limit, which is a good approximation for the temperature range relevant for 
recombination.

The main challenge is to accurately calculate the fraction $f_\text{eff}$ of energy deposited in the plasma, which depends both 
on the DM mass and the final state. Fortunately, the public code \hazma v2.0~\cite{Coogan:2019qpu,Coogan:2022cdd} provides 
the yields of injected $\gamma$-rays and positrons for annihilations via a vector mediator with kinetic mixing as a function of 
the DM mass. These spectra have been obtained under the assumption of vector meson dominance from dedicated simulations 
of hadron production in DM annihilations using \textsc{HERWIG4DM}~\cite{Plehn:2019jeo}. We have tabulated these spectra 
and {made them available in \darksusy, as part of the v6.4 release,} in order to provide them 
to \darkages~\cite{Stocker:2018avm} and calculate $f_\text{eff}$ at 
every point in parameter space (see ref.~\cite{GAMBIT:2021rlp} for further details).

One could in principle calculate the impact of a given $p_\text{ann}$ on the CMB temperature anisotropies while varying the 
cosmological parameters. However, there are no significant degeneracies {between $p_\text{ann}$ and other parameters}, 
which makes it possible to calculate a marginalised 
Planck likelihood that depends only on $p_\text{ann}$. Such a likelihood has been obtained in ref.~\cite{GAMBIT:2021rlp} and 
can be directly used in the present context. The resulting constraint is shown in the right panel of 
figure~\ref{fig:relic}. The various features above 100 MeV are a direct consequence of the varying hadronic branching ratio of 
the dark photon, which leads to variations in $f_\text{eff}$.

Finally, we note that for the smallest DM masses that we consider, there may be relevant constraints from DM annihilations after 
the end of BBN, leading to the photodisintegration of deuterium and helium-3~\cite{Depta:2019lbe}. However, it was argued in 
ref.~\cite{Bernreuther:2020koj} that this constraint is only relevant for parameter points with very large resonant enhancement 
(corresponding to $\epsilon_R < 10^{-3}$)~\cite{Braat:2024khe}. These parameter regions are not the focus of the present study, and we therefore 
leave a detailed study of constraints from photodisintegration to future work.

\subsection{Astrophysical constraints}
\label{sec:astro}

\subsubsection*{X-ray constraints}

Assuming that DM is not fully asymmetric, it may annihilate to SM particles, producing detectable gamma-ray or X-ray signals. 
Annihilation directly to photons would lead to a distinctive monochromatic line signal at the DM mass, while annihilation to quarks, 
leptons and mesons can result in a broad photon spectrum resulting from the subsequent particle fragmentation and decay. 
However, gamma-ray searches for 
sub-GeV dark matter are limited by the so-called ``MeV gap" between keV-range experiments like SPI aboard the INTEGRAL 
satellite~\cite{2003A&A...411L..63V}, and Fermi-LAT at GeV energies~\cite{Fermi-LAT:2009ihh}. Numerous experiments have 
been proposed to fill in this gap in coverage; see ref.~\cite{Aramaki:2022zpw} for a detailed review. However, for the models that we consider in this work, in which DM particles annihilate primarily into charged leptons and mesons, while neutral mesons -- and hence photons -- are sub-dominant, MeV gamma-ray telescopes are not expected to improve on the CMB constraints discussed above~\cite{ODonnell:2024aaw}.

{There are, however, also secondary}
 signals of dark matter annihilation: 
 {energetic charged annihilation products can undergo inverse Compton scattering on}
 starlight or CMB photons, thereby producing keV-scale X-rays. Thus, data from INTEGRAL can be used to constrain 
 MeV-scale dark matter by searching for this flux of 
upscattered photons, as studied in ref.~\cite{Cirelli:2020bpc}. This analysis was extended in ref.~\cite{Cirelli:2023tnx} to 
include data from not just INTEGRAL~\cite{2011ApJ...739...29B}, but also 
NuSTAR~\cite{Mori:2015vba, Hong:2016qjq, Roach:2019ctw, Krivonos:2020qvl}, 
XMM-Newton~\cite{Dessert:2018qih, Foster:2021ngm}, and Suzaku~\cite{2009PASJ...61..805Y}. The resulting analysis places 
stringent limits on dark matter annihilation to electrons down to 1 MeV, and annihilation to muons and charged pions down to the 
respective threshold. The authors of ref.~\cite{Cirelli:2023tnx} have generously provided the 
likelihoods used in their 
analysis for each of these annihilation channels. We do not attempt to combine these likelihoods, i.e.~we conservatively
only include the 
final state giving the strongest constraint, which is typically $e^+ e^-$.

We note that ref.~\cite{DelaTorreLuque:2023olp} recently updated the results of ref.~\cite{Cirelli:2023tnx} using a more realistic 
treatment of cosmic ray propagation {including spatial diffusion and reacceleration}, and ref.~\cite{DelaTorreLuque:2023cef} studied additional constraints obtained from INTEGRAL measurements of the 511 keV line. {While the resulting limits can be substantially stronger, especially at low mass, they depend on the model and parameters used to describe cosmic ray transport, which carry large uncertainties}. {We therefore conservatively} use the earlier result~\cite{Cirelli:2023tnx} in the present work. These constraints are still stronger than the ones obtained from Voyager 1~\cite{Boudaud:2016mos}.

The resulting constraint is 
shown in the right panel of figure~\ref{fig:relic}. Similar to  the CMB case discussed above, we can see the various features 
resulting from the hadronic branching ratio of the 
dark photon. 

\subsubsection*{Bullet cluster constraints}

Merging galaxy cluster 1E0657-56, otherwise known as the Bullet Cluster, serves as one of the best test beds for (velocity-independent) DM 
self-interactions. It comprises a subcluster that passed through the main cluster's central region along the plane of the sky. X-ray 
and lensing observations have identified gas and mass {distributions with well-separated peaks}, 
implying 
that feebly interacting DM {dominates the clusters' masses}~\cite{Clowe:2003tk,Clowe:2006eq,Bradac:2006er,Paraficz:2012tv}. 
The presence of DM self-interactions would lead to the subcluster DM component experiencing friction as it passes through the 
main cluster. This would 
result in decelerating the DM component of the subcluster, such that it lags behind the collisionless galaxies, and cause the 
subcluster to lose part of its DM mass.

These observables can be used to constrain self-interaction. The offset between DM and galaxy centroids has been used to 
place the limit $\sigma_0/m_{\text{DM}} < 1.25\, \unit{cm^2 g^{-1}}$~\cite{Randall:2008ppe}, although 
ref.~\cite{Robertson:2016xjh} pointed out that the measured offsets are quite sensitive to the methods used, and more 
recent analyses~\cite{Wittman:2017gxn} have obtained slightly weaker bounds. Ref.~\cite{Randall:2008ppe} also used 
subcluster survival to place the most stringent constraint, $\sigma_0/m_{\text{DM}} < 0.7\, \unit{cm^2 g^{-1}}$. The subcluster 
mass loss is determined by comparing the observed mass-to-light ratios ($\gamma$) of the main and the subcluster assuming 
that they start out with identical $\gamma$. However, it has been pointed out that $\gamma$ increases slowly with total cluster 
mass~\cite{Holland:2015dia} and that there is significant scatter in the mass-luminosity 
relationship~\cite{Popesso:2006uv,Proctor:2015bua,Tempel:2017dhe}. 

In this work, we constrain DM self-interactions by investigating the subcluster survival rather than the DM-galaxy offset. The 
reason is that the latter can only be studied using numerical simulations, whereas for the former it is possible to derive an 
analytic estimate for the \textit{evaporation rate}~\cite{Kahlhoefer:2013dca}, which can be directly applied to asymmetric 
DM and DM sub-component to predict the fraction of total mass lost ($\Delta_\text{M}$) by the subcluster , see appendix~\ref{app:bc_constraints} for details. The theoretical prediction for the final subcluster 
mass-to-light ratio, $\gamma^{\text{sub}}_\text{f} = \gamma^{\text{sub}}_\text{i} (1-\Delta_\text{M}, $) is then compared to the measured subcluster mass-to-light ratio, $\gamma^{\text{sub}}_\text{obs} = 179 \pm 11$~\cite{Randall:2008ppe}. The novelty in our constraint 
is that we choose to treat the initial subcluster mass-to-light ratio $\gamma^{\text{sub}}_\text{i}$ as a nuisance parameter and calculate a marginalized likelihood,
\begin{align}
    \log \mathcal{P} \propto& \log \int \dd \gamma^{\text{sub}}_\text{i} \mathcal{L}(\gamma^{\text{sub}}_\text{f}) \Pi(\gamma^{\text{sub}}_\text{i}) \nonumber \\ 
    \propto& \log \sum_n \exp \left(-\frac{1}{2}\frac{(\gamma^{\text{sub}}_\text{obs} - \gamma^{\text{sub}}_\text{i,n} (1-\Delta_\text{M}))^2}{({\sigma^{\text{sub}}_\text{obs})^2 + \sigma_\text{theory}^2}}\right)
\end{align}
We have implemented two different priors: (\textit{i}) a Gaussian prior centred around the measured main cluster mass-to-light ratio, $\gamma^{\text{main}}_\text{obs} = 214 \pm 13$~\cite{Randall:2008ppe}, and (\textit{ii}) a log-normal prior 
fitted to the $i$-band sample extracted from ref.~\cite{Proctor:2015bua}. The likelihoods are constructed using mass-to-light 
ratios of the subcluster and main cluster within $150\, \unit{kpc}$ of the total mass peak, in the photometric $i$-band, taken from 
ref.~\cite{Randall:2008ppe}. The Gaussian prior assumes that the subcluster and main cluster $\gamma$'s are initially 
correlated. Since the measured values are quite close to each other, this approach leads to strong constraints. On the other 
hand, the log-normal prior makes no assumption of such a correlation and allows much larger initial subcluster mass-to-light 
ratios, thus leading to an {overly} conservative constraint. 

\begin{figure}[t]
    \centering
    \includegraphics[scale=0.6]{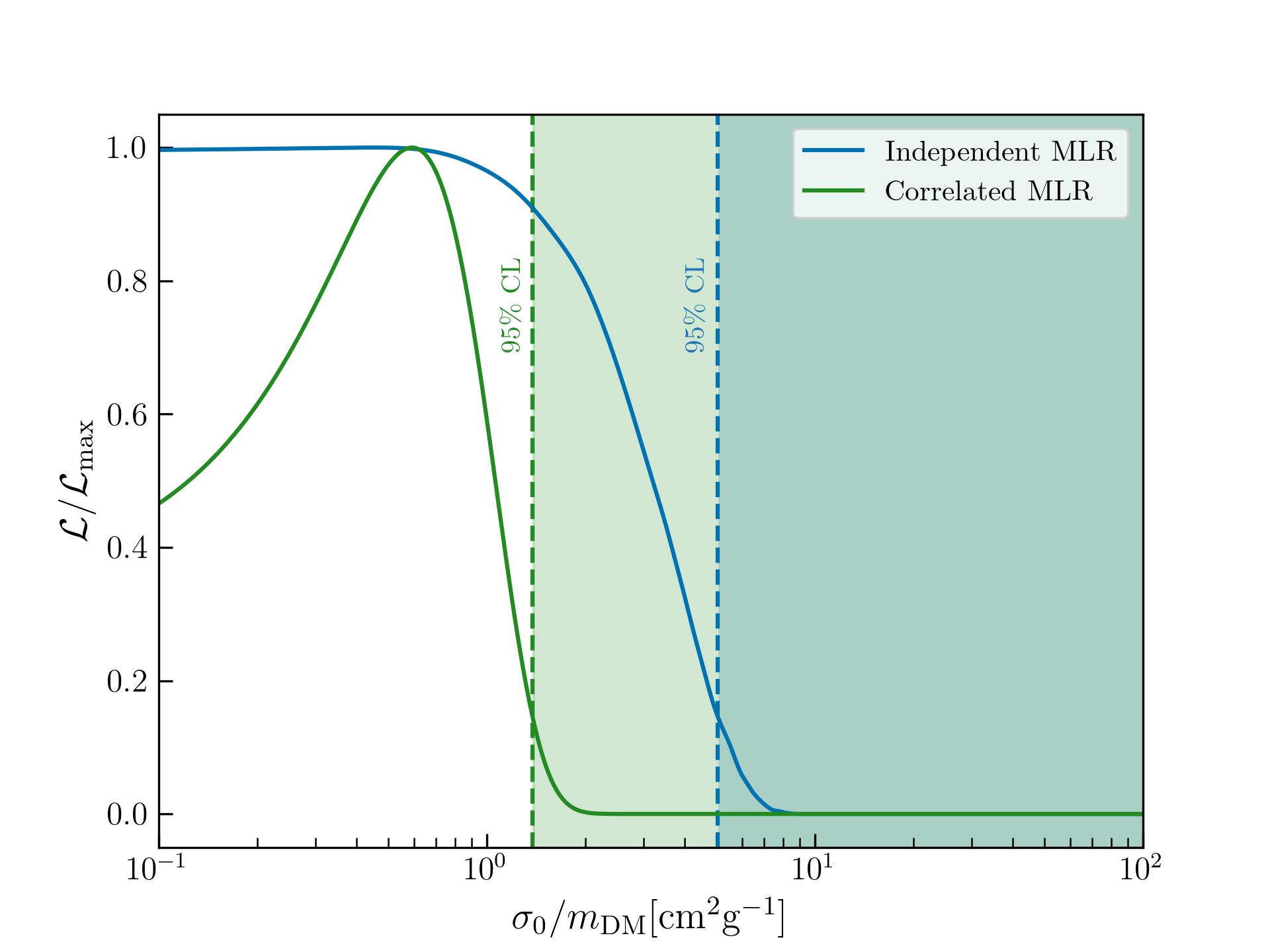}
    \caption{The $95\%$ CL upper limits on the isotropic and velocity-independent self-interaction
    cross-section, $\sigma_0/m_{\text{DM}}$, obtained 
    from subcluster survival in the Bullet Cluster. Two different priors are used to marginalize the likelihood over the initial mass-to-light ratio (MLR): 
    the Gaussian prior (green) assumes correlations between the main cluster and sub-cluster initial MLR values; the log-normal 
    prior (blue) assumes no such correlation.}
    \label{fig:constant cross-section likelihoods}
\end{figure}

The marginalized likelihoods for a single DM component with isotropic and velocity-independent self-interaction cross-section 
are shown in figure~\ref{fig:constant cross-section likelihoods}.
The Gaussian prior centered around the main cluster's mass-to-light ratio gives a 
limit $\sigma_0/m_{\text{DM}} < 1.4 \, \unit{cm^2 g^{-1}}$ ($\Delta \chi^2 < 3.84$), very similar to the limit obtained in 
ref.~\cite{Markevitch:2003at} based on  similar arguments and assumptions. The log-normal prior is more conservative, as it 
assumes no correlation between the clusters and instead only considers the general scatter 
in the mass-luminosity relationships for galaxy clusters. This prior gives a limit 
$\sigma_0/m_{\text{DM}} < 5.0 \, \unit{cm^2 g^{-1}}$ ($\Delta \chi^2 < 3.84$). In the following, we will use the  
likelihood {with correlations, because it resembles more closely the treatment found elsewhere in the literature}. 
We note that this likelihood leads to a slight preference for non-zero self-interaction cross section, with a best-fit value 
around $\sigma_0 / m_\text{DM} = 0.5 \, \mathrm{cm^2 \, g^{-1}}$. However, this preference is not significant and the constraint 
is consistent with vanishing self-interactions at the $1\sigma$ level.

\subsection{Accelerator experiments}
\label{sec:acc}

\begin{figure}[t]
    \centering
    \includegraphics[width=0.7\linewidth]{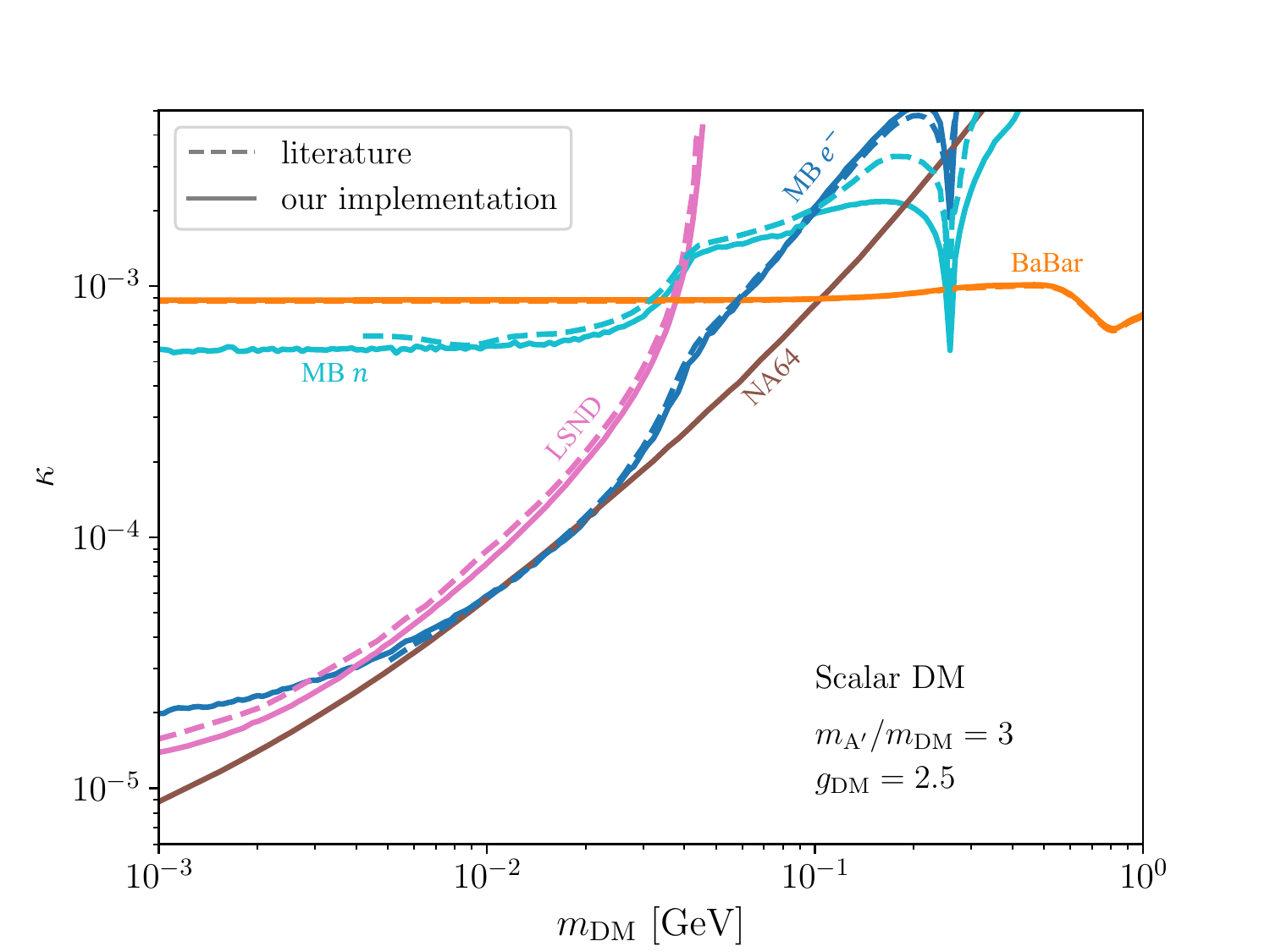}
    \caption{The 90\%\,C.L.~exclusion bounds from NA64, LSND, MiniBooNE electron and nucleon scattering, and BaBar 
    experiments which we consider in this analysis, for the case of complex scalar DM with $m_{A'}/m_\text{DM} = 3$ and $g_\text{DM} = 2.5$. 
    The exclusion bounds are plotted in the $\kappa$ vs $m_\text{DM}$ plane, for interpolated simulations used in these global scans 
    (solid curves) in addition to literature 
    values~\cite{LSND:2001akn,deNiverville:2011it,PhysRevD.98.112004,NA64:2023wbi,BaBar:2017tiz} for comparison 
    (dashed curves). The NA64 literature values agree exactly with 
    the interpolated values.}
    \label{fig:beam_dumps}
\end{figure}

As discussed in section~\ref{sec:decays}, we focus our analysis on parameter regions where the dark photon decays almost 
exclusively into pairs of DM particles. Constraints on this scenario can come from two types of experiments: searches for 
missing energy and searches for the scattering of DM particles produced in the dark photon decays. Figure~\ref{fig:beam_dumps} 
shows the 90 \% C.L. exclusion bounds from LSND~\cite{LSND:2001akn,deNiverville:2011it}, MiniBooNE electron and nucleon 
scattering~\cite{PhysRevD.98.112004}, NA64~\cite{NA64:2023wbi} and BaBar~\cite{BaBar:2017tiz} for a complex scalar DM 
candidate with an {example benchmark choice} of model parameters. These exclusion bounds are shown from both previous 
literature results (dashed) 
and from using the interpolations and scaling techniques implemented in this work (solid). The rest of this section describes 
these experiments in detail. Note that we do not consider the recent MicroBooNE analysis~\cite{MicroBooNECollaboration:2023kmx}, which employs a convolutional neural network and therefore cannot easily be reinterpreted in our context. For similar reasons, we also do not include the first results from the COHERENT experiment~\cite{COHERENT:2021pvd}.

\subsubsection*{LSND and MiniBooNE}

Beam dump experiments such as LSND~\cite{LSND:2001akn,deNiverville:2011it} and MiniBooNE~\cite{PhysRevD.98.112004} 
aim to produce a relativistic flux of DM particles from interactions between a proton beam and nucleons in a dense target. 
Neutral mesons such as $\pi^0$ and $\eta$ are produced in these interactions, which then decay producing DM via the  
chain $\pi^0,\eta \to \gamma + A'$, $A' \to \chi \bar{\chi}$ with either $\chi = \psi$ for fermionic DM or $\chi = \Phi$ for scalar DM. 
The DM particles can then be detected in a downstream detector through DM-nucleon or DM-electron scattering, analogous to 
underground DM direct detection experiments searching for Galactic DM. The number of expected signal events 
in the detector is proportional to model parameters $g_\text{DM}$ and $\kappa$, and branching ratios $\text{BR}_{X \to A' \gamma}$ and 
$\text{BR}_{A' \to \chi \bar{\chi}}$:
\begin{equation}
    N_{\chi} \propto \text{BR}_{X \to A' \gamma} \text{BR}_{A' \to \chi \bar{\chi}} \kappa^2 g_\text{DM}^2 \propto \text{BR}_{A' \to \chi \bar{\chi}} \kappa^4 g_\text{DM}^2\,,
\end{equation}
where $X$ is a meson produced from beam-target interactions, $\kappa^2$ comes from dark photon production, and 
$\kappa^2 g_\text{DM}^2$ from DM scattering in the detector.
The number of expected DM signal events is calculated for each model using Monte Carlo simulation 
software for beam dump experiments, BdNMC~\cite{deNiverville:2016rqh}.
We consider the case where $m_{A'} > 2m_\text{DM}$, therefore on-shell contributions to DM production dominate. These simulations 
are performed on a log-spaced grid of $m_{A'}$ and $m_\text{DM}$ values at constant values for $\kappa$ and $g_\text{DM}$, which we 
denote $\tilde{g}_\text{DM}$ and $\tilde{\kappa}$. From this given set of values, we calculate $N_\text{DM}$ by scaling the simulated number 
of events, $\tilde{N}_\text{DM}={N}_\text{DM}(g_\text{DM} = \tilde{g}_\text{DM}, \kappa = \tilde{\kappa})$, in the following way:
\begin{equation}
    N_\text{DM}(m_{A'},m_\text{DM},\kappa,g_\text{DM}) = \tilde{N}_{\chi} \left( \frac{\kappa}{\tilde{\kappa}} \right)^4  \left( \frac{g_\text{DM}}{\tilde{g}_\text{DM}} \right)^2 \text{BR}_{A' \to \chi \bar{\chi}}(m_{A'},m_\text{DM},\kappa,g_\text{DM})\,,
\end{equation}
where $\tilde{\kappa}$ and $\tilde{g}_\text{DM}$ are chosen such that $\tilde{g}_\text{DM} \gg \tilde{\kappa}$, and hence 
$\text{BR}_{A' \to \chi \bar{\chi}}(m_{A'},m_\text{DM},\tilde{\kappa},\tilde{g}_\text{DM}) \approx 1$. 

We take LSND and MiniBooNE as counting experiments, modelling the likelihood of observing $n$ events, 
with predicted $s$ signal and $b$ background events, as a Poisson distribution:

\begin{equation}
    \mathcal{L} = e^{-(s+b)} \frac{(s+b)^n}{n!} \,.
    \label{eq:poisson}
\end{equation}
For the case of MiniBooNE, with no excess above the background predictions, we take $n=0$ and $b=0$, and $s$ given 
by the simulated signal value~\cite{PhysRevD.98.112004}. LSND reported $n=242$ events which include elastic 
scattering by neutrinos and potentially DM, with an expected SM background 
$b = 229 \pm 28$~\cite{LSND:2001akn,deNiverville:2011it}. We include the background uncertainty by treating $b$ as a nuisance parameter with a Gaussian likelihood that is marginalised over at each parameter point. Figure~\ref{fig:beam_dumps} shows the Mini-BooNE electron scattering (blue), 
MiniBooNE nucleon scattering (cyan), and LSND electron scattering (pink) 90 \% C.L. exclusion bounds. 
Note that the literature 
MiniBooNE nucleon scattering limits are taken from ref.~\cite{PhysRevD.98.112004}, where the treatment of the $\rho/\omega$ 
resonance is different from our approach.

\subsubsection*{NA64}

The NA64 experiment~\cite{NA64:2023wbi} searches for missing energy events of high energy electron collisions with a fixed 
target, $Z$, resulting from the production of dark photons via dark bremsstrahlung, $e^- Z \to e^- Z A'$ and the prompt decay of 
the dark photon into invisible final states, $A' \to \chi \bar{\chi}$. 

We take results from ref.~\cite{NA64:2023wbi} to calculate the expected number of signal events, $N_\text{DM}$, as a function of 
$\kappa$ and $m_{A'}$. As long as $m_{A'} > 2 m_\text{DM}$, dark bremsstrahlung to invisible final states occurs through an on-shell 
dark photon, thus the process is independent of $g_\text{DM}$ and $m_\text{DM}$. The number of missing energy events is proportional to 
$\kappa^2$, therefore the results from ref.~\cite{NA64:2023wbi} are re-scaled for the desired $\kappa$ value. To account for 
cases where $g_\text{DM} \sim \kappa$, when dark photons can also decay visibly, the results are scaled by $\text{BR}_{A' \to \chi \bar{\chi}}$. 

Similarly to the beam dump experiments discussed above, the likelihood of observing $n$ events at NA64 is given by 
eq.~\eqref{eq:poisson}. No signal events were observed at NA64~\cite{NA64:2023wbi}, hence $n=0$ and $b=0$. Exclusion 
bounds from NA64 at 90 \% C.L. are shown in figure~\ref{fig:beam_dumps}.

\subsubsection*{BaBar single-photon search}

The BaBar collaboration has used a data set of $53 \, \mathrm{fb}^{-1}$ of $e^+ e^-$ collisions to search for events in which an 
invisibly decaying dark photon is produced in association with a mono-energetic SM photon~\cite{BaBar:2017tiz}. While the 
event selection cannot be reproduced in detail, ref.~\cite{BaBar:2017tiz} provides detailed information on the likelihood for the 
kinetic mixing parameter $\kappa$ as a function of the dark photon mass. Moreover, BaBar provides both a Bayesian limit 
(imposing the prior boundary $\kappa > 0$) and a frequentist limit. With the available information it is possible to implement a 
detailed likelihood function that accurately reproduces both limits. We note that for dark photons with general branching ratios 
we need to make the replacement
\begin{equation}
   \kappa \to \kappa \sqrt{\text{BR}_{A' \to \chi \bar{\chi}}}
\end{equation}
inside the likelihood function, which conservatively assumes that all visible decays have been vetoed by BaBar.
We indicate also these limits in figure~\ref{fig:beam_dumps}.

\subsubsection*{Electroweak precision observables}

In principle the kinetic mixing parameter $\kappa$ is constrained by electroweak precision 
data~\cite{Curtin:2014cca,Loizos:2023xbj,Bento:2023flt}. However, for dark photons that decay dominantly invisibly, the 
constraints discussed above are much stronger than the direct constraints on $\kappa$. We therefore do not include these 
constraints in our analysis.

\subsubsection*{Constraints on visibly decaying dark photons}

Although our analysis focuses on the case that $g_\text{DM} > \kappa$, there are regions of parameter space where the visible 
branching ratio of the dark photon is non-negligible. This is because the decay of the dark photon into DM is phase-space 
suppressed in the resonant region ($\epsilon_R \ll 1$). One might worry that in such a case the likelihood 
implementations described above 
are insufficient and that additional constraints may arise from searches for prompt or displaced decays of dark photons into 
visible final states. We therefore check for all points passing all constraints that
\begin{itemize}
    \item The proper dark photon decay length $c \tau_{A'}$ is shorter than $1\,\mathrm{mm}$, such that searches for long-lived 
    dark photons (for example in beam-dump experiments such as NA62~\cite{NA62:2023qyn}) are insensitive.
    \item The visible branching fraction satisfies the constraint $\kappa^2 \text{BR}_{A' \to SM} < 10^{-7}$, such that prompt 
    decays into SM final states are out of reach of the leading experiments such as LHCb~\cite{LHCb:2020ysn}, BaBar~\cite{BaBar:2014zli} and 
    NA48/2~\cite{NA482:2015wmo}.
\end{itemize}
We find that even though these requirements are very conservative, they are always satisfied in the parameter regions of 
interest.

\subsection{Direct detection}
\label{sec:dd}

Strong constraints on sub-GeV DM result from searches for DM scattering in ultra-low-background detectors. Only few direct 
detection experiments achieve sufficiently low energy thresholds to search for sub-GeV DM in nuclear recoils. However, the reach of 
these experiments can be extended to much smaller masses by searching for electron recoils. Moreover, additional sensitivity 
can be gained by searching for electrons produced from nuclear recoils via the so-called Migdal effect~\cite{Ibe:2017yqa}.

\subsubsection*{Electron recoils}

The reference cross section {at fixed momentum transfer $q = \alpha m_e$} for DM-electron scattering is given by~\cite{Essig:2011nj}
\begin{align}
\sigma_e = \frac{4 \mu_{\chi, e}^2 \,\alpha \kappa^2 g_\text{DM}^2}{(m_{A^\prime}^2 + \alpha^2 m_e^2)^2}
\end{align}
for both scalar and fermionic DM, where $\mu_{\chi,e} \equiv m_\text{DM} m_e / (m_\text{DM} + m_e) \approx m_e$ is the reduced mass of 
the DM-electron system. 

The leading bounds on $\sigma_e$ stem from XENON1T~\cite{XENON:2019gfn}, SENSEI~\cite{SENSEI:2020dpa}, 
DarkSide50~\cite{DarkSide:2022knj}, PandaX-4T~\cite{PandaX:2022xqx}, DAMIC-M~\cite{DAMIC-M:2023gxo} and 
SuperCDMS HV~\cite{SuperCDMS:2020ymb}.\footnote{The recent results from LZ~\cite{LZ:2023poo} are difficult to reinterpret 
in terms of a full likelihood, and are therefore not included in our analysis.} To obtain likelihood functions for these experiments, we 
have interfaced \gambit with the \obscura library~\cite{Emken2021}, which provides the atomic response functions following 
ref.~\cite{Catena:2019gfa} for liquid noble gases and following ref.~\cite{Essig:2015cda} for semiconductors. Likelihoods are 
obtained by comparing the background expectation and signal prediction to the observed number of events in each bin and 
calculating the resulting Poisson likelihood.  Details of the likelihood functions and the new \gambit interface can be found in 
appendix~\ref{app:backends}.

\begin{figure}[t]
    \centering
    \includegraphics[width=0.45\linewidth]{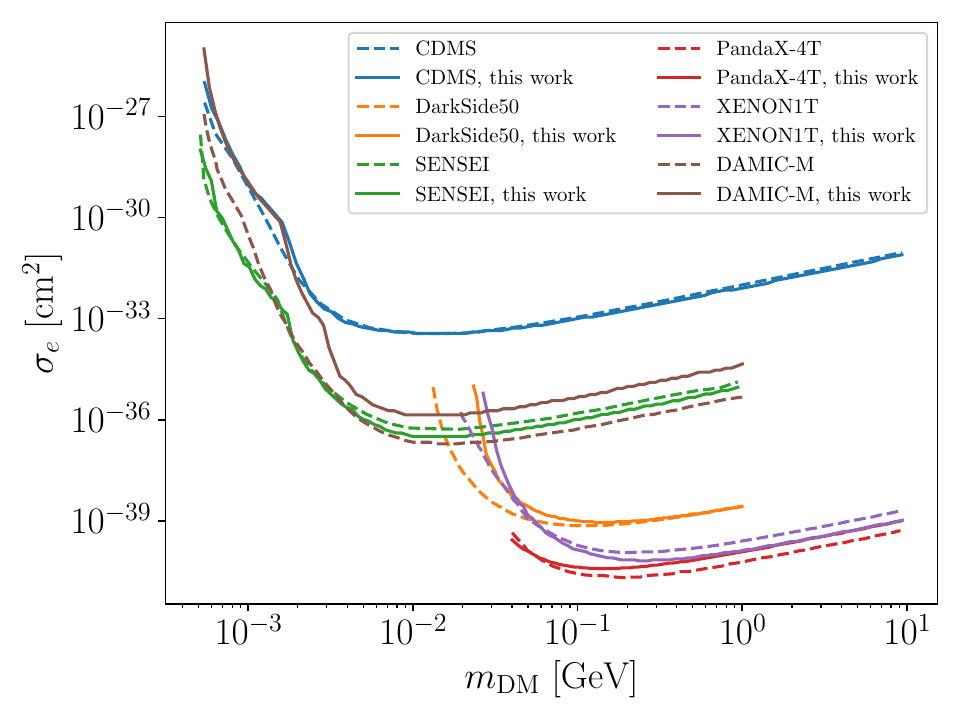}
    \includegraphics[width=0.45\linewidth]{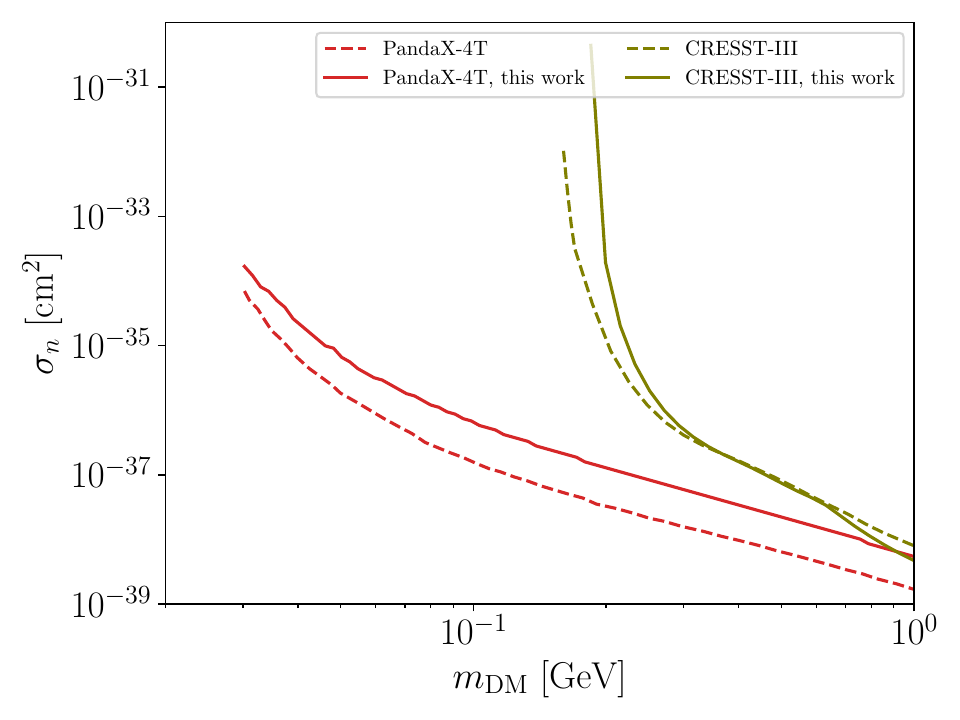}
    \caption{The 90\% C.L. exclusion bounds from direct detection experiments with electron recoils on the left and nuclear recoils 
    (including Migdal scattering) on the right. All limits are computed for the case of contact interactions. Dashed lines are the 90\% C.L. reported by the experiments, solid lines correspond to our implementation in \obscura (electron recoil and Migdal effect) and \ddcalc (nuclear recoil).}
    \label{fig:direct_detection}
\end{figure}

We show a comparison of the published limits and our implementation in the left panel of figure~\ref{fig:direct_detection}. With 
the publicly available information it is not possible to reproduce all limits perfectly, in particular close to threshold, where 
complicated detector effects play an important role. Nevertheless, the leading constraints from SENSEI and PandaX-4T are 
accurately captured in our implementation.

\subsubsection*{Nuclear recoils}

The DM-proton scattering cross section {for vanishing momentum transfer} is given by
\begin{align}
\sigma_p = \frac{4 \mu_{\chi, p}^2 \,\alpha \kappa^2 g_\text{DM}^2}{m_{A^\prime}^4}
\end{align}
for both scalar and fermionic DM, while the DM-neutron scattering cross section vanishes because the dark photon couples to SM particles proportionally to their charge. As a result, the DM-nucleus cross 
section is proportional to $Z^2$ (instead of the commonly  assumed scaling proportional to $A^2$).

The leading constraint on sub-GeV DM from nuclear recoils stems from CRESST-III~\cite{CRESST:2019jnq}, which we 
implement using \ddcalc~\cite{GAMBIT:2018eea}. However, substantially stronger constraints on the DM-proton scattering cross 
section can be obtained by exploiting the Migdal effect, which has been implemented in \obscura for liquid noble gas detectors 
following refs.~\cite{Dolan:2017xbu,Baxter:2019pnz,Essig:2019xkx}. We use this implementation and the electron recoil data 
discussed above to calculate likelihoods for DarkSide50, XENON1T and PandaX-4T~\cite{PandaX:2023xgl}, finding that PandaX-4T gives the strongest 
constraint.

We show a comparison between the published PandaX-4T bound and our implementation in the right panel of 
figure~\ref{fig:direct_detection}. For the purpose of comparing to the published bound from PandaX-4T, we set the neutron coupling equal to the proton coupling in this figure. We find that our implementation leads to a considerably weaker constraint than the published result. This is 
somewhat surprising, given that the underlying data set is the same as for the electron-recoil search, which we are able to 
reproduce quite well (left panel). This suggests some differences in the theoretical predictions for the Migdal effect, which is 
affected by large uncertainties in the determination of the electron wave functions that enter in the calculation of the ionisation probabilities~\cite{Baxter:2019pnz,Ibe:2017yqa}. To reflect these uncertainties, we use our (weaker) bound instead of the published one. We also show the CRESST-III bound, which is not subject to these large theoretical uncertainties and can be accurately reproduced away from the threshold.

We mention in passing that in the parameter space that we consider (in particular for $m_{A'} > 2 m_\text{DM}$) the scattering 
cross sections can never become large enough for Earth shielding~\cite{Emken:2017qmp,Emken:2018run} or cosmic-ray 
upscattering~\cite{Bringmann:2018cvk} to become relevant. Hence these effects are neglected in our analysis.

\section{Results}
\label{sec:results}

In this section we present the results from our global fits of fermionic and scalar sub-GeV DM using the likelihoods discussed 
above. For the fermionic model we consider the case that the DM particle under consideration constitutes a sub-dominant DM 
component ($\Omega_\text{DM} h^2 \leq 0.12$) as well as the case that it is required to reproduce the observed DM abundance 
within observational and theoretical errors ($\Omega h^2 = 0.12$). In the latter case we either set the asymmetry parameter 
$\eta_\text{DM}$ to zero (symmetric) or include it as an independent parameter in our scans (asymmetric). For the scalar model, 
indirect detection constraints are much weaker, and we therefore only consider the symmetric case with saturated relic density.

For each of these scenarios we carry out both a frequentist and a Bayesian analysis. The parameter ranges and priors have 
already been discussed in section~\ref{sec:priors}. The exploration of the parameter space is done with the \gambit 
global fitting framework, using Diver~\cite{Martinez:2017lzg} with a population size of 38,000 and a convergence threshold of $10^{-6}$ for frequentist 
scans and Polychord~\cite{Handley:2015fda} with 1000 live points and a tolerance of $10^{-10}$ for Bayesian scans. We have checked that these settings are sufficient to ensure that the scans converge. The results of all scans along with relevant \gambit configuration files and an example plotting script are available on Zenodo~\cite{Zenodo:2024}.

\subsection{Fermionic dark matter}

\subsubsection*{Frequentist results}

\begin{figure}[t]
    \centering
    \includegraphics[width=0.49 \textwidth]{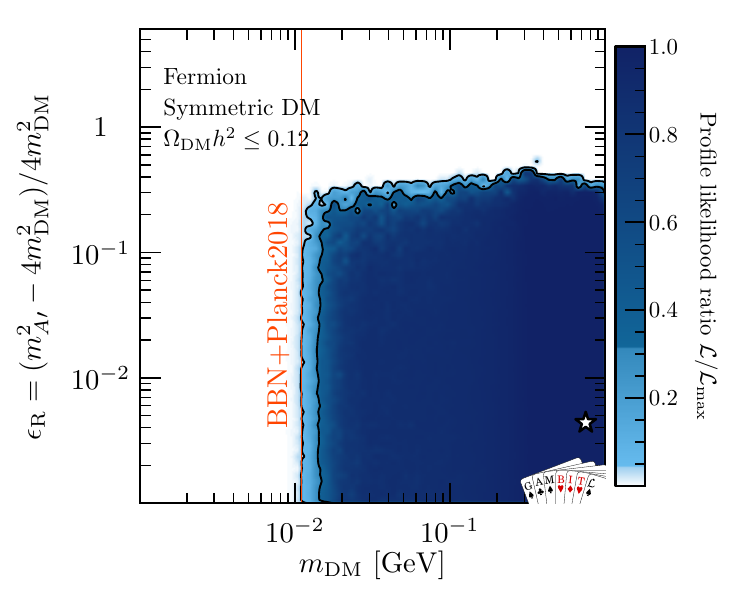}
    \includegraphics[width=0.49 \textwidth]{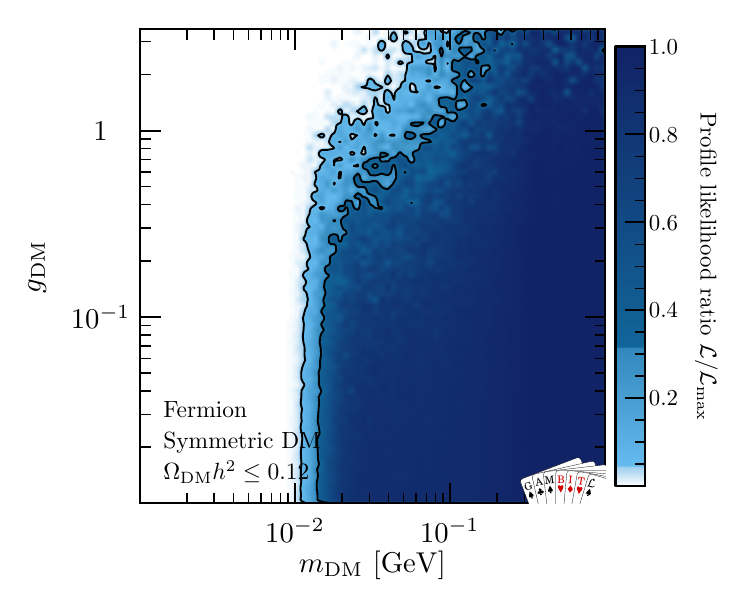}

    \includegraphics[width=0.49 \textwidth]{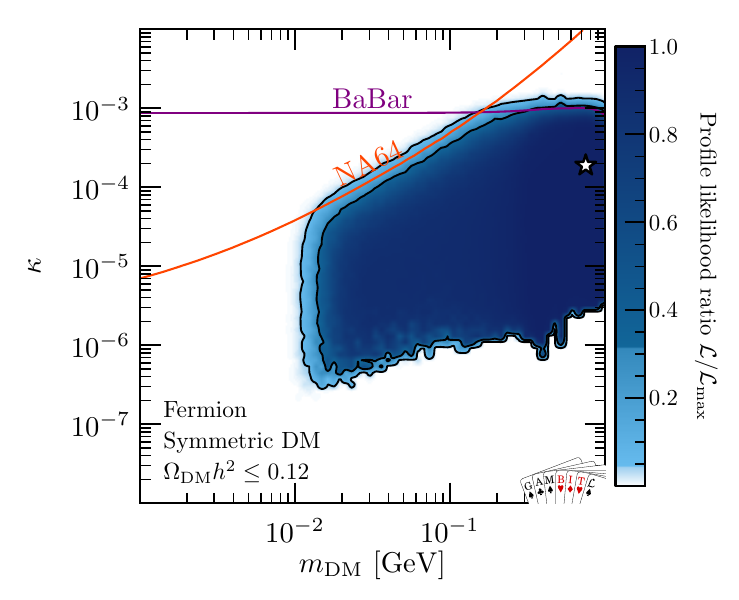}
    \includegraphics[width=0.49 \textwidth]{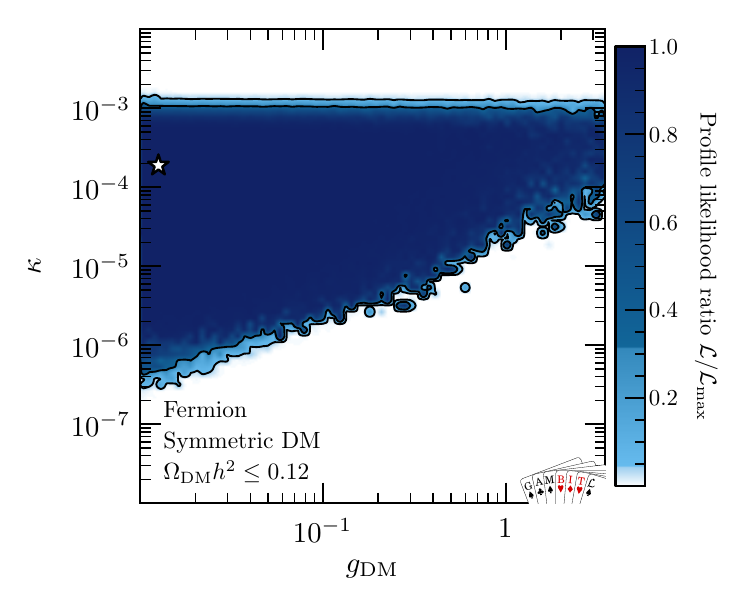}

    \caption{Allowed parameter regions for symmetric fermionic dark matter with $\Omega_\text{DM} h^2 \leq 0.12$ in terms of 
    the different model parameters. The color shading indicates the profile likelihood relative to the best-fit point (indicated by a 
    star). Black lines indicate the $1\sigma$ and $2\sigma$ confidence regions. Coloured lines indicate various experimental and 
    observational constraints (see section~\ref{sec:likelihoods} for details).}
    \label{fig:results_partDM_sym}
\end{figure}

Our results for the case of sub-dominant symmetric fermion DM are shown in figure~\ref{fig:results_partDM_sym}. 
The colour scale represents the likelihood (relative to the maximum $\mathcal{L}_\text{max}$), the lines indicate the allowed parameter 
regions at 68\% and 95\% confidence level. In each panel, the parameters not shown have been profiled over and the best-fit point is indicated by a white star. We emphasize that the likelihood around the best-fit point is very flat and it's precise position is a result of small numerical fluctuations without deeper physical significance. 
To guide the eye, we further show the various constraints described in detail in section~\ref{sec:likelihoods}.
We make the 
following observations:
\begin{itemize}
    \item In the top-left panel we clearly see the impact of the cosmological likelihoods ruling out 
    $m_\text{DM} \lesssim 10 \, \mathrm{MeV}$
    and the indirect detection constraints ruling out $\epsilon_R \gtrsim 0.4$. In other 
    words, the dark photon mass must be tuned to the resonance condition $m_{A'} = 2 m_\text{DM}$ with approximately 20\% 
    precision.
    \item In the bottom-left panel we see the upper bound $\kappa < 10^{-3}$ imposed by BaBar as well as an increasingly 
    stronger bound towards small DM masses stemming from NA64 
    (note that the published bounds, which we show for comparison here, are given at 90\%\,C.L.)

    We also see that $\kappa$ cannot be arbitrarily small, which is a consequence of the relic density requirement.
    \item In the top-right panel we see that the dark sector coupling $g_\text{DM}$ is not strongly constrained and can take values up to the perturbativity 
    bound. For small DM masses, beam-dump experiments become relevant and impose $g_\text{DM} \lesssim 1$.
    \item Finally, in the bottom-right panel, we can see that for increasing $g_\text{DM}$ the allowed range 
    of $\kappa$ shrinks towards the upper bound. This is because larger values of $g_\text{DM}$ correspond to a larger dark photon width, which implies less resonant enhancement even for small $\epsilon_R$.
\end{itemize}
\begin{figure}[t]
    \centering
    \includegraphics[width=0.49 \textwidth]{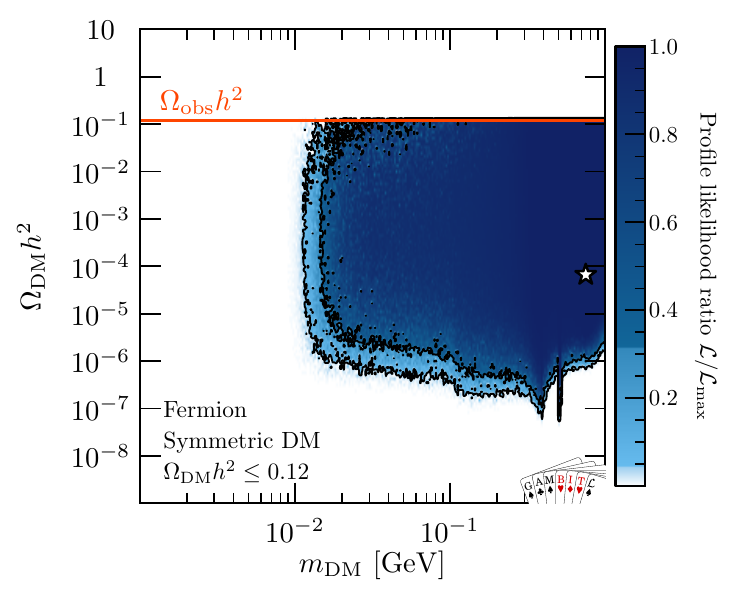}
    \includegraphics[width=0.49 \textwidth]{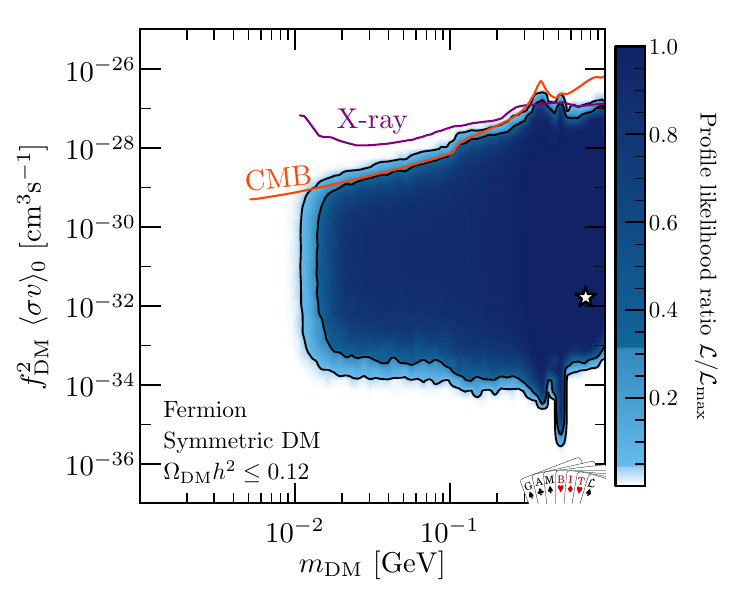}

    \includegraphics[width=0.49 \textwidth]{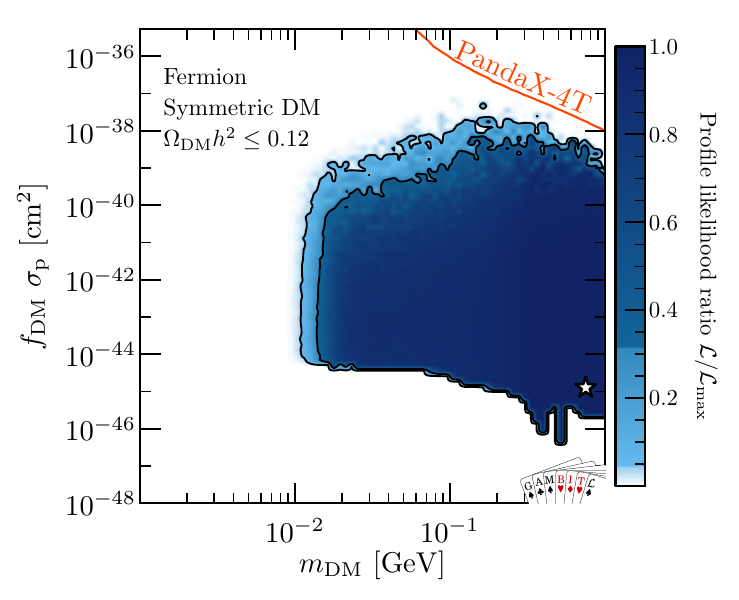}
    \includegraphics[width=0.49 \textwidth]{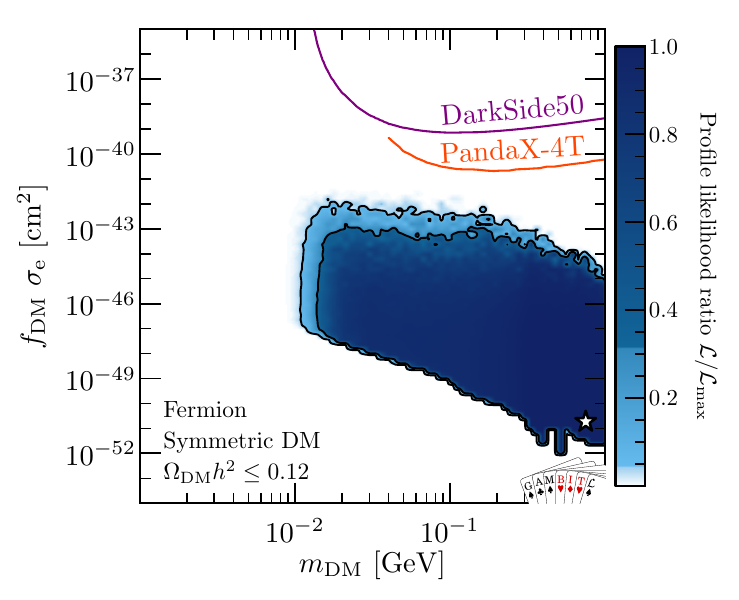}

    \caption{Allowed parameter regions for symmetric fermionic dark matter with $\Omega_\text{DM} h^2 \leq 0.12$ in terms of 
    various observables as a function of the DM mass. 
    See caption of figure \ref{fig:results_partDM_sym} for details on the various line styles and symbols.}
    \label{fig:observables_partDM_sym}
\end{figure}
We also point out that the lower bound on $\kappa$ relaxes slightly for $m_\text{DM} \approx 400 \, \mathrm{MeV}$ and 
$m_\text{DM} \approx 500 \, \mathrm{MeV}$, leading to noticeable features in the bottom-left panel. 
The reason is that the 
centre-of-mass energy of the annihilation process then becomes close to the masses of hadronic bound states (the $\omega$ 
meson and the $\phi$ meson, respectively), such that the annihilation cross section into hadrons receives a resonant 
enhancement.

It is furthermore instructive to plot various observables as a function of the DM mass, as shown in 
figure~\ref{fig:observables_partDM_sym}. We can see that in the bulk of the parameter space the DM relic abundance is well 
below the observed value, such that indirect detection signals (which scale proportional to $f_\text{DM}^2$ for symmetric DM) 
and direct detection signals (which scale proportional to $f_\text{DM}$) can be strongly suppressed. Nevertheless, the observed 
DM relic abundance can be saturated across the entire DM mass range. Both direct and indirect detection experiments place 
relevant constraints on the parameter space, which can however be evaded by many orders of magnitude when considering 
sub-dominant dark matter components. However, it turns out that the predicted DM-electron scattering cross section is below 
current exclusion limits in the entire allowed parameter space. We have also checked explicitly that the visible branching ratio of the dark photon into Standard Model particles is sufficiently small to satisfy all experimental constraints.

\begin{figure}[t]
    \centering
    \includegraphics[width=0.49 \textwidth]{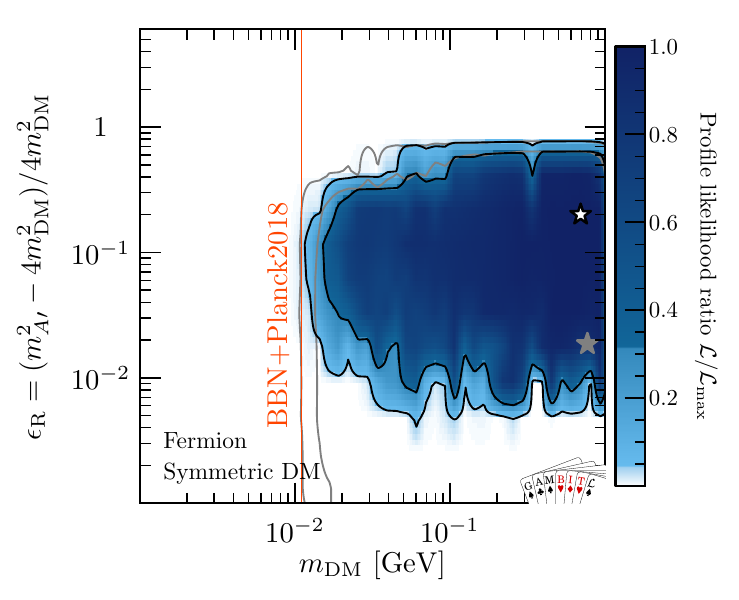}
    \includegraphics[width=0.49 \textwidth]{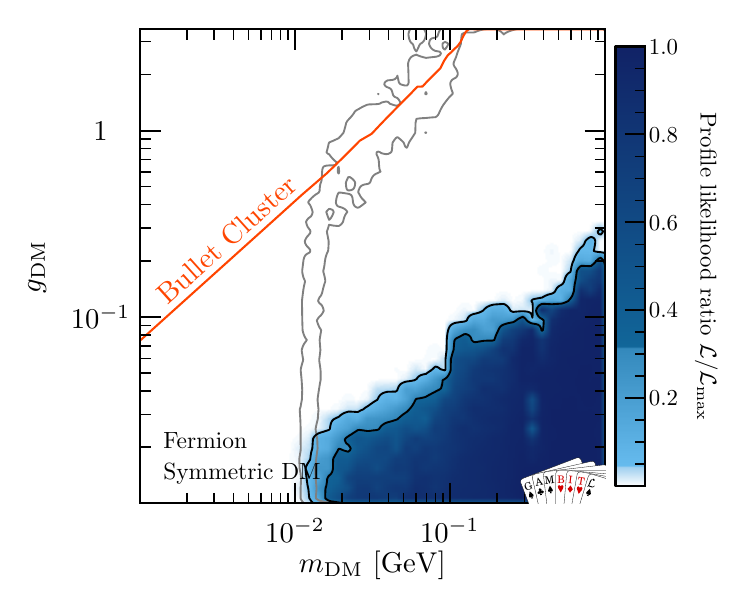}

    \includegraphics[width=0.49 \textwidth]{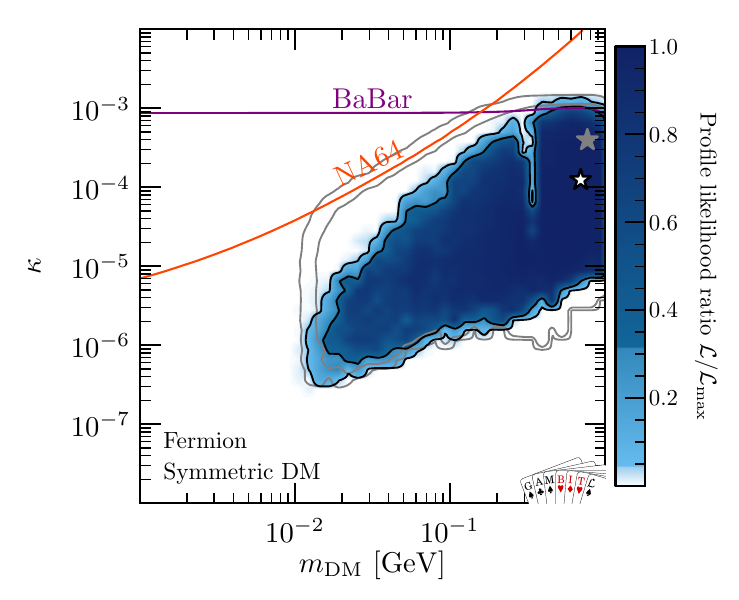}
    \includegraphics[width=0.49 \textwidth]{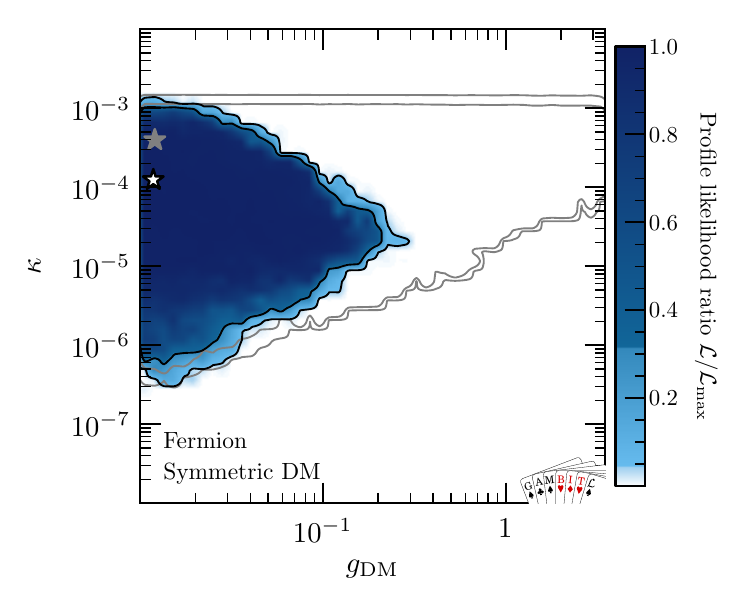}

    \caption{Allowed parameter regions for symmetric fermionic dark matter, when requiring that the observed DM relic 
    abundance is saturated ($\Omega_\text{DM} h^2 \approx 0.12)$. For comparison, we indicate with gray lines the allowed 
    parameter regions for a sub-dominant DM component and with a gray star the corresponding best-fit point. 
    See caption of figure \ref{fig:results_partDM_sym} for details on the various line styles and symbols.}
    \label{fig:results_allDM_sym}
\end{figure}

Let us now turn to the case where we require DM to saturate the observed DM relic density. The results of this scan are shown 
in figure~\ref{fig:results_allDM_sym}. For comparison, we indicate the allowed parameter regions from the previous scan 
(allowing a sub-dominant DM component) with grey lines. The main difference compared to the previous scan is that {there is now a lower bound on the resonance parameter $\epsilon_R \gtrsim 10^{-2}$ in order to evade indirect detection constraints (see figure~\ref{fig:epsR}). Moreover,}
$g_\text{DM}$ is now constrained to much smaller values in order to avoid very large annihilation cross sections, which would 
lead to an underabundant DM component with $\Omega_\text{DM} h^2 < 0.12$. For the same reason also the largest values of 
$\kappa$ are disfavoured. However, in the remaining parameter regions it is possible to satisfy all experimental constraints simultaneously, such that the likelihood of the best-fit point remains largely unchanged relative to the previous case. Correspondingly, the difference in log-likelihood is found to be insignificant: $-2 \Delta \log \mathcal{L} = -0.9$.

Because of the small couplings, the remaining allowed regions of parameter space are extremely difficult to probe with 
laboratory experiments. Both the DM-proton and DM-electron scattering cross sections are orders of magnitude below current 
limits, and the predicted number of events in the beam-dump experiments that we consider is much smaller than unity. The only 
laboratory experiment that places relevant constraints on the parameter space is the BaBar single-photon search.

\begin{figure}[t]
    \centering
    \includegraphics[width=0.49 \textwidth]{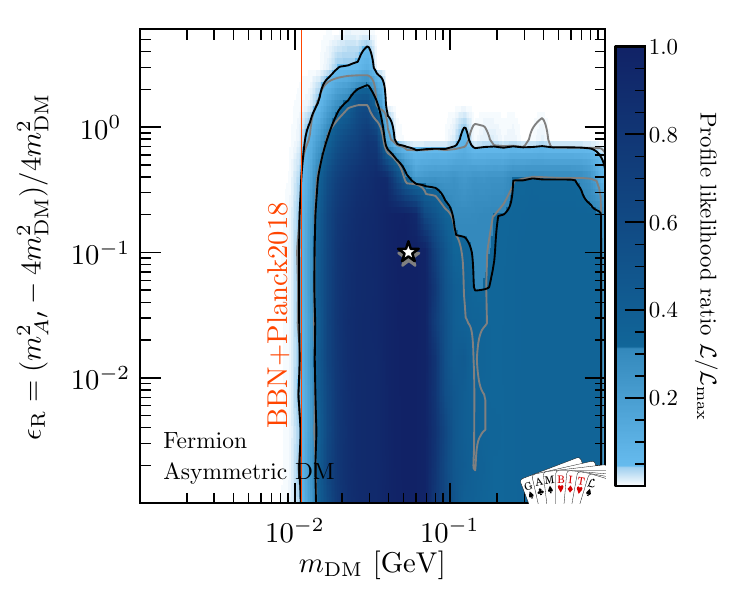}
    \includegraphics[width=0.49 \textwidth]{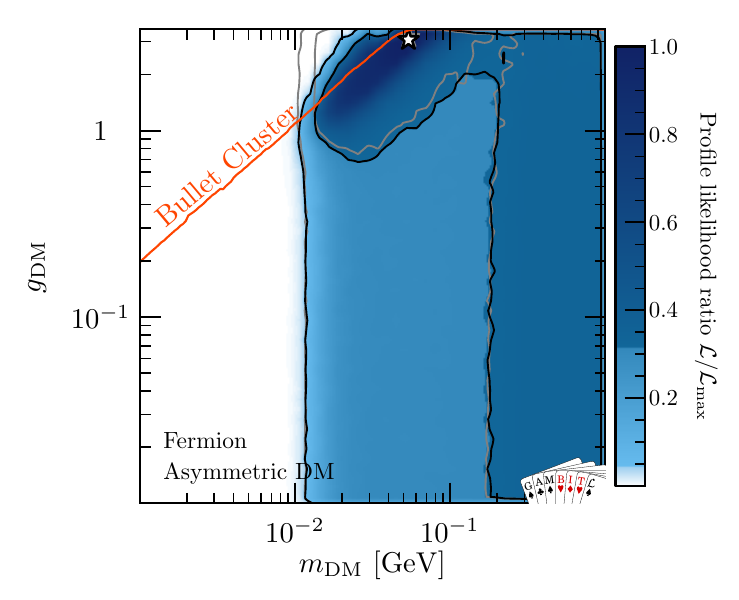}

    \includegraphics[width=0.49 \textwidth]{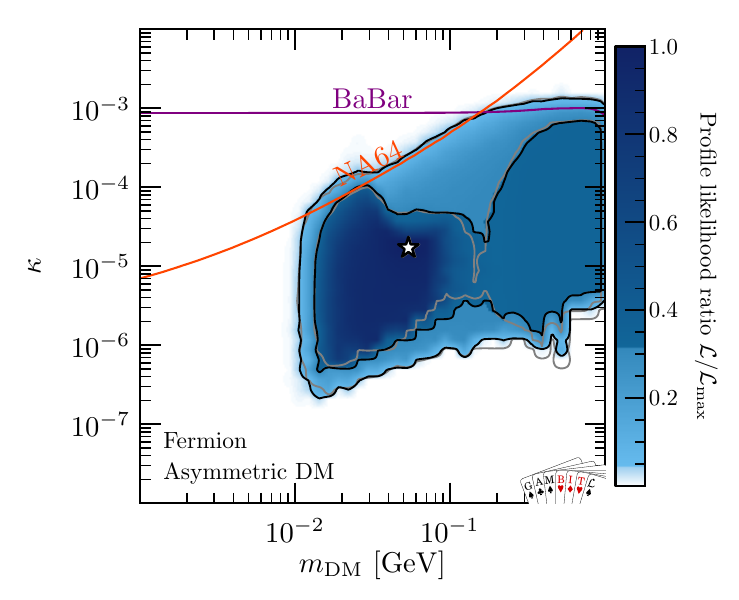}
    \includegraphics[width=0.49 \textwidth]{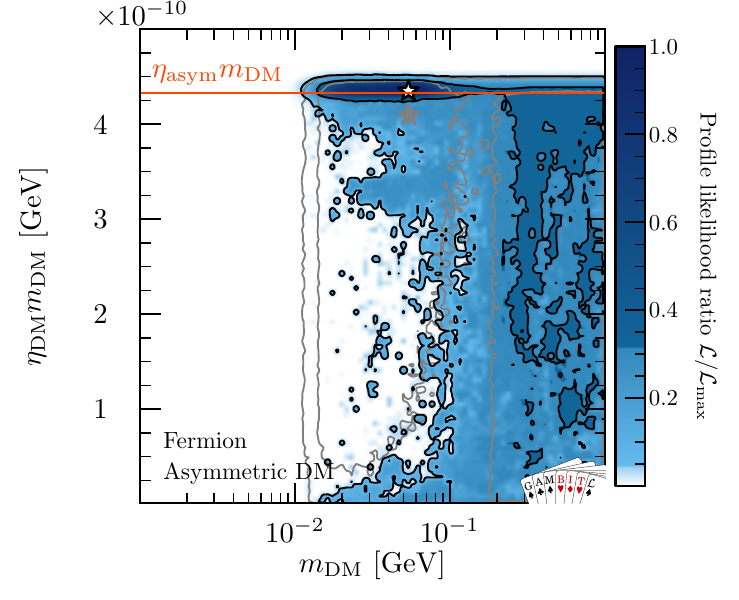}

    \caption{Allowed parameter regions for asymmetric fermionic dark matter with $\Omega_\text{DM} h^2 \approx 0.12$. 
    As in figure \ref{fig:results_allDM_sym}, 
    we indicate with gray lines the allowed parameter regions for a sub-dominant DM component and with a gray star the 
    corresponding best-fit point. The red line in the bottom-right panel indicates the value of $\eta_\text{DM} m_\text{DM}$ that gives $\Omega_\text{DM} h^2 = 0.12$ for the case of a negligible symmetric component, see eq.~\eqref{eq:eta_asym}.}
    \label{fig:results_allDM_asym}
\end{figure}

Next we turn to the question whether the constraints on the parameter space can be relaxed when including a DM asymmetry 
$\eta_\text{DM}>0$. We expect that a sizeable asymmetry will make it possible to saturate the relic density requirement 
without violating indirect detection constraints. Hence we expect to find viable parameter space at larger couplings and 
correspondingly larger cross sections.

We show the allowed parameter regions in figure~\ref{fig:results_allDM_asym}. Note that compared to previous figures we have 
replaced the panel with $g_\text{DM}$ versus $\kappa$ by a panel with $m_\text{DM}$ versus $\eta_\text{DM}m_\text{DM}$. As before, grey 
lines indicate the allowed parameter space when allowing for sub-dominant DM components, while black lines and 
shaded regions indicate the allowed parameter regions when requiring the relic abundance to be saturated. We see that the 
difference between the two cases is minimal, indicating that indeed the asymmetry makes it much easier to satisfy the relic 
density requirement.

Compared to previous figures, we observe two new features. First, the bottom-right panel clearly shows a slight 
preference for $\eta_\text{DM} m_\text{DM} \approx 4 \cdot 10^{-10} \, \mathrm{GeV}$, which corresponds to the case where 
DM is highly asymmetric. 

In this case, a relevant constraint on the parameter space stems from the Bullet Cluster bound on DM 
self-interactions. The Bullet Cluster likelihood that we have implemented prefers a self-interaction cross section around 
$0.5 \, \mathrm{cm^2/g}$. This preference is the origin of the mild preference for small DM masses and large couplings seen in 
figure~\ref{fig:results_allDM_asym}. 
In terms of the log-likelihood of the best-fit point, we find a difference of $-2 \Delta \log \mathcal{L} = 2.2$ compared to the symmetric case.
We emphasize, however, that this preference at the $1\sigma$ level is not significant.

\begin{figure}[t]
    \centering

    \includegraphics[width=0.49 \textwidth]{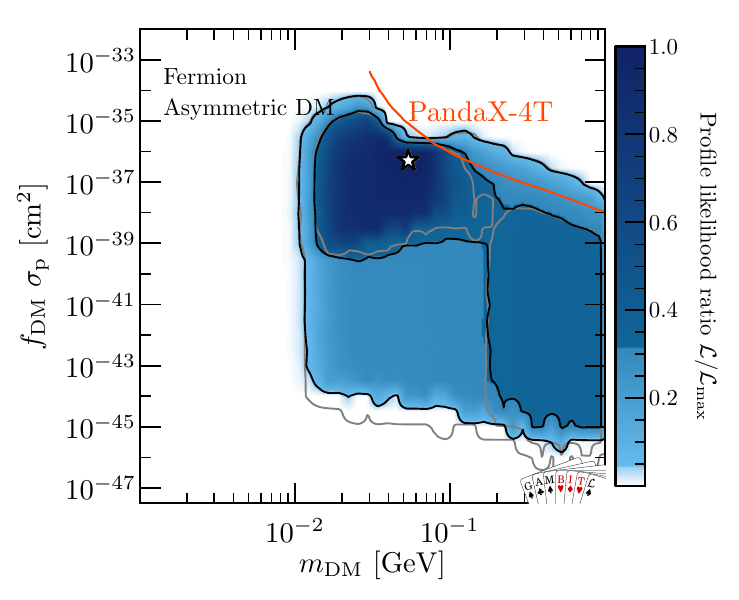}
    \includegraphics[width=0.49 \textwidth]{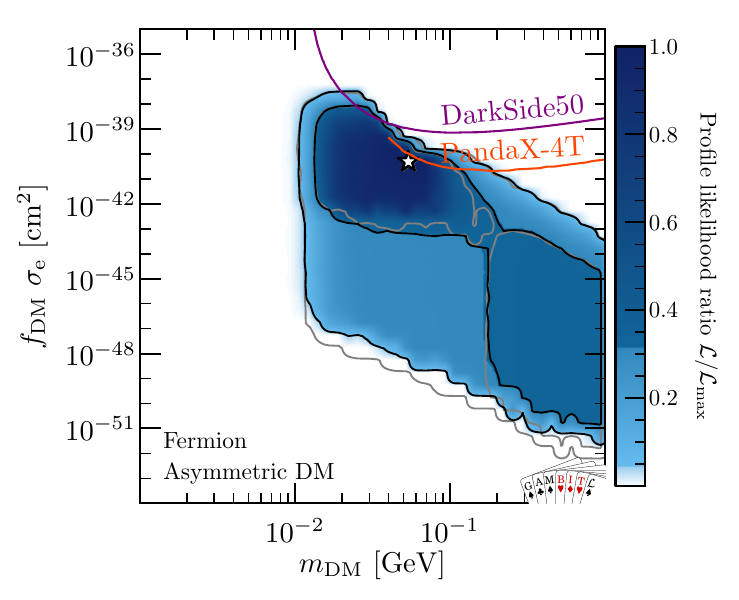}

    \caption{Allowed parameter regions for asymmetric fermionic dark matter with $\Omega_\text{DM} h^2 \approx 0.12$ in terms 
    of various observables as a function of the DM mass.  See caption of figure \ref{fig:results_partDM_sym} for details on the various line styles and symbols.}
    \label{fig:observables_allDM_asym}
\end{figure}

We also show in figure~\ref{fig:observables_allDM_asym} the DM-proton and DM-electron scattering cross sections 
corresponding to the viable regions of parameter space. In contrast to the case of symmetric DM, we find that now the DM-
electron scattering cross section can be large enough for direct detection experiments to place relevant constraints on the 
model. Indeed, the best-fit point is only slightly below current exclusion limits\footnote{Note that we show published bounds, 
which are given at 90\% confidence level and, in contrast to our treatment, do not include uncertainties in the local DM density and velocity distribution.} and 
within reach of the next generation of experiments. 

\subsubsection*{Bayesian results}

As we have seen above, including the asymmetry parameter increases the allowed parameter ranges for the fermionic DM 
model, but it does not significantly improve the likelihood of the best-fit point. From the frequentist perspective, there is hence no preference for including the additional 
parameter. Let us now revisit this conclusion in the Bayesian framework, where not only the likelihood of the best-fit point 
matters, but the volume of the allowed parameter region. In other words, the Bayesian approach penalizes fine-tuning and 
rewards models that can fit all observations in large fractions of parameter space.

\begin{figure}[t]
    \centering
    \includegraphics[width=\textwidth]{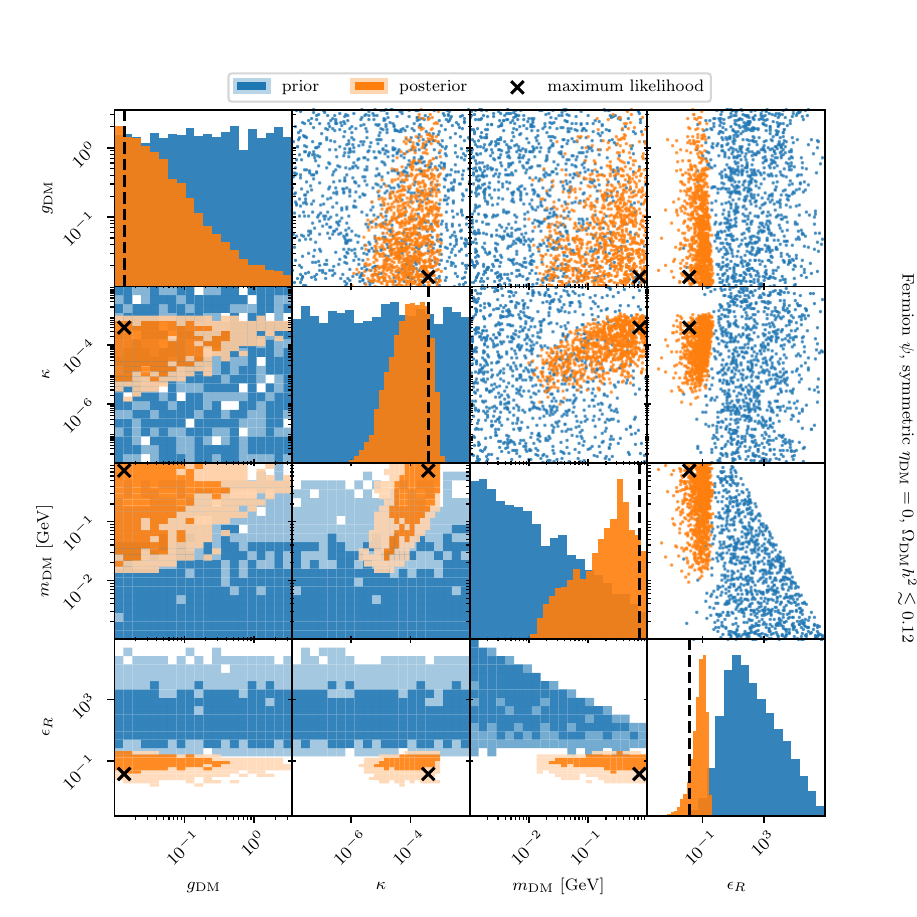}
    \caption{Prior (blue) and posterior (orange) probabilities for the symmetric fermionic DM model with 
    $\Omega_\text{DM} h^2 \leq 0.12$. Panels above the diagonal show scatter plots of a representative set of the sampled 
    parameter points, panels below the diagonal show the regions of highest probability (dark and light shading corresponding to 
    68\% and 95\% probability, respectively). The panels along the diagonal show histograms of the prior and marginalised 
    posterior for each parameter under consideration. The maximum likelihood point is indicated by a cross in the 2-dimensional 
    plots and by a vertical dashed line in the 1-dimensional histograms.}
    \label{fig:bayesian_partDM}
\end{figure}

We first consider the case of no asymmetry with the relic density imposed as upper bound. Our results are shown in 
figure~\ref{fig:bayesian_partDM}, produced using \textsc{anesthetic}~\cite{Handley:2019mfs}. This figure shows in each panel 
the prior probability in blue and the posterior probability in orange, thus providing a direct visual impression of how the prior
belief is updated given the data. Panels above the diagonal show scatter plots of a 
representative set of the sampled parameter points, panels below the diagonal show the regions of highest probability (dark and 
light shading corresponding to 68\% and 95\% probability, respectively). The panels along the diagonal show histograms of the 
prior and marginalised posterior for each parameter under consideration. Note that the prior for $m_\text{DM}$ and $\epsilon_R$ 
is not flat due to the requirement $m_{A'} > 2 m_\text{DM}$ imposed at prior level.

We find good agreement between the Bayesian results and the frequentist results shown in figure~\ref{fig:results_partDM_sym}. In particular, we find that while $\epsilon_R$ can be as large as $10^7$ at the prior level, the posterior is tightly constrained by the relic density and indirect detection constraints to values below unity. However, while in the frequentist analysis the profile likelihood was found to be approximately constant for $\epsilon_R < 0.4$, we find that the posterior probability rapidly decreases for $\epsilon_R < 0.1$ due to the required fine-tuning between $m_{A'}$ and $m_\text{DM}$, leading to a posterior that is strongly peaked around $10^{-1}$. 
Similarly, the marginalised posteriors for the other model parameters exhibit some clear preferences, which are not visible in the profile likelihoods obtained in the frequentist scans. Specifically, there is a clear preference for small values of $g_\text{DM}$ and large values of $m_\text{DM}$. As for $\epsilon_R$ these preferences are a result of the volume effect, i.e.\ the fine-tuning penalty inherent in the Bayesian approach, rather than a result of a preference in the likelihood.

\begin{figure}[t]
    \centering
    \includegraphics[width=\textwidth]{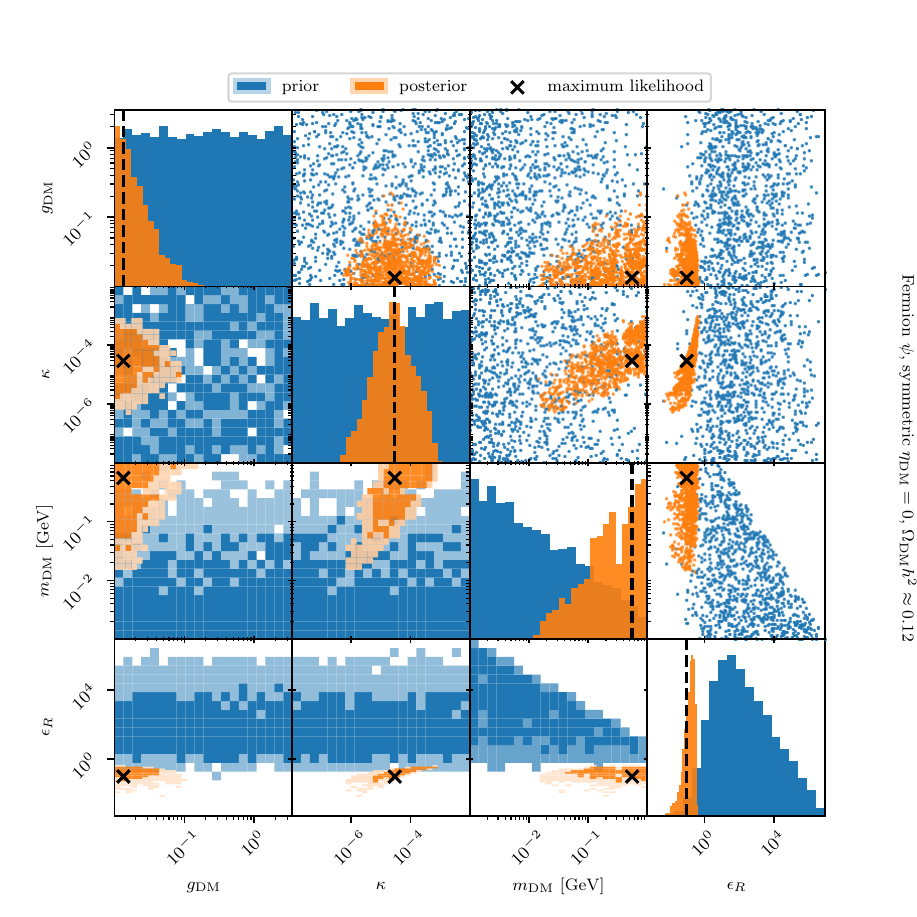}
    \caption{Same as figure~\ref{fig:bayesian_partDM}, but for the symmetric fermionic DM model with $\Omega_\text{DM} h^2 \approx 0.12$.}
    \label{fig:bayesian_allDM}
\end{figure}

These effects become even more pronounced when requiring the DM particle to constitute all of DM, see figure~\ref{fig:bayesian_allDM}. For small DM masses, significant tuning in both $\kappa$ and $g_\text{DM}$ is required in order to satisfy all constraints. As a result, the posterior probability for the DM mass peaks strongly at the largest values considered in the scan, while the posterior probability of $g_\text{DM}$ peaks strongly at the smallest values. The posterior for $\kappa$ is less tightly constrained, with a broad peak in the range $10^{-5}$--$10^{-4}$.

\begin{figure}[t]
    \centering
    \includegraphics[width=\textwidth]{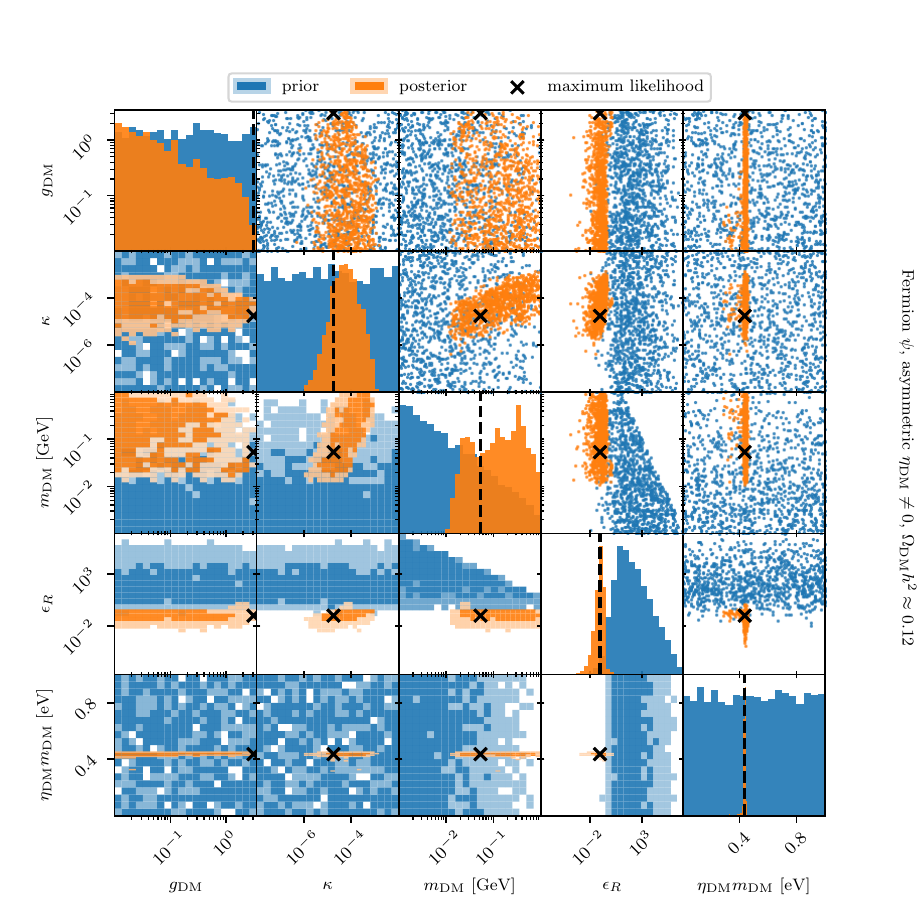}
    \caption{Same as figure~\ref{fig:bayesian_partDM}, but for the {\it asymmetric} fermionic DM model with $\Omega_\text{DM} h^2 \approx 0.12$.
  }
    \label{fig:bayesian_asymmetric}
\end{figure}

Finally, when we include the asymmetry parameter, we obtain the results shown in figure~\ref{fig:bayesian_asymmetric}. We see 
that there is a strong preference for $\eta_\text{DM} m_\text{DM}$ to be close to $4 \cdot 10^{-10} \, \mathrm{GeV}$, which 
corresponds to the case where DM is highly asymmetric. More precisely, the (equal-tailed) credible interval for $\eta_\text{DM} m_\text{DM}$ is $[4.28, 4.35] \cdot 10^{-10} \, \mathrm{GeV}$ at 68\% confidence level (or $[3.78, 4.41] \cdot 10^{-10} \, \mathrm{GeV}$ at 95\% confidence level). For comparison, when we allow for a sub-dominant DM component, the credible intervals broaden significantly to $[0.12,	4.30] \cdot 10^{-10} \, \mathrm{GeV}$ at 95\% confidence level. Including the asymmetry significantly relaxes the constraints on the 
other model parameters, leading to much broader posteriors for $m_\text{DM}$ and $g_\text{DM}$, which can now go up to the 
perturbativity bound.

As in the frequentist case, the best fit point lies at small DM masses and large coupling, corresponding to a self-interaction cross 
section around $0.5 \, \mathrm{cm^2/g}$. However, the best-fit value of $g_\text{DM}$ does not coincide with the maximum of 
the marginalised posterior, which lies at much smaller values. In other words, the Bayesian analysis prefers the large parameter 
region with negligible self-interaction cross section over the small parameter region with sizeable self-interaction cross section, 
even though the latter has a slightly higher likelihood.

Given that including an asymmetry substantially relaxes the constraints on the other model parameters, it is interesting to 
perform a Bayesian model comparison between the models with and without asymmetry parameter. For this purpose, we can 
calculate the Bayesian evidence
\begin{equation}
    \mathcal{Z} = \int \mathcal{L}(\theta) \pi(\theta) \mathrm{d} \theta \, ,
\end{equation}
where $\theta$ denotes the model parameters, $\mathcal{L}(\theta)$ the likelihood function and $\pi(\theta)$ the prior 
probabilities. It is furthermore possible~\cite{Hergt:2021qlh,Balazs:2022tjl} to decompose the Bayesian evidence into the 
posterior-weighted log-likelihood
\begin{equation}
    \langle \log \mathcal{L} \rangle_\mathcal{P} = \int \mathcal{P}(\theta) \log \mathcal{L}(\theta) \mathrm{d} \theta 
\end{equation}
and the Kullback-Leibler divergence of prior and posterior probability
\begin{equation}
    \mathcal{D}_\text{KL} = \int \mathcal{P}(\theta) \log \frac{\mathcal{P}(\theta)}{\pi(\theta)} \mathrm{d} \theta \, ,
\end{equation}
in the sense that
\begin{equation}
    \log \mathcal{Z} = \langle \log \mathcal{L} \rangle_\mathcal{P} - \mathcal{D}_\text{KL} \, .
\end{equation} 
The first term becomes large if the model can fit all available data in the most probable regions of parameter space and small if the entire parameter space is in tension with data. The second term is large if the available data can only be fitted in very special regions of parameter space, and small if the model predictions are generically in agreement with data.
This decomposition makes it possible to attribute the preference between two models either to the likelihood (i.e.\ a genuine preference in the data) or to the volume effect (i.e.\ a fine-tuning penalty for parameters that need to lie in very narrow regions of parameter space in order for the model to agree with data). For the model without asymmetry we find
\begin{equation}
\log \mathcal{Z}_\text{sym} = -376.93 \pm  0.12  = -358.59  - 18.34
\end{equation}
whereas the model with asymmetry gives
\begin{equation}
\log \mathcal{Z}_\text{asym} = -374.18 \pm  0.11  = -358.69  - 15.49 \, .
\end{equation}
As expected, there is no preference between the two models in the likelihood term, but the fine-tuning penalty is much smaller in the second model than in the first one, leading to an overall Bayes factor of
\begin{equation}
    \frac{\mathcal{Z}_\text{asym}}{\mathcal{Z}_\text{sym}} = 15.6 \, ,
\end{equation}
which constitutes a ``strong'' preference according to the Jeffreys' scale.

We emphasize, however, that the fine-tuning penalty is strongly determined by the chosen priors for the model parameters. If for 
example we were to use a logarithmic prior for $\eta_\text{DM}$ in the range $[10^{-13}, 10^{-6}]$ 
instead of the linear prior on 
$\eta_\text{DM} m_\text{DM}$, the Bayes factor between the two models decreases from 15.6 to 2.9, which corresponds to a 
model preference that is ``barely worth mentioning''.

\subsection{Scalar dark matter}

For the case of scalar DM, indirect detection constraints play a negligible role, because the annihilation cross section is 
$p$-wave 
suppressed for small velocities. As a result, it is much easier to saturate the observed DM relic abundance even without 
introducing an asymmetry parameter. We therefore restrict ourselves to the case of symmetric DM.

\begin{figure}[t]
    \centering
    \includegraphics[width=0.49 \textwidth]{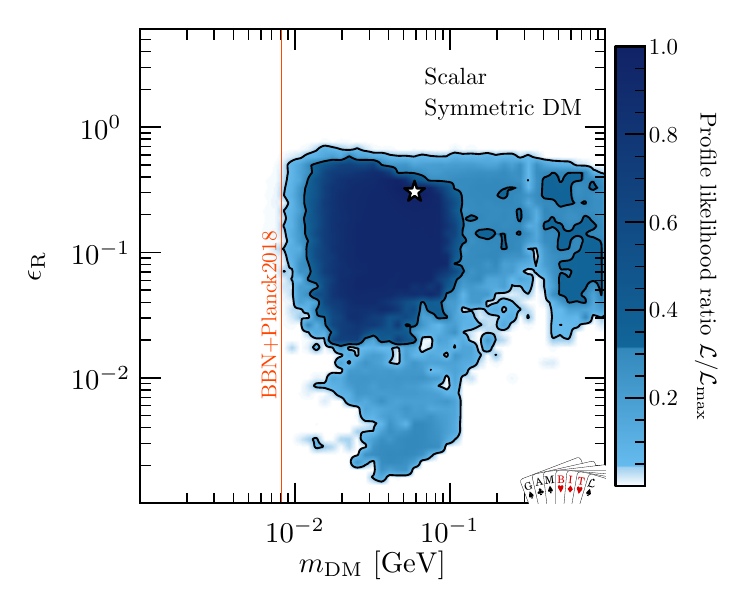}
    \includegraphics[width=0.49 \textwidth]{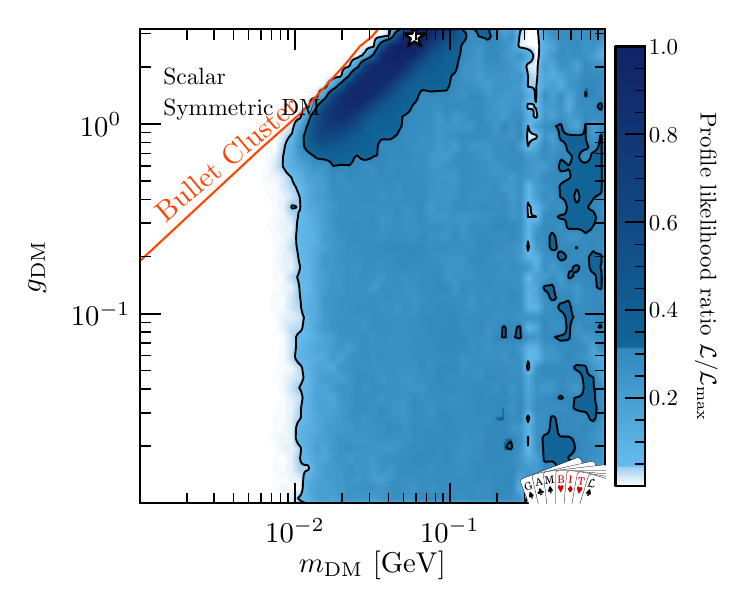}
    \includegraphics[width=0.49 \textwidth]{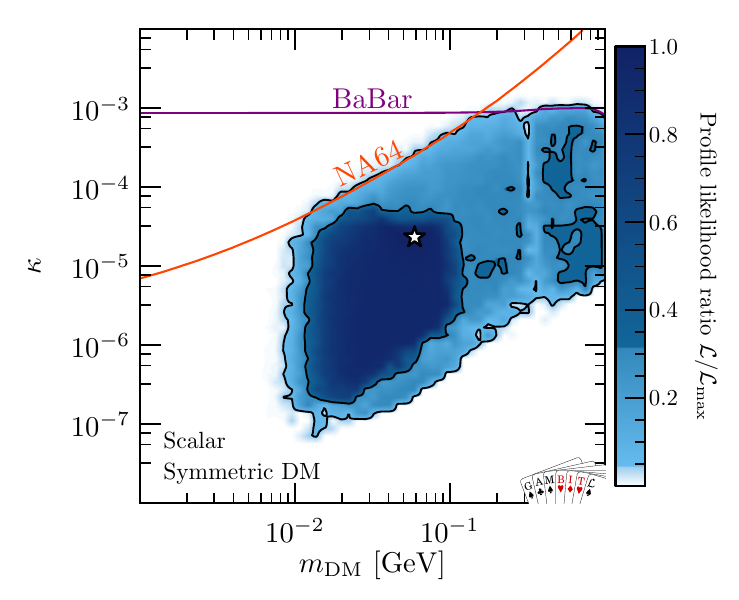}
    \includegraphics[width=0.49 \textwidth]{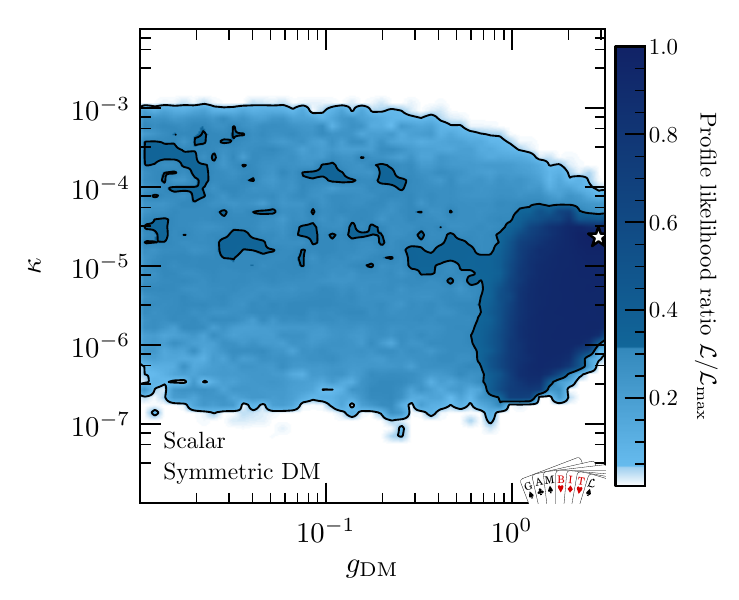}

    \caption{Allowed parameter regions for symmetric scalar dark matter with $\Omega_\text{DM} h^2 \approx 0.12$.   See caption of figure \ref{fig:results_partDM_sym} for details on the various line styles and symbols.}
    \label{fig:frequentist_scalar}
\end{figure}

\begin{figure}[t]
    \centering
    \includegraphics[width=\textwidth]{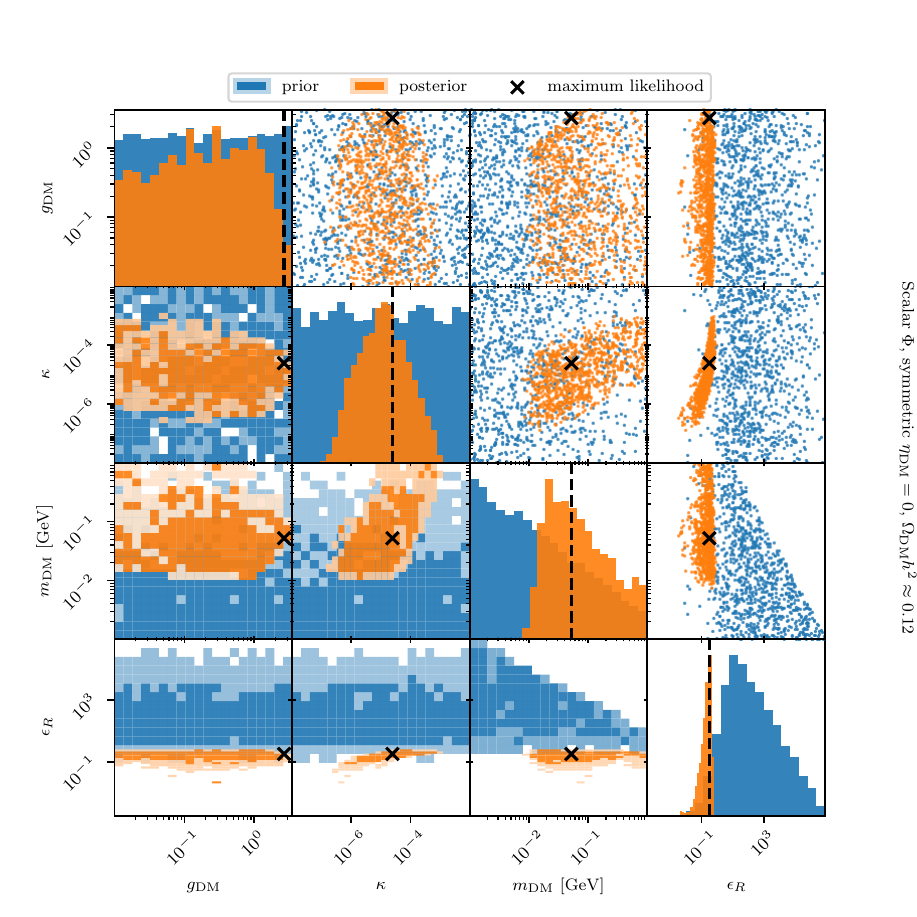}
    \caption{Prior (blue) and posterior (orange) probabilities for the symmetric scalar DM model with $\Omega_\text{DM} h^2 \approx 0.12$.}
    \label{fig:bayesian_scalar}
\end{figure}

The results from our frequentist scan are shown in figure~\ref{fig:frequentist_scalar}. The allowed parameter regions look very similar to the 
case of highly asymmetric fermionic dark matter. This finding is expected, since in both cases indirect detection constraints are 
absent and the phenomenology is otherwise very similar. We also observe that the allowed parameter regions change only very 
slightly when considering a sub-dominant DM component (gray lines). The only notable differences are that for a DM 
sub-component the Bullet Cluster constraint can be evaded, which opens up additional parameter space at large $g_\text{DM}$ 
and small $m_\text{DM}$, and that the BaBar bound can be evaded by having simultaneously $g_\text{DM} \approx 10^{-2}$ 
and $\kappa > 10^{-3}$, such that $\text{BR}_{A' \to \chi \bar{\chi}} \ll 1$. The latter possibility would however likely be 
constrained by additional likelihoods not implemented in the present work, such as searches for dark photons decaying into 
muon pairs at LHCb~\cite{LHCb:2019vmc}.

The corresponding Bayesian results are shown in figure~\ref{fig:bayesian_scalar}. We find that the posterior probabilities for 
$m_\text{DM}$ and $g_\text{DM}$ now follow the prior probabilities much more closely (apart from the lower bound 
$m_\text{DM} \gtrsim 10 \, \mathrm{MeV}$ imposed by cosmological constraints). Nevertheless, the relic density requirement 
still imposes a clear upper bound on $\epsilon_R$ and a lower bound on $\kappa$, while collider and beam-dump experiments 
rule out large values of $\kappa$.

\begin{figure}[t]
    \centering
    \includegraphics{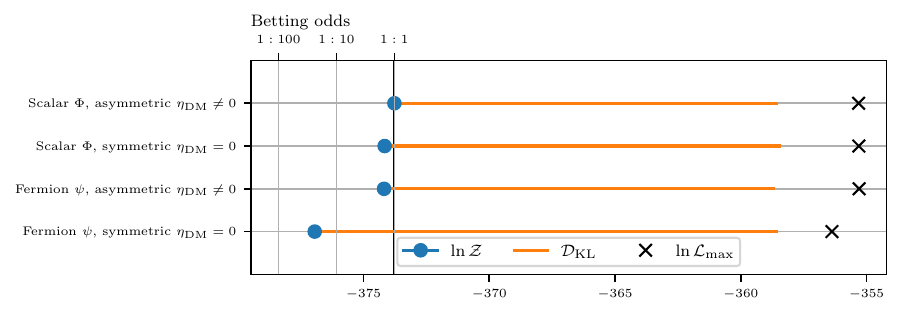}
    \caption{Bayes factors for the various models considered in this work. The blue dots indicate the logarithm of the Bayesian 
    evidence, which can be translated into betting odds relative to the most favoured model (see vertical lines and labels at the 
    top). The orange bars show the Kullback-Leibler divergence of prior and posterior for each model, with longer bars indicating 
    a larger fine-tuning penalty. This penalty is largest for the symmetric fermionic DM model, accounting for the smaller Bayesian 
    evidence, whereas the log-likelihoods of the respective best-fit points (indicated by the black crosses) are similar for all 
    models.}
    \label{fig:evidences}
\end{figure}

As expected, we find that including the asymmetry parameter does not change the picture substantially. There is no strong 
preference for a large asymmetry, and the posteriors of the other model parameters are not significantly modified. The Bayesian 
evidences are found to be
\begin{equation}
    \log \mathcal{Z}_\text{sym} = -374.16 \pm  0.11  = -358.46  - 15.70
\end{equation}
and
\begin{equation}
    \log \mathcal{Z}_\text{asym} =  - 373.77 \pm  0.11  = -358.60  - 15.18 \, ,
\end{equation}
which are very similar to each other and to the fermionic DM model with asymmetry, such that none of these models are 
preferred in a Bayesian sense. We illustrate the various Bayesian evidences in figure~\ref{fig:evidences}.

\section{Discussion}
\label{sec:discussion}

In this section we discuss the implications of our results for the experimental search programme for sub-GeV DM.

\subsection{Sensitivity projections}

\begin{figure}[h!]
    \centering
    \includegraphics[width=0.49\textwidth]{fermion_asym_mAp_kappa_sensitivities.pdf}
    \includegraphics[width=0.49\textwidth]{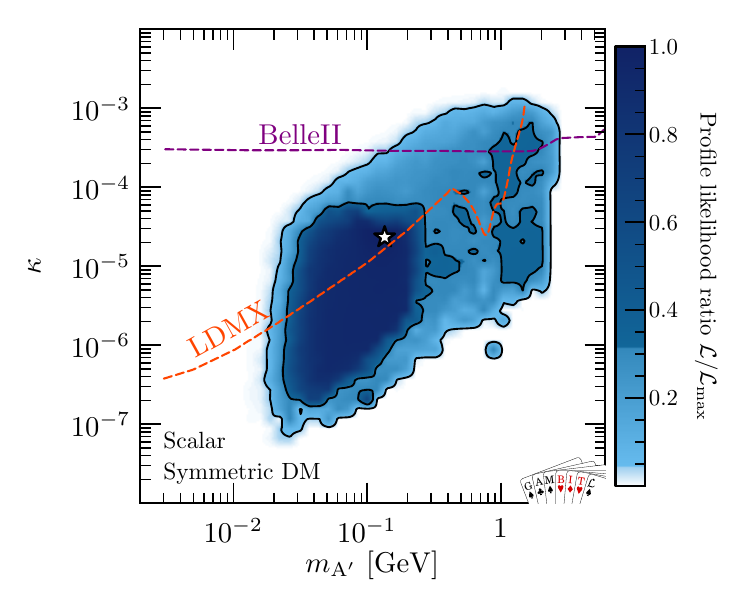} \\
    \includegraphics[width=0.49\textwidth]{fermion_asym_mDM_fsigmap_sensitivities.pdf}
    \includegraphics[width=0.49\textwidth]{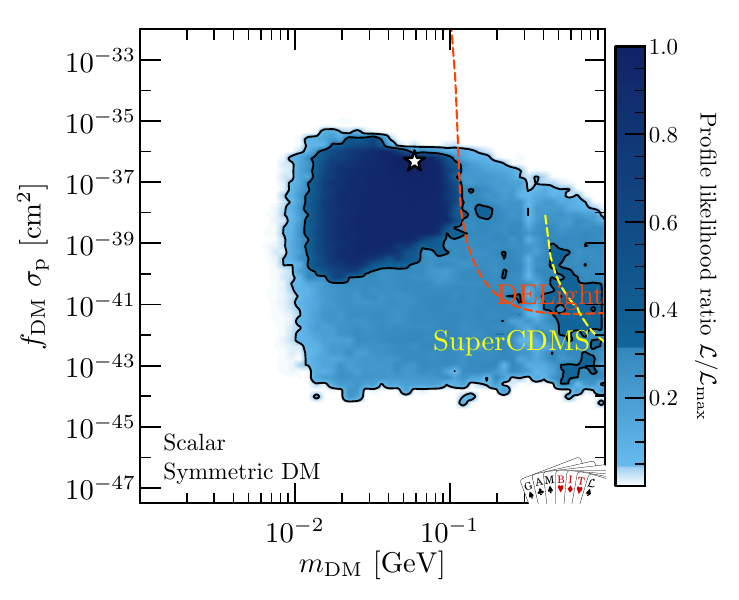} \\
    \includegraphics[width=0.49\textwidth]{fermion_asym_mDM_fsigmae_sensitivities.pdf}
    \includegraphics[width=0.49\textwidth]{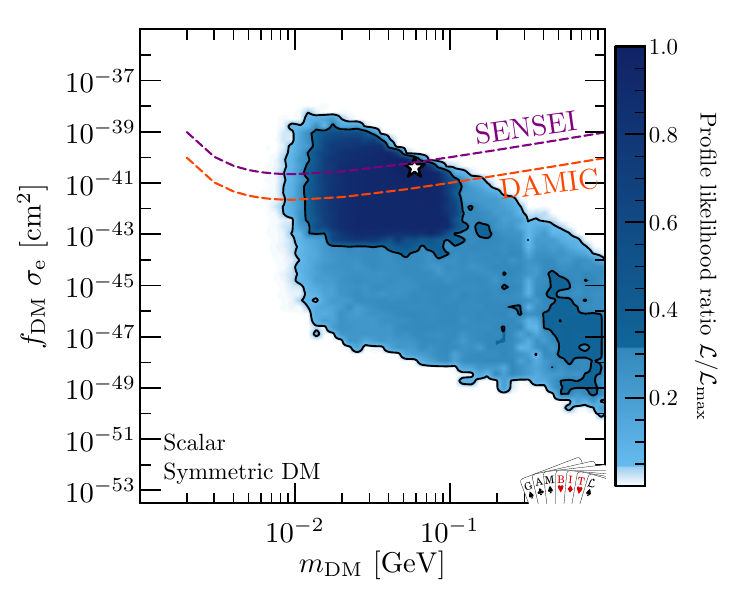}
    \caption{Allowed parameter regions of the asymmetric fermionic DM model (left column) and the symmetric scalar DM model 
    (right column) in terms of the quantities that are most directly relevant for observations: kinetic mixing versus dark photon 
    mass (top row), effective DM-nucleon scattering cross section versus DM mass (middle row) and effective DM-electron 
    scattering cross section versus DM mass (bottom row). In each panel we show the projected sensitivities for various 
    near-future experiments.}
    \label{fig:projections}
\end{figure}
In figure~\ref{fig:projections} we compare the allowed parameter regions of the two preferred models, namely asymmetric 
fermionic DM and symmetric scalar DM, with the projected sensitivities of near-future experiments. For accelerator experiments 
we consider the single-photon search of Belle II with 20 $\mathrm{fb^{-1}}$~\cite{Belle-II:2018jsg} and the missing energy 
search of LDMX with $10^{16}$ electrons from an 8 GeV beam~\cite{Akesson:2022vza}. For DM-nucleus scattering we consider 
SuperCDMS~\cite{SuperCDMS:2016wui}, and the recently proposed DELight experiment~\cite{vonKrosigk:2022vnf}, which 
plans to use superfluid helium to reach lower DM masses.\footnote{Very similar sensitivities are expected to be achieved by the HeRALD detector within the TESSERACT programme~\cite{SPICE:2023aqd}.} For DM-electron scattering we show the projections from DAMIC 
and SENSEI (both taken from ref.~\cite{Essig:2015cda}).

We find that near-future experiments can probe significant parts of the allowed parameter spaces, but will not be able to explore 
the two models comprehensively. The reason is in particular that the relic density requirement can be satisfied for rather small 
values of $g_\text{DM}$ and $\kappa$, given sufficient resonant enhancement (i.e.\ sufficiently small values of $\epsilon_R$).

The corresponding plots in the Bayesian framework is shown in appendix~\ref{app:bayesian}. As discussed above, in the Bayesian treatment very small values of $\epsilon_R$ are disfavoured due to the required 
fine-tuning between $m_\text{DM}$ and $m_{A'}$. This leads to a general preference for larger couplings and cross sections 
than in the frequentist analysis. Out of the experiments that we consider, we find that LDMX has the greatest chance of 
discovery, probing 64\% of the posterior volume.

\subsection{A new benchmark scenario}

Models of sub-GeV DM coupled to dark photons are often studied under simplifying assumptions in the 
literature~\cite{Izaguirre:2015yja,Alexander:2016aln,Berlin:2018bsc,Beacham:2019nyx}. In particular, it has become 
conventional to fix the ratio of dark photon and DM mass to 3, corresponding to $\epsilon_R = 5/4$. Furthermore, 
$\alpha_D = g_\mathrm{DM}^2 / (4\pi)$ is often fixed to either 0.1 or 0.5, corresponding to $g_\text{DM} = 1.1$ and 
$g_\text{DM} = 2.5$, respectively.

\begin{figure}[t]
    \centering
    \includegraphics[width=0.49\textwidth]{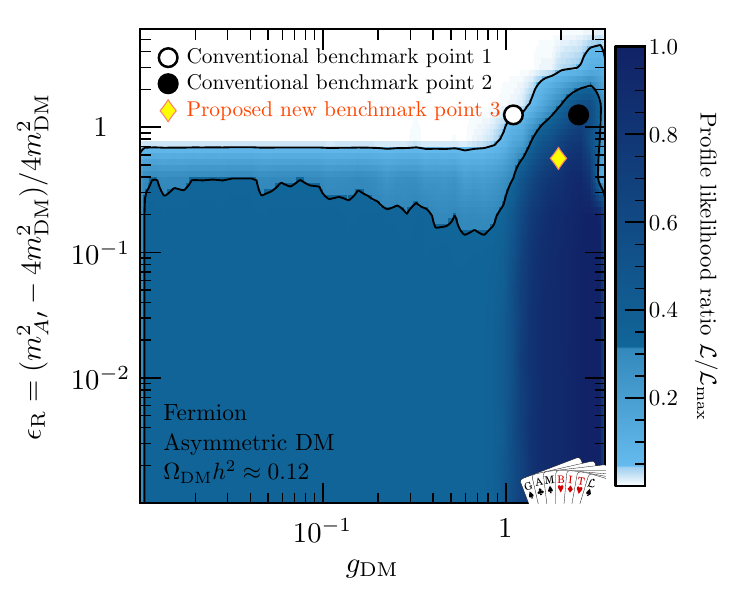}
    \includegraphics[width=0.49\textwidth]{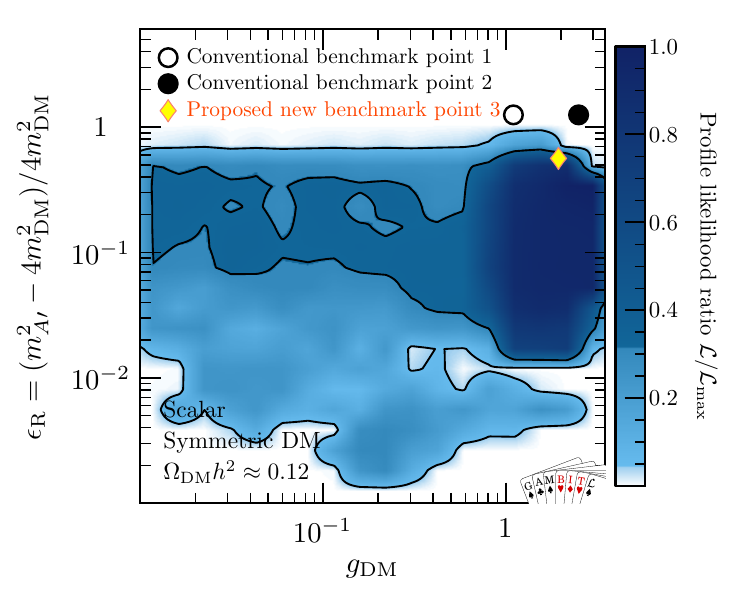}
    \caption{Allowed parameter space for the asymmetric fermion DM model (left) and the symmetric scalar DM model (right) in 
    the $\epsilon_R$ -- $g_\text{DM}$ parameter plane. The conventional benchmark points with $m_{A'} / m_\text{DM} = 3$ and 
    either $\alpha_D = 0.1$ (BP1) or $\alpha_D = 0.5$ (BP2) are indicated by a white and a black dot, respectively. Our proposed 
    benchmark point with $m_{A'} / m_\text{DM} = 2.5$ and $\alpha_D = 0.3$ (BP3) is indicated by a yellow diamond.}
    \label{fig:benchmark}
\end{figure}

In figure~\ref{fig:benchmark}, we show these benchmark choices (called BP1 and BP2) in the $g_\text{DM}$ versus 
$\epsilon_R$ parameter plane together with the allowed parameter regions that we have identified for asymmetric fermionic DM 
(left) and scalar DM (right) in our scans. We find that in both models BP1 lies outside of the allowed parameter region at 
95\%\,C.L., while BP2 lies barely within the 68\% C.L. for asymmetric fermionic DM and outside of the 95\% C.L. region for 
symmetric scalar DM. We therefore propose a new benchmark point (BP3), which is given by
\begin{align}
    m_{A'} = \tfrac{5}{2} m_\text{DM} \quad & \text{or} \quad \epsilon_R = \tfrac{9}{16} \nonumber \\
    \alpha_\text{DM} = 0.3 \quad & \text{or} \quad g_\text{DM} = 1.94 \, .
\end{align}
We emphasize that this benchmark point does not suffer from the uncertainties in relic density calculations illustrated in 
figure \ref{fig:epsR}, both because $\epsilon_R$ is not small and because $g_\text{DM}$ (and
hence the invisible decay width) is larger than what is used there.

Once these parameters have been fixed, the allowed ranges of the remaining parameters shrink substantially. In fact, for the 
case of scalar DM, the relic density requirement implies a very tight relation between $\kappa$ and $m_\text{DM}$, such that 
effectively all observables are uniquely predicted as a function of the DM mass (see for example ref.~\cite{Feng:2017drg}). The case of 
asymmetric fermionic DM is more interesting, since the relic density requirement only imposes a lower bound on $\kappa$ 
(corresponding to the case where $\eta_\text{DM} = 0$), whereas larger values of $\kappa$ can still be viable for 
$\eta_\text{DM} > 0$.

\begin{figure}[t]
    \centering
    \includegraphics[width=0.49\textwidth]{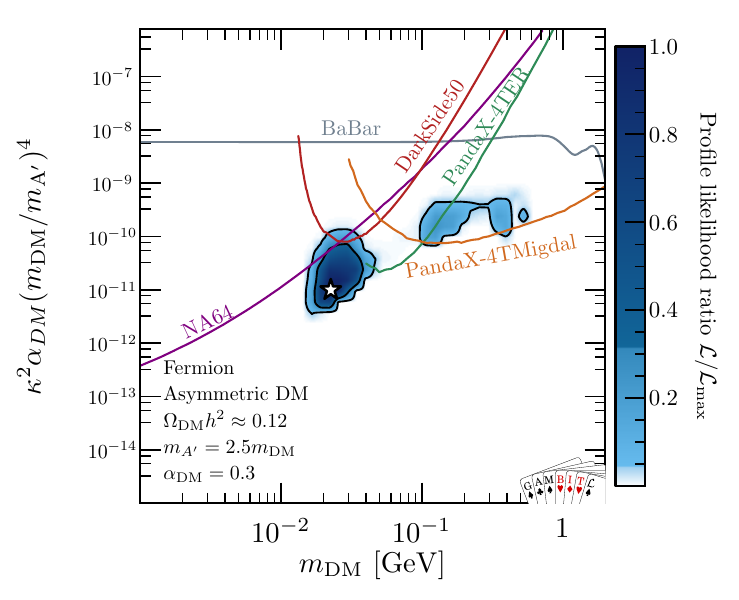}
    \includegraphics[width=0.49\textwidth]{Fermion_asym_allDM_benchmark_sensitivities.pdf}
    \caption{Allowed parameter regions for asymmetric fermionic DM with fixed $\epsilon_R = 0.5625$ and $g_\text{DM} = 1.94$ (BP3). In the left panel we show the various constraints implemented in this work, whereas the right panel shows projected sensitivities.}
    \label{fig:BP3}
\end{figure}

We show the allowed parameter regions for asymmetric fermionic DM with $\epsilon_R$ and $g_\text{DM}$ fixed to BP3 in 
figure~\ref{fig:BP3}. To facilitate the comparison with other results in the literature, we show on the y-axis the effective coupling
\begin{equation}
    y = \kappa^2 \alpha_\mathrm{D} \frac{m_\text{DM}^4}{m_{A'}^4} = 0.00768 \kappa^2 \, ,
\end{equation}
where the second equality holds specifically for BP3. Since we have fixed $\epsilon_R$ and $g_\text{DM}$, it is possible to 
superimpose exclusion limits from both accelerator and direct detection experiments in the same parameter 
plane. Indirect detection constraints, on the other hand, depend in a non-trivial way on the asymmetry parameter and do therefore not appear in this parameter plane. In the right panel of figure~\ref{fig:BP3}, we show the corresponding sensitivity projections, which cover the entire allowed parameter space.

\section{Conclusions}
\label{sec:conclusions}

In this work we have explored several different models of sub-GeV  DM coupled to a dark photon with kinetic mixing. These 
models face much weaker constraints from bounds on the DM scattering cross section from direct detection experiments than 
traditional WIMPs, but are potentially in tension with bounds on the DM annihilation cross section from indirect detection 
experiments and cosmological data. The central focus of our study has therefore been to understand how these bounds can be 
evaded and to explore the implications for laboratory experiments,
thus providing a general status update for this much-discussed model class.

For fermionic DM, where annihilation proceeds via $s$-wave, the simplest possibility to satisfy all experimental constraints is to 
consider resonant freeze-out, which requires a tight relation between the DM mass and the dark photon mass: 
$2 < m_{A'} / m_\text{DM} \lesssim 2.5$. An attractive alternative is to allow for a particle-antiparticle asymmetry $\eta_\text{DM}$, 
which increases the relic abundance relative to the symmetric case, for a given annihilation rate, 
but suppresses indirect detection signals. For scalar DM, on 
the other hand, annihilation proceeds via $p$-wave and is therefore velocity-suppressed in the present universe and effectively 
unconstrained.

To study these possibilities in detail, we have implemented state-of-the-art likelihoods for all relevant observables and 
measurements. Many of these are calculated using external codes, which we have interfaced using the \gambit global fitting 
framework. In particular, we calculate the DM relic density using \darksusy, direct detection constraints using \ddcalc and 
\obscura, dark photon branching ratios using \darkcast, BBN constraints using \alterbbn and CMB energy injection constraints 
using \hazma and \darkages. Furthermore, we have performed several calculations not previously available in the literature, in 
particular regarding DM production in beam-dump experiments using a version of \BdNMC extended for this analysis
and DM self-interactions in the Bullet Cluster. Finally, we reinterpret interpolated likelihoods from the literature to constrain dark 
photon production at accelerators as well as X-rays from DM annihilations.

For the case of symmetric fermionic DM we have found large viable parameter regions close to resonance when allowing for the 
DM particle to constitute only a fraction of the observed DM abundance, {see figure~\ref{fig:results_partDM_sym}}. When we instead require the DM particles to saturate the 
observed DM abundance, the dark sector couplings are tightly constrained by astrophysical and cosmological observations, 
such that laboratory experiments are not currently sensitive to the allowed parameter regions, {see figure~\ref{fig:results_allDM_sym}}.

This conclusion changes when including an asymmetry parameter, which makes it possible to saturate the DM relic abundance 
for larger dark sector couplings without violating astrophysical bounds. {As shown in figure~\ref{fig:results_allDM_asym}}, this opens up additional parameter regions that are 
constrained primarily by laboratory experiments (as well as by the DM self-interaction constraint from the Bullet Cluster). We 
emphasize, however, that the likelihood of the best-fit point does not change significantly when including the asymmetry 
parameter, such that in the frequentist analysis there is no preference for the asymmetric model over the symmetric one. 

We have also performed a Bayesian interpretation of our results, finding general agreement with the frequentist analysis. 
However, the Bayesian analysis penalises parameter regions that require a finely tuned mass ratio of 
$m_{A'} / m_\text{DM} \approx 2$. In the case of fermionic DM, this fine-tuning penalty leads to a clear  preference for a 
non-zero asymmetry, {see figure~\ref{fig:bayesian_asymmetric}}. To quantify this preference, we have calculated the Bayesian evidence for both symmetric and asymmetric 
fermionic DM, finding a Bayes factor (i.e.\ an evidence ratio) of around 15 in favour of the asymmetry.

For the case of scalar DM, on the other hand, we have found no preference for introducing an asymmetry. Indeed, the 
phenomenology and the allowed parameter regions for symmetric scalar DM are very similar to the ones for asymmetric 
fermionic DM, {see figure~\ref{fig:frequentist_scalar}}. Correspondingly the Bayes factor between these two models is very close to unity.

Recognizing the vibrant experimental activity to explore the sub-GeV scale, we have also studied the discovery prospects for 
the next generation of laboratory experiments. We found that these 
will be able to explore 
large parts of allowed parameter space, even though they cannot probe the preferred models comprehensively, {see figure~\ref{fig:projections}}. 
Since the commonly adopted benchmark points are already disfavoured, we have also
proposed a new benchmark scenario, defined as $m_{A'} / m_\text{DM} = 2.5$ and $\alpha_\text{DM} = 0.3$, which can serve 
as an appealing target for upcoming searches. 
Indeed, we expect that this benchmark can be fully probed within 
the next decade, highlighting the tantalizing possibility for a discovery in the near future.

\acknowledgments

We thank Jordan Koechler for providing the interpolated likelihoods from ref.~\cite{Cirelli:2023tnx}, and Pieter Braat, Jan Conrad, Adam Coogan, Lukas Hergt, Patrick deNiverville, Ruth Pottgen and all members of the \gambit community for discussions. SB acknowledges the support by the Doctoral School 'Karlsruhe School of Elementary and Astroparticle Physics: Science and Technology'.  The research of CB is supported by Australian Research Council grants DP210101636, DP220100643 and LE210100015. CC was supported by the Arthur B. McDonald Canadian Astroparticle Physics Research Institute. Research at the Perimeter Institute is supported by the Government of Canada through the Department of Innovation, Science, and Economic Development, and by the Province of Ontario. RC acknowledges support from an individual research grant from the Swedish Research Council (Dnr.~2018-05029). RC, TE, and TG research was partly financed via the Knut and Alice Wallenberg project ``Light Dark Matter'' (Dnr. KAW 2019.0080).
TE~was also supported by the Knut and Alice Wallenberg Foundation (PI, Jan Conrad).
TE~thanks the Theoretical Subatomic Physics group at Chalmers University of Technology for its hospitality. TEG and FK acknowledge funding by the Deutsche Forschungsgemeinschaft (DFG) through the Emmy Noether Grant No.\ KA 4662/1-2 and Grant No.\ 396021762 -- TRR 257. 
This work was performed using the Cambridge Service for Data Driven Discovery (CSD3), part of which is operated by the University of Cambridge Research Computing on behalf of the STFC DiRAC HPC Facility (\url{www.dirac.ac.uk}). The DiRAC component of CSD3 was funded by
BEIS capital funding via STFC capital grants ST/P002307/1 and ST/R002452/1 and STFC operations
grant ST/R00689X/1. DiRAC is part of the National e-Infrastructure. The plots were made using pippi v2.1~\cite{Scott:2012qh} and anesthetic~\cite{Handley:2019mfs}.

\appendix

\section{Relic density calculations}
\label{app:darksusy}

\noindent
In this appendix we document the implementation of asymmetric DM relic density calculations 
in \darksusy, which has been made available with release v6.4 of the code (another update in that release, also
developed in the context of this work, is the inclusion of interpolated \hazma yield tables from 
ref.~\cite{Coogan:2022cdd}, cf.~section \ref{sec:energy_injection}). 
We further discuss the limitations of the common Boltzmann approach that the 
current \darksusy implementation rests on.

Neglecting oscillations and $CP$ violations, the Boltzmann equation for the number densities of non-relativistic DM ($\chi$) 
and anti-DM ($\bar\chi$) particles are given by
\begin{align}
\label{eq:Boltz_standard}
 \dot n_\chi + 3H n_\chi= \dot n_{\bar \chi} + 3H n_{\bar \chi}=-\langle \sigma v\rangle\left (n_\chi n_{\bar\chi}-n_{\rm eq}^2\right)\,,
\end{align}
where $\langle \sigma v\rangle$ is the total thermally averaged annihilation rate  and $n_{\rm eq}$ is the number density in 
equilibrium. Introducing $Y\equiv n_{\bar \chi}/s$ and $x\equiv m_{\rm DM}/T$, and assuming entropy conservation, 
this can be re-written in full analogy to the case with $\eta_{\rm DM}=0$ (see for example refs.~\cite{Gondolo:1990dk, Bringmann:2018lay} for details) as
\begin{align}
\label{eq:Boltzmann_aDM}
 \frac{dY}{dx} = -\lambda \left(Y^2 + Y\eta_{\rm DM} - Y_{\rm eq}^2 \right)\,,
\end{align}
with 
\begin{align}
\lambda \equiv \sqrt{\frac{\pi}{45G}}\frac{g_*^{1/2}m_{\rm DM}}{x^2}\langle \sigma v\rangle\,.
\end{align}
Here, $G=M_{\rm Pl}^{-2}$ is Newton's gravitational constant, and $g_*$ is defined in terms
of the effective number of energy ($g_{\rm eff}$) and entropy ($h_{\rm eff}$) degrees of freedom as
\begin{align}
g_*^{1/2} \equiv \frac{h_{\rm eff}}{\sqrt{g_{\rm eff}}}\left(
1+\frac13 \frac{T}{h_{\rm eff}}\frac{dh_{\rm eff}}{dT}
\right) \,. 
\end{align}

\subsection{Asymmetric dark matter with \darksusy}
\label{app:rd_ds64}

We numerically solve eq.~(\ref{eq:Boltzmann_aDM}) by means of an implicit trapezoidal method with adaptive stepsize $h$: 
when going from $x_i$ to $x_{i+1}=x_i+h$, we estimate $Y_{i+1}\equiv Y(x_{i+1})$ as
\begin{align}
 Y_{i+1} &= Y_i + \frac{h}{2}\left(Y_i'+Y'_{i+1}\right) 
 =\frac{\tilde C_i}{2}-\frac12\tilde\lambda_{i+1}Y_{i+1}^2 -\frac12\tilde\eta_{i+1}Y_{i+1}\,,
\end{align}
where $\tilde \lambda_i\equiv h\lambda(x_i)$, $\tilde \eta_{i} \equiv \eta\tilde\lambda_{i}$ and 
$\tilde C_i \equiv 2Y_i(1-\frac{\tilde\eta_i}{2}) - \tilde\lambda_i\left( Y_i^2\!-\!Y_{\mathrm{eq},i}^2\right) + \tilde\lambda_{i+1}Y_{\mathrm{eq},i+1}^2$\,.
This constitutes a quadratic equation for $Y_{i+1} $ with solution
\begin{align}
Y_{i+1} &= \frac{\tilde C_i/\left[1+\tilde\eta_{i+1}/2\right]}{
1+\sqrt{1+\tilde\lambda_{i+1}\tilde C_i / \left[1+\tilde\eta_{i+1}/2\right]^{2}}}\,. \label{eq:BE_trapez}
\end{align} 
In order to estimate the local relative error we also calculate the abundance at $x_{i+1}$ with a modified Euler
method. We denote this alternative estimate for $Y_{i+1}$ with a lower-case $y_{i+1}$:
\begin{align}
y_{i+1} &\equiv Y_i + h Y_{i+1}'
= Y_i+\tilde\lambda_{i+1} Y_{{\rm eq}, i+1}^2 - \tilde\lambda_{i+1} y_{i+1}^2 - \tilde \eta_{i+1} y_{i+1}\,,
\end{align}
which we solve for
\begin{align}
y_{i+1} &=\frac{\tilde c_i/(2\left[1+\tilde\eta_{i+1}\right])}{1+\sqrt{1+\tilde\lambda_{i+1} \tilde c_i  / \left[1+\tilde\eta_{i+1}\right]^{2} }}\,,
\label{eq:newton}
\end{align}
with $\tilde c_i \equiv 4Y_i +4\tilde \lambda_{i+1} Y_{{\rm eq}, i+1}^2$\,.
When solving the Boltzmann equation (\ref{eq:Boltzmann_aDM}), as implemented in eq.~(\ref{eq:BE_trapez}), 
we adaptively decrease the step size $h$  if
$\left(Y_{i+1}-y_{i+}\right)/Y_{i+1}$ exceeds a given tolerance (contained in the common block variable  \term{compeps}; the initial step size for $h$
is stored in the common block variable  \term{hstep}). 

A new routine \term{dsrdomega\_aDM} returns the total DM density $\Omega_\text{DM} h^2$ thus calculated -- in full analogy to the 
corresponding routine \term{dsrdomega} for the symmetric case (but taking an additional input parameter $\eta_{\rm DM}$). 
It also returns the fraction of symmetric DM that 
contributes to the relic density, 
\begin{align}
r\equiv \frac{2\Omega_{\bar \chi}h^2}{\Omega_{\rm DM}h^2} =\frac{f_{\rm sym}}{f_{\rm DM}}\,.
\end{align}

\subsection{Limitations of the standard Boltzmann approach}
\label{app:RD_cBE}

One of the main assumptions leading to the formulation of the Boltzmann equation in its standard form,
 eq.~(\ref{eq:Boltz_standard}),
is that kinetic equilibrium between the DM particles and the heat bath is maintained throughout the entire freeze-out
process. While kinetic decoupling generically indeed happens much later than chemical decoupling~\cite{Bringmann:2009vf},
very early kinetic decoupling can occur exactly in the situation we are interested in here, namely in the vicinity
of narrow resonances~\cite{Binder:2017rgn}. In other words,  eq.~(\ref{eq:Boltz_standard}) runs risk of becoming invalid,
and the goal of this appendix is to discuss alternatives and  consequences.

One option is to solve the full Boltzmann equation at the phase-space level,
which however is computationally much more challenging. Sizeable 
DM self-interactions are even more difficult to model in this framework
than the elastic scattering of DM on relativistic heat bath particles (though significant progress has recently been made in this 
direction~\cite{Hryczuk:2022gay}). Alternatively, one can truncate the Boltzmann hierarchy at second order
and only consider the much 
simpler coupled 
system of equations describing the evolution of DM number density and velocity dispersion (or `temperature'), respectively.
Notably, these coupled Boltzmann equations actually become an exact description of the system in the limit of {\it large} 
DM self-interactions.
We refer to ref.~\cite{Binder:2021bmg} for further references and a more detailed discussion of these two approaches.

For the specific case of resonant annihilations, both approaches give comparable 
results~\cite{Binder:2017rgn,Binder:2021bmg}. For parameter points with $\epsilon_R\sim0.1$, in particular, 
it turns out that the actual relic density can be an order of magnitude above the one naively inferred from
solving eq.~(\ref{eq:Boltz_standard}). This difference of course significantly exceeds the uncertainty typically associated 
with relic density calculations. 
Since the true relic density is larger than the one obtained from the naive approach, a correspondingly larger value of 
$\langle\sigma v\rangle$ is needed during freeze-out to compensate for this effect and avoid DM overproduction.
For example, this can be achieved by decreasing $\epsilon_R$ or increasing $\kappa$. As illustrated in Fig.~\ref{fig:epsR},
notably, the former option can typically even be combined with a slightly {\it lower} value of $\kappa$, resulting
in a smaller uncertainty in the relic density calculations and at the same time avoiding more stringent
indirect detection constraints. 
Another option to suppress the relic density is to decrease the DM mass. 
In practice,  the required shift in the model parameters is often found to be somewhat smaller than suggested by a simple 
estimate based on the scaling  $\Omega_{\rm DM} h^2\propto 1/\langle\sigma v\rangle$~\cite{Binder:2017rgn}.

In the context of global scans, we are less interested in the shift of individual parameter points than in the 
behaviour of the allowed parameter space as a whole. Crucially, when showing two-dimensional confidence regions, we profile 
over all other directions in parameter space. As long as this projection includes at least some parameter points with accurate 
relic density calculations, the final result will also be accurate. 
Even in parameter regions where large uncertainties remain, these can usually be compensated for by shifts in the values of
parameters that are being profiled over, as discussed above.
Finally, even though there may in principle remain small parts of the two-dimensional confidence regions where 
such shifts are penalized by complementary constraints in {\it all} available directions of parameter space, these regions
will be very small compared to the orders of magnitude spanned by the total allowed 
parameter space. Taken together, these effects substantially reduce

the impact of the uncertainty in the relic density calculation on our main results. 
We therefore conclude that the results of our global 
scans are generally robust. In the following, we discuss in more detail how our findings are expected to change with a more 
accurate relic density calculation.

First, we note that for the case of a sub-dominant DM component, the likelihood typically does not depend sensitively on 
the precise value of the relic density. This is because direct and indirect detection constraints can be evaded by many orders of 
magnitude (see figure~\ref{fig:observables_partDM_sym}), while constraints from accelerator experiments do not depend on 
$\Omega_\text{DM} h^2$. Inaccuracies in the relic density calculation therefore do not significantly modify the allowed regions of 
parameter space. For the case of asymmetric DM, on the other hand, the most interesting regions of parameter space turn out 
to be those with asymmetry parameter $\eta_\text{DM}$ close to its upper bound $\eta_\text{asym}$. In these regions, the relic 
density is determined primarily by $\eta_\text{DM}$ and is therefore less sensitive to the precise value of the annihilation rate 
than in the symmetric case.

In particular, as shown in figure~\ref{fig:relic}, 
 tuning $\eta$ to be closer to $\eta_\text{asym}$ would allow to avoid
 indirect detection constraints even for significantly larger values of $\kappa$. 
 We therefore do not expect our scan results for asymmetric DM
 to change in any visible way.

\begin{figure}[t]
    \centering
    \includegraphics[width=0.49 \textwidth]{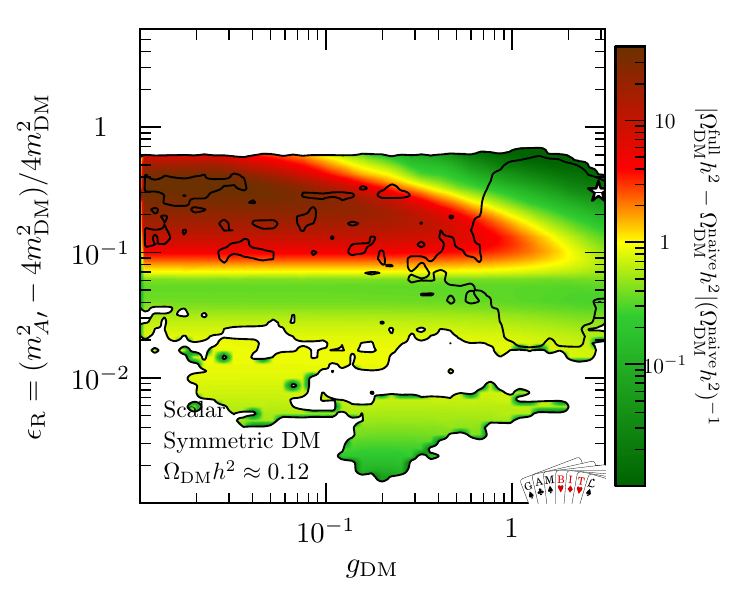}
    \includegraphics[width=0.49 \textwidth]{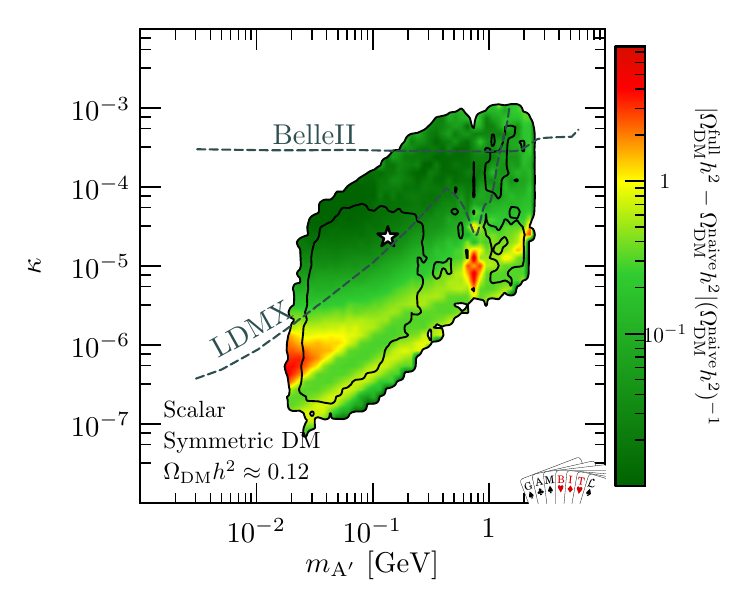}

    \includegraphics[width=0.49 \textwidth]{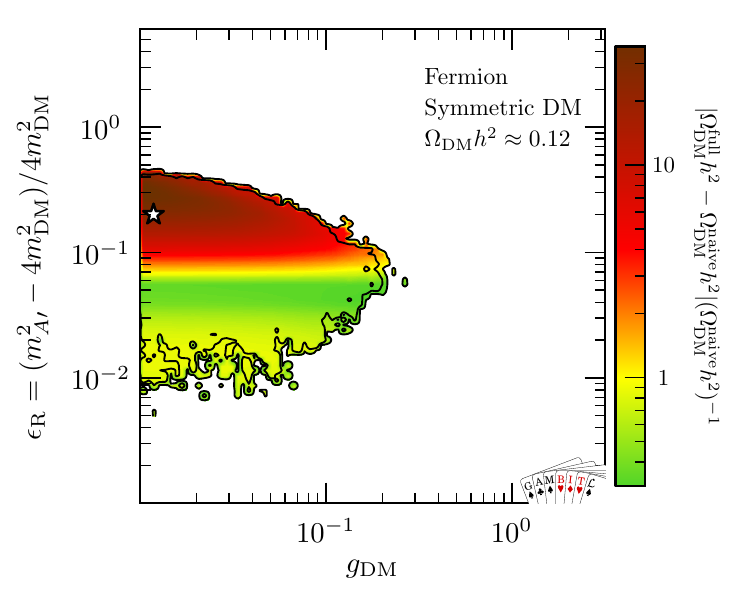}
    \includegraphics[width=0.49 \textwidth]{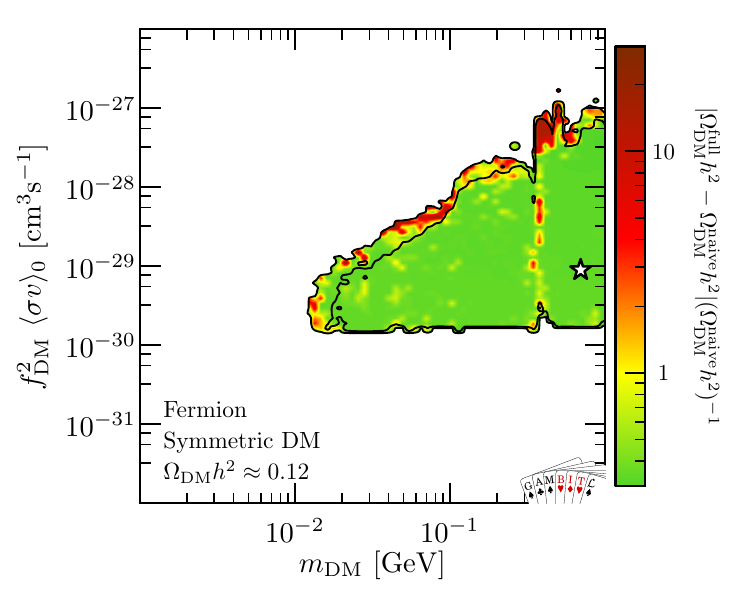}

    \caption{Estimated uncertainty in the relic density calculation for symmetric scalar DM (top row) and symmetric fermionic DM (bottom row), obtained by comparing the result from 
    solving  eq.~(\ref{eq:Boltz_standard}), called $\Omega_\text{DM}^\text{naive} h^2$, to the results from 
    ref.~\cite{Binder:2021bmg}, called $\Omega_\text{DM}^\text{full} h^2$. For each pair of model parameters, we minimise the 
    uncertainty over the remaining model parameters. 
    }
    \label{fig:uncertainties}
\end{figure}

For the purpose of this discussion, we therefore focus on the case of symmetric DM with 
$\Omega_\text{DM} h^2 \approx 0.12$,
considering in figure \ref{fig:uncertainties} both scalar DM (upper panels) and fermion DM (lower panels).
The colour coding in each panel corresponds to our estimate of the inaccuracy of the standard method to calculate the relic 
density, minimised over all parameters not shown explicitly. In other words, regions coloured in green are 
not affected by the uncertainties in the 
relic density calculation, whereas regions coloured in red are {\it potentially} affected and
require in principle
a more accurate calculation to 
determine whether or not they remain allowed.

In order to interpret the figure and its implications, let us first recall that the inaccuracy in the relic 
density calculation is by far most pronounced around $\epsilon_R \approx 0.1$, and that it becomes sizeable for  resonance 
widths $\Gamma_{A'}/m_{A'}\lesssim10^{-3}$~\cite{Binder:2021bmg}.
This is exactly what is seen in the left panels of the figure. For fermion DM, in particular, the upper bound on $g_{\rm DM}$
implies that the invisible decay rate is always rather small  -- while for scalar DM the resonance can be so wide that there is no 
significant issue with relic density calculations even for  $\epsilon_R \approx 0.1$.

However, when projecting the parameter space in terms of the observables 
that we are interested in 
for which we provide examples in the right panels of figure~\ref{fig:uncertainties}), 
we find that

most of the parameter regions are actually very robust and not affected by the inaccuracy of our relic 
density calculation.
Correcting for these inaccuracies may move individual parameter points from red `islands' well within the 
$2\sigma$ contours -- but without changing the overall boundaries of the allowed parameter region.
In that sense, the small red parameter regions at the upper edge of the displayed contours are of potentially
greatest concern. However, as explained above, the relic density can be reduced by both a larger value of $\kappa$ 
and a smaller value $\epsilon_R$. The latter would not leave a visible imprint in these plots, justifying our claim that
the results we have presented in the main text are robust.  

An interesting open question is whether it may be possible in certain cases to increase $\kappa$ while keeping all other parameters fixed, which might bring the model closer to being in reach of future experiments. We leave a more detailed study of this possibility for future work.

\section{Bullet Cluster constraints} \label{app:bc_constraints}

The Bullet Cluster comprises a main cluster and a subcluster, each of which consists dominantly of gas and DM, such that $M^\text{main,sub} = M^\text{main,sub}_\text{tot} + M^\text{main,sub}_\text{gas}$. Here $M_\text{tot}$ denotes the total DM mass, which for the case of multi-component DM may be larger than the mass $M_\text{DM}$ of the DM component under consideration. In this case, we define the fraction of DM particles $f_\chi$ and antiparticles $f_{\bar{\chi}}$ relative to $M_\text{tot}$  such that $f_\chi + f_{\bar{\chi}} = M_\text{DM} / M_\text{tot} \equiv f_\text{DM} \leq 1$. 

For an asymmetric DM model, the self-interactions that can lead to expulsion of DM particles from the sub-cluster include those between $\chi \chi$, $\bar{\chi} \bar{\chi}$ and $\chi \bar{\chi}$. The loss of DM mass as a function of time is then given by the sum of loss in DM particles and antiparticles,
\begin{equation}
    M^\text{sub}_\text{DM}(t) - M^\text{sub}_\text{DM}(0) = - M^\text{sub}_\text{tot} \left[ f_\chi \left( 1 - \exp^{-\int \dd{t} \, \Gamma_{\chi}} \right) + f_{\bar{\chi}} \left( 1 - \exp^{-\int \dd{t} \, \Gamma_{\bar{\chi}}} \right) \right] \, ,
\end{equation}
where $\Gamma_{\chi(\bar{\chi})} = R_{\text{imd},\chi(\bar{\chi})} + R_{\text{cml},\chi(\bar{\chi})}$ is the rate of evaporation of DM particles (antiparticles), which is a sum of immediate evaporation (i.e.\ expulsion of individual particles) and cumulative evaporation (i.e.\ mass loss due to heating). We adopt the results from ref.~\cite{Kahlhoefer:2013dca}:
\begin{align}
    R_\text{imd} =& \frac{\bar{\rho}_\text{main}}{m_\text{DM}}v_0 \sigma_\text{imd} \equiv \frac{\bar{\rho}_\text{main}}{m_\text{DM}} \ v_0 \ \int \dd{\phi} \int_{2 \bar{v}_\text{esc,sub}^2/v_0^2 - 1}^{1-2 \bar{v}_\text{esc,sub}^2/v_0^2} \dd{\cos{\theta}} \dv{\sigma}{\Omega}   \\
    R_\text{cml} =& \frac{\bar{\rho}_\text{main}}{m_\text{DM}} v_0 \sigma_{T} \equiv  \frac{\bar{\rho}_\text{main}}{m_\text{DM}} v_0 \int \dd{\phi} \int_{-1}^{1} \dd{\cos{\theta}} \ (1-|\cos \theta|) \frac{d \sigma}{d \Omega} \, ,
\end{align}
where $\bar{\rho}_\text{main} = \unit[2.955 \cdot 10^6]{M_{\odot}kpc^{-3}}$ is the average DM density within $150 \, \unit{kpc}$ of the main cluster, $m_\text{DM}$ is the DM mass, $\bar{v}_\text{esc,sub} = 2408 \ \unit{km/s}$ is the average escape velocity within $150 \, \unit{kpc}$ of the subcluster, $v_0 = 3900 \ \unit{km/s}$ is the constant collision velocity and $\sigma_{T}$ is the (corrected) momentum-transfer cross section (cluster and merger parameters obtained from ref.~\cite{Robertson:2016xjh}).

The total evaporation rate for particles and antiparticles are 

\begin{align}
    \Gamma_{\chi} =& \frac{\bar{\rho}_\text{main}}{m_\text{DM}} v_0 \ \left[f_{\chi} (\sigma_{\text{imd},\chi\chi} +\sigma_{T,\chi\chi}) + f_{\bar{\chi}}(\sigma_{\text{imd},\chi\bar{\chi}}+\sigma_{T,\chi\bar{\chi}})\right] = \frac{\bar{\rho}_\text{main}}{m_\text{DM}} \ v_0 \ \sigma_{\text{eff},\chi}\\
    \Gamma_{\bar{\chi}} =& \frac{\bar{\rho}_\text{main}}{m_\text{DM}} v_0 \ \left[ f_{\bar{\chi}} (\sigma_{\text{imd},\bar{\chi}\bar{\chi}}+\sigma_{T,\bar{\chi}\bar{\chi}}) + f_{\chi}(\sigma_{\text{imd},\chi\bar{\chi}}+\sigma_{T,\chi\bar{\chi}}) \right]  = \frac{\bar{\rho}_\text{main}}{m_\text{DM}} \ v_0 \ \sigma_{\text{eff},\bar{\chi}}.
\end{align}

We further make two simplifying assumptions: (\textit{i}) Evaporative collisions occur only after pericenter passage, (\textit{ii}) a particle ejected with velocity $\bar{v}_\text{esc,sub}$ takes time $\Delta t = 150 \, \mathrm{kpc} / \bar{v}_\text{esc,sub}$ to leave the central region of the subcluster. With these assumptions, the relative mass loss $\Delta_\text{DM} \equiv (M^\text{sub}_\text{DM,i} - M^\text{sub}_\text{DM,f}) / M^\text{sub}_\text{tot,i}$ (with subscript i (f) denoting the initial (final) value) can be written as
\begin{align}
        \Delta_\text{DM} = (f_\text{DM}-&f_{\bar{\chi}}) \left( 1 - \exp\left(- \bar{\Sigma}_\text{main} \frac{\sigma_{\text{eff},\chi}}{m_\text{DM}} \right)\right)
        + f_{\bar{\chi}} \left( 1 - \exp\left( -\bar{\Sigma}_\text{main}\frac{\sigma_{\text{eff},\bar{\chi}}}{m_\text{DM}}\right)\right)
\end{align}
 where
 \begin{equation}
     \bar{\Sigma}_\text{main} = \bar{\rho}_\text{main} \left(Z-{v_0} \Delta t\right) = \unit[1.41 \cdot 10^9]{M_{\odot}kpc^{-2}}
\end{equation}
 and $Z=\unit[720]{kpc}$ is the observed separation between the two clusters.  This expression gives us an analytic prediction for the DM mass loss. But during major mergers, gas is also stripped away from the colliding cluster. Thus, the measured total mass loss $\Delta_{\text{M}} = (M^\text{sub}_\text{f} - M^\text{sub}_\text{i}) / M^\text{sub}_\text{i}$  will include a contribution from gas. The DM mass loss and the total mass loss are related by,
 \begin{equation}
     \Delta_\text{M} = \Delta_{\text{DM}} (1-R^\text{sub}_i) + R^\text{sub}_i (1-x)
 \end{equation}
where $R^\text{sub}_i = M^\text{sub}_\text{gas,i}/M^\text{sub}_\text{i}$ is the ratio of initial gas mass to initial total cluster mass and $x = M^\text{sub}_\text{gas,f} / M^\text{sub}_\text{gas,i}$, such that $(1-x)$ is the fraction of gas lost during the collision. While $R_{\text{f}}^\text{sub}$ can be obtained from observations, $R_\text{i}^\text{sub}$ is more difficult to estimate. Here we use the observed value from the main cluster after the collision as approximation, i.e.\ we set $R_\text{i}^\text{sub} = R_\text{f}^\text{main}$ to the observed values of main cluster and subcluster, respectively, we have 
\begin{equation}x=\left(\frac{R^{\text{sub}}_\text{f}}{1-R^{\text{sub}}_\text{f}}\right)\left(\frac{1-R^{\text{main}}_\text{f}}{R^{\text{main}}_\text{f}}\right)(1-\Delta_{\text{DM}}) \, .
\end{equation} We use $R_{\text{f}}^{\text{main}}=0.09 \pm 0.01$ and $R_{\text{f}}^{\text{sub}}=0.04 \pm 0.01$ from ref.~\cite{Bradac:2006er}.

\section{Relativistic cross sections for beam dump DM searches}
\label{app:relcs}
In this appendix, we provide analytic expressions for the differential cross sections for relativistic DM-electron and -nucleon scattering for both the complex scalar and Dirac DM models investigated in this work. As explained in section~\ref{sec:likelihoods}, we use these cross sections to predict the number of DM signal events in beam dump experiments with \BdNMC~\cite{deNiverville:2016rqh}. We have obtained them by implementing the models of section~\ref{sec:models} in \FeynRules~\cite{Alloul:2013bka}, and then using \CalcHEP~\cite{Belyaev:2012qa} to calculate the squared modulus of the corresponding scattering amplitudes. We have validated the outcome of this symbolic computations through direct analytical calculations. While in the case of complex scalar DM the relativistic cross sections for DM-electron and -nucleon scattering were already implemented in \BdNMC, the implementation in the \BdNMC code of the analogous cross sections for the case of Dirac DM has been performed within this work.

For the differential cross section for DM-electron scattering, in the laboratory frame we find
\begin{align}
\frac{{\rm d} \sigma_{\rm eDM}(E_{\vec p}, E_{\vec k'})}{{\rm d} E_{\vec k'}} = \frac{1}{32 \pi m_e (E_{\vec p}^2-m_{\rm DM}^2)} \overline{|\mathcal{M}_{\rm eDM}(E_{\vec p}, E_{\vec k'})|^2} \,,
\label{eq:sigma_e}
\end{align}
where 
\begin{align}
\overline{|\mathcal{M}_{\rm eDM}(E_{\vec p}, E_{\vec k'})|^2} = 8 g_{\rm DM}^2 \kappa^2 e^2 \, \frac{2m_e^2 E^2_{\vec p} - m_e (E_{\vec k'}-m_e) f(E_{\vec p},E_{\vec k'})}{(m_{A'}^2+2 m_e E_{\vec k'} - 2 m_e^2)^2} \,,
\end{align}
and 
\begin{align}
f(E_{\vec p},E_{\vec k'}) = \left\{
\begin{array}{ll}
2m_e E_{\vec p} + m_{\rm DM}^2 & \quad\quad\quad\quad\quad\textrm{Complex scalar DM}\\
2m_e E_{\vec p} - m_e E_{\vec k'} + 2m_e^2 + m_{\rm DM}^2 & \quad\quad\quad\quad\quad \textrm{Dirac DM} \,,
\end{array}
\right.
\end{align}
while $E_{\vec p}$ ($E_{\vec k'}$) is the initial (final) state DM (electron) energy. Eq.~\eqref{eq:sigma_e} agrees with eq.~(3) in ref.~\cite{Batell:2014mga}. For the differential cross section for DM-nucleon scattering, in the laboratory frame we find
\begin{align}
\frac{{\rm d} \sigma_{\rm nDM}(E_{\vec p}, E_{\vec p'})}{{\rm d} E_{\vec p'}} = \frac{1}{32 \pi m_n (E_{\vec p}^2-m_{\rm DM}^2)} \overline{|\mathcal{M}_{\rm nDM}(E_{\vec p}, E_{\vec p'})|^2} \,,
\label{eq:sigma_n}
\end{align}
where 
\begin{align}
\overline{|\mathcal{M}_{\rm nDM}(E_{\vec p}, E_{\vec p'})|^2} = g_{\rm DM}^2 \, \frac{F_1^2\,\mathcal{A}(E_{\vec p}, E_{\vec p'})  + F_2^2\,\mathcal{B}(E_{\vec p}, E_{\vec p'})  + F_1 F_2\,\mathcal{C}(E_{\vec p}, E_{\vec p'}) }{[m_{A'}^2+2 m_n(E_{\vec p}-E_{\vec p'})]^2}\,,
\end{align}
and
\begin{align}
\mathcal{A}(E_{\vec p}, E_{\vec p'})  &= \left\{
\begin{array}{ll}
8  m_n [E_{\vec p'} \left(2 E_{\vec p} m_n+m_{\rm DM}^2\right)-E_{\vec p} m_{\rm DM}^2]  & \\
 8  m_n [E_{\vec p}^2 m_n+E_{\vec p '}
   (m_n E_{\vec p '}+m_n^2+m_{\rm DM}^2)-E_{\vec p}
   (m_n^2+m_{\rm DM}^2)] & 
\end{array}
\right. \nonumber\\
\mathcal{B}(E_{\vec p}, E_{\vec p'})  &= \left\{
\begin{array}{ll}
2  m_n \left(E_{\vec p}-E_{\vec p'}\right) [E_{\vec p}^2+2 (E_{\vec p}+m_n)
   E_{\vec p'} +E_{\vec p'}^2-2 E_{\vec p} m_n-4 m_{\rm DM}^2]  & \\
4 m_n (E_{\vec p}-E_{\vec p '}) [(2
   E_{\vec p}-m_n) E_{\vec p '}+E_{\vec p} m_n-2 m_{\rm DM}^2] & 
\end{array}
\right. \nonumber\\
\mathcal{C}(E_{\vec p}, E_{\vec p'})  &= \left\{
\begin{array}{ll}
-8 m_n (E_{\vec p}-E_{\vec p'}) (-m_n E_{\vec p'}+E_{\vec p} m_n+2
   m_{\rm DM}^2) & \,\,\,\quad\quad\quad\quad \textrm{Complex scalar DM}\\
16 m_n (E_{\vec p}-E_{\vec p'})
   (-m_n E_{\vec p'}+E_{\vec p} m_n-m_{\rm DM}^2) & \,\,\,\quad\quad\quad\quad \textrm{Dirac DM} \,.
\end{array}
\right.
\end{align}
Here, $E_{\vec p'}$ is the outgoing DM particle energy, $m_n$ is the nucleon mass, while $F_1$ and $F_2$ are momentum-dependent nucleon form factors~\cite{Batell:2014yra}. Eq.~(\ref{eq:sigma_n}) agrees with eq.~(B.10) in ref.~\cite{Batell:2014yra}, and includes the interference terms (i.e.~terms proportional to $F_1 F_2$) that are missing in eq.~(14) of ref.~\cite{deNiverville:2011it}.

\section{Observable predictions from Bayesian scans}
\label{app:bayesian}

\begin{figure}[t]
    \centering
    \includegraphics[width=\textwidth]{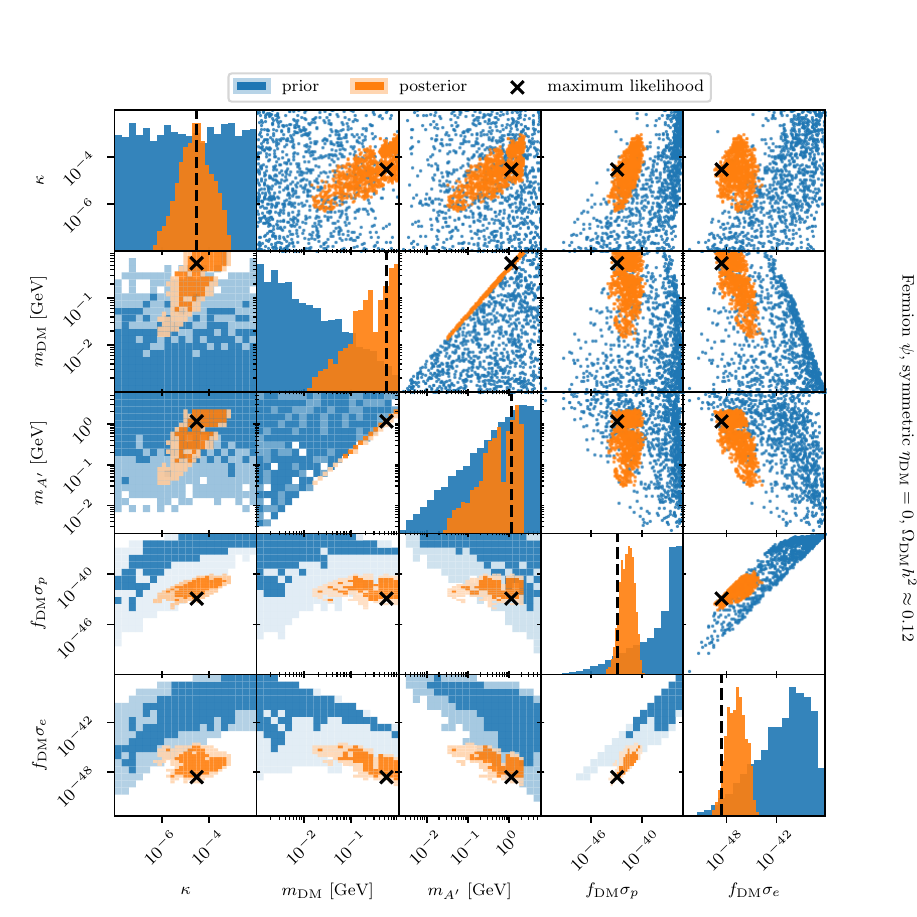}
    \caption{Prior (blue) and posterior (orange) probabilities for symmetric fermionic DM in terms of the most relevant parameters and observables.}
    \label{fig:projections_Bayesian_fermion}
\end{figure}

\begin{figure}[t]
    \centering
    \includegraphics[width=\textwidth]{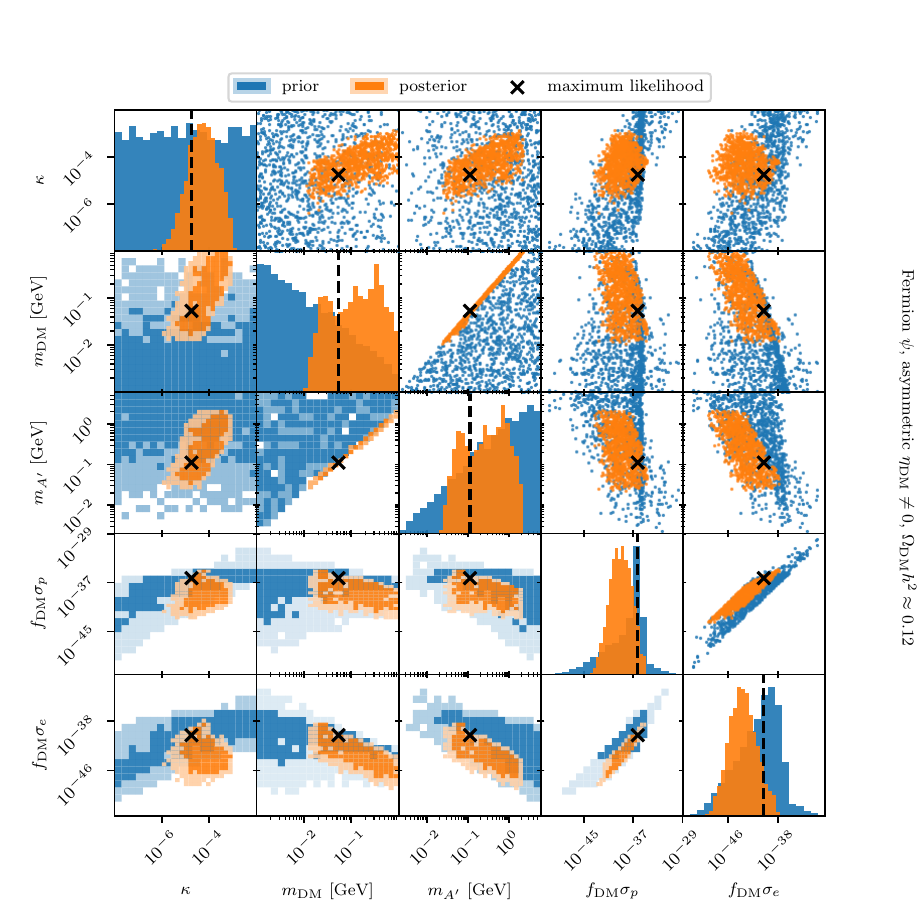}
    \caption{Prior (blue) and posterior (orange) probabilities for asymmetric fermionic DM in terms of the most relevant 
    parameters and observables.}
    \label{fig:projections_Bayesian_asymmetric}
\end{figure}

\begin{figure}[t]
    \centering
    \includegraphics[width=\textwidth]{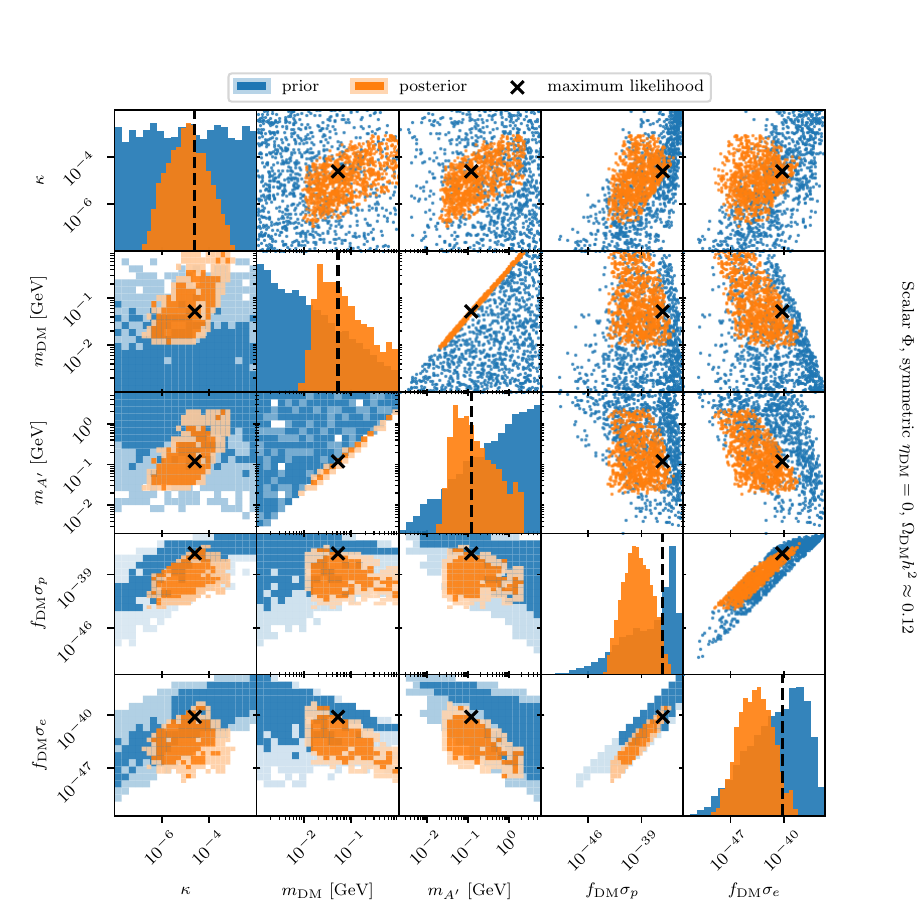}
    \caption{Prior (blue) and posterior (orange) probabilities for scalar DM in terms of the most relevant parameters and observables.}
    \label{fig:projections_Bayesian_scalar}
\end{figure}

In this appendix we provide additional results from our Bayesian scans, showing posterior probabilities for the parameters and observables relevant for various experiments. These figures allow us to identify promising targets for future experiments and to quantify the (Bayesian) probability of a detection, i.e.\ the fraction of the credible region that can be probed. The case of symmetric fermionic DM is shown in figure~\ref{fig:projections_Bayesian_fermion}, while the case of asymmetric fermionic DM is shown in figure~\ref{fig:projections_Bayesian_asymmetric}. As expected, the latter case allows for much larger couplings and therefore for more promising detection prospects. Finally the case of symmetric scalar DM is shown in figure~\ref{fig:projections_Bayesian_scalar}.

\section{GAMBIT implementation} \label{app:gambit}

\subsection{New models}

Two new model trees have been added to the \gambit model hierarchy, according to the fermionic and scalar sub-GeV models described in this work. The top of the model trees are \doublecrosssf{SubGeVDM\_fermion}{SubGeVDM_fermion} and \doublecrosssf{SubGeVDM\_scalar}{SubGeVDM_scalar}, respectively. Each of the trees contain a resonant DM model where the mass splitting $\epsilon_R = (m^2_{A'} - 4 m^2_{\text{DM}})/(4 m^2_\text{DM})$ replaces the dark photon mass as a model parameter, \doublecrosssf{Resonant\_SubGeVDM\_fermion}{Resonant_SubGeVDM_fermion} and \doublecrosssf{Resonant\_SubGeVDM\_scalar}{Resonant_SubGeVDM_scalar}. Furthermore, for each of the four models above, there is a companion model where the asymmetry parameter $\eta_\text{DM}$ is substituted by the combination $\eta_\text{DM}m_\text{DM}$ called, respectively, \doublecrosssf{SubGeVDM\_fermion\_RDprior}{SubGeVDM_fermion_RDprior}, \doublecrosssf{Resonant\_SubGeVDM\_fermion\_RDprior}{Resonant_SubGeVDM_fermion_RDprior}, for fermionic DM, and \doublecrosssf{SubGeVDM\_scalar\_RDprior}{SubGeVDM_scalar_RDprior} and \doublecrosssf{Resonant\_SubGeVDM\_scalar\_RDprior}{Resonant_SubGeVDM_scalar_RDprior} for scalar DM. Finally, two reparametrisations of the general \doublecrosssf{SubGeVDM\_fermion}{SubGeVDM_fermion} are also available, where the dark matter coupling $g_\text{DM}$ is replaced by the DM-electron cross section $\sigma_e$, \doublecrosssf{SubGeVDM\_fermion\_sigmae}{SubGeVDM_fermion_sigmae}, or the DM-nucleon cross section $\sigma_N$, \doublecrosssf{SubGeVDM\_fermion\_sigmaN}{SubGeVDM_fermion_sigmaN}.

\begin{description}

\item[\textbf{\textsf{SubGeVDM\_fermion}}: \label{SubGeVDM_fermion}]\term{mDM,mAp,gDM,kappa,etaDM}

Sub-GeV DM fermion model parametrised with the DM mass \term{mDM}, the dark photon mass \term{mAp}, the DM-dark photon coupling \term{gDM}, the kinetic mixing parameter \term{kappa} and the asymmetry parameter \term{etaDM}.

\item[\textbf{\textsf{SubGeVDM\_fermion\_sigmae}}: \label{SubGeVDM_fermion_sigmae}] \term{mDM,mAp,sigmae,kappa,etaDM}

Child model of \doublecrosssf{SubGeVDM\_fermion}{SubGeVDM_fermion} with the DM-electron cross section \term{sigmae} as a parameter instead of the coupling \term{gDM}.

\item[\textbf{\textsf{SubGeVDM\_fermion\_sigmaN}}: \label{SubGeVDM_fermion_sigmaN}] \term{mDM,mAp,sigmaN,kappa,etaDM}

Child model of \doublecrosssf{SubGeVDM\_fermion}{SubGeVDM_fermion} with the DM-nucleon cross section \term{sigmaN} as a parameter instead of the coupling \term{gDM}.

\item[\textbf{\textsf{SubGeVDM\_fermion\_RDprior}}:\label{SubGeVDM_fermion_RDprior}] \term{mDM,mAp,gDM,kappa,etaDM\_mDM}

Child model of \doublecrosssf{SubGeVDM\_fermion}{SubGeVDM_fermion} with a relic density prior, i.e.\ substituting the asymmetry parameter \term{etaDM} with the combination \term{etaDM\_mDM}.

\item[\textbf{\textsf{Resonant\_SubGeVDM\_fermion}}:\label{Resonant_SubGeVDM_fermion}] \term{mDM,epsR,gDM,kappa,etaDM}

Child model of \doublecrosssf{SubGeVDM\_fermion}{SubGeVDM_fermion} with in the resonance $m_\text{DM} \sim 2m_{A'}$, so substituting the parameter \term{mAp} by the mass splitting \term{epsR} defined in eq.\eqref{eq:resparam}.

\item[\textbf{\textsf{Resonant\_SubGeVDM\_fermion\_RDprior}}:\label{Resonant_SubGeVDM_fermion_RDprior}]\term{mDM,epsR,gDM,kappa,etaDM\_mDM}

Child model of \doublecrosssf{SubGeVDM\_fermion}{SubGeVDM_fermion} using both the resonance parameter \term{epsR} and the relic density prior parameter \term{etaDM\_mDM}.

\item[\textbf{\textsf{SubGeVDM\_scalar}}: \label{SubGeVDM_scalar}] \term{mDM,mAp,gDM,kappa,etaDM}

Sub-GeV DM scalar model parametrised with the DM mass \term{mDM}, the dark photon mass \term{mAp}, the DM-dark photon coupling \term{gDM}, the kinetic mixing parameter \term{kappa} and the asymmetry parameter \term{etaDM}.

\item[\textbf{\textsf{SubGeVDM\_scalar\_RDprior}}:\label{SubGeVDM_scalar_RDprior}] \term{mDM,mAp,gDM,kappa,etaDM\_mDM}

Child model of \doublecrosssf{SubGeVDM\_scalar}{SubGeVDM_scalar} with a relic density prior, i.e.\ substituting the asymmetry parameter \term{etaDM} with the combination \term{etaDM\_mDM}.

\item[\textbf{\textsf{Resonant\_SubGeVDM\_scalar}}:\label{Resonant_SubGeVDM_scalar}] \term{mDM,epsR,gDM,kappa,etaDM}

Child model of \doublecrosssf{SubGeVDM\_scalar}{SubGeVDM_scalar}  with in the resonance $m_\text{DM} \sim 2m_{A'}$, so substituting the parameter \term{mAp} by the mass splitting \term{epsR} defined in eq.\eqref{eq:resparam}.

\item[\textbf{\textsf{Resonant\_SubGeVDM\_scalar\_RDprior}}] \label{Resonant_SubGeVDM_scalar_RDprior}\term{mDM,epsR,gDM,kappa,etaDM\_mDM}
\end{description}

Child model of \doublecrosssf{SubGeVDM\_scalar}{SubGeVDM_scalar} using both the resonance parameter \term{epsR} and the relic density prior parameter \term{etaDM\_mDM}.

To complement the model descriptions above, a capability called \cpp{SubGeVDM_spectrum} was added to the \specbit module~\cite{GAMBITModelsWorkgroup:2017ilg} to set up relevant spectrum details.

\subsection{Updates to \darkbit, \cosmobit and \colliderbit}

\begin{table*}[p]
\centering
  \caption{Capabilities added to \darkbit associated with this work. \cpp{LL} is shorthand for \cpp{LogLikelihood}.}
  \scriptsize
   \makebox[\linewidth]{
   \begin{tabular}{p{3.4cm} p{6.3cm} p{6.5cm} }
    \toprule
   \textbf{Capability}  & \textbf{Function} (\textbf{type}) & \textbf{Dependencies [type] / \newline Backend reqs [type (args)]} \\ \midrule

    \cpp{TH_ProcessCatalog}
      &\cpp{TH_ProcessCatalog_SubGeVDM_fermion}(\cpp{TH_ProcessCatalog}) \newline \cpp{TH_ProcessCatalog_SubGeVDM_scalar}(\cpp{TH_ProcessCatalog}) & {\cpp {SubGeVDM_spectrum [Spectrum]}\newline \cpp{decay_rates [DecayTable] } }  \\
   \midrule

   \cpp{DD_couplings} & 
   \cpp{DD_couplings_SubGeVDM_fermion} (\cpp{DM_nucleon_couplings}) \newline \cpp{DD_couplings_SubGeVDM_scalar} (\cpp{DM_nucleon_couplings}) & {}  \\
   \midrule

   \cpp{DarkMatter_ID} & 
   \cpp{DarkMatter_ID_SubGeVDM_fermion} (\cpp{string}) \newline \cpp{DarkMatter_ID_SubGeVDM_scalar} (\cpp{string}) & {}  \\
   \midrule

  \cpp{DarkMatterConj_ID} & 
   \cpp{DarkMatterConj_ID_SubGeVDM_fermion} (\cpp{string}) \newline \cpp{DarkMatterConj_ID_SubGeVDM_scalar} (\cpp{string}) & {}  \\
   \midrule
 
   \cpp{RD_oh2_aDM}
      &\cpp{RD_oh2_DS_general_aDM} (\cpp{ddpair}) & {\cpp{RD_spectrum_ordered [RD_spectrum_type]} \newline \cpp{RD_eff_annrate [fptr_dd]} \newline \cpp{RD_oh2_DS6_ini [int]} \newline \cpp{dsrdstart [void (...)]} \newline \cpp{dsrdens [void (...)]} \newline \cpp{rdpars [DS_RDPARS]} \newline \cpp{adm_com [DS_ADM_COM]}}\\
   \midrule
   
   \cpp{ID_suppression},
      &\cpp{ID_suppression_aDM} (\cpp{double}) \newline \cpp{ID_suppression_symDM} (\cpp{double}) & {\cpp{RD_oh2_aDM [ddpair]} \newline \cpp{RD_fraction [double]} \newline \cpp{DM_process [std::string]}} \\
   \midrule
   
   \cpp{DM_mass_loss} 
      &\cpp{TH_ProcessCatalog_SubGeVDM_scalar} (\cpp{TH_ProcessCatalog}) & {\cpp {SubGeVDM_spectrum [Spectrum]}\newline \cpp{decay_rates [DecayTable] } }  \\
   \midrule
   
   \cpp{BulletCluster_lnL}
      &\cpp{calc_bullet_cluster_DMmassLoss} (\cpp{double}) & {\cpp {SubGeVDM_spectrum [Spectrum]}\newline     \cpp{RD_fraction [double]} \newline \cpp {RD_oh2_aDM [ddpair]} \newline \cpp{decay_rates [DecayTable] } }  \\
   \midrule
   
   \cpp{Xray_loglikelihoods}
      &\cpp{TH_ProcessCatalog_SubGeVDM_scalar} (\cpp{TH_ProcessCatalog}) & {\cpp {SubGeVDM_spectrum [Spectrum]}\newline \cpp{decay_rates [DecayTable] } }  \\
   \midrule
   
   \cpp{set_gamLike_GC_halo} 
      &\cpp{Xray_loglikes_Cirelli} (\cpp{double}) & {\cpp{WIMP_properties [WIMPprops]} \newline \cpp{LocalHalo [LocalMaxwellianHalo]} \newline \cpp{TH_ProcessCatalog [TH_ProcessCatalog]} \newline \cpp{ID_suppression [double]}}  \\
   \midrule
   
   \cpp{LocalHalo_GeV} 
      &\cpp{ExtractLocalMaxwellianHalo_GeV} (\cpp{LocalMaxwellianHalo}) & {}  \\
   \midrule

   \cpp{XENON1T_ER_LL}
    & \cpp{calc_XENON1T_ER_LL} (\cpp{double}) & {\cpp{
    LocalHalo_GeV [LocalMaxwellianHalo]} \newline \cpp {RD_fraction [double]} \newline \cpp{sigma_e [double]} \newline \cpp{XENON1T_S2_ER [obscura::DM_Detector_Ionization_ER]}}  \\
   \midrule

    \cpp{DarkSide50_ER_LL}
    & \cpp{calc_DarkSide50_ER_LL} (\cpp{double}) & {\cpp{
    LocalHalo_GeV [LocalMaxwellianHalo]} \newline \cpp {RD_fraction [double]} \newline \cpp{sigma_e [double]} \newline \cpp{DarkSide50_S2_ER [obscura::DM_Detector_Ionization_ER]}}  \\
   \midrule

    \cpp{DarkSide50_ER_2023_LL}
    & \cpp{calc_DarkSide50_ER_2023_LL} (\cpp{double}) & {\cpp{
    LocalHalo_GeV [LocalMaxwellianHalo]} \newline \cpp {RD_fraction [double]} \newline \cpp{sigma_e [double]} \newline \cpp{DarkSide50_S2_ER_2023 [obscura::DM_Detector_Ionization_ER]}}  \\
   \midrule

   \end{tabular}
}
  \label{tab:capabilities_1}
\end{table*}

\begin{table*}[p]
\centering
  \caption{Capabilities added to \darkbit associated with this work (continued). \cpp{LL} is shorthand for \cpp{LogLikelihood}.}
  \scriptsize
   \makebox[\linewidth]{
   \begin{tabular}{p{4.0cm} p{6.3cm} p{5.9cm} }
    \toprule
   \textbf{Capability}  & \textbf{Function} (\textbf{type}) & \textbf{Dependencies [type] / \newline Backend reqs [type (args)]} \\ \midrule

    \cpp{PandaX_4T_ER_LL}
    & \cpp{calc_PandaX_4T_ER_LL} (\cpp{double}) & {\cpp{
    LocalHalo_GeV [LocalMaxwellianHalo]} \newline \cpp {RD_fraction [double]} \newline \cpp{sigma_e [double]} \newline \cpp{PandaX_S2_4T_ER [obscura::DM_Detector_Ionization_ER]}}  \\
   \midrule 

    \cpp{SENSEI_at_MINOS_LL}
    & \cpp{calc_SENSEI_at_MINOS_LL} (\cpp{double}) & {\cpp{
    LocalHalo_GeV [LocalMaxwellianHalo]} \newline \cpp {RD_fraction [double]} \newline \cpp{sigma_e [double]} \newline \cpp{SENSEI_at_MINOS [obscura::DM_Detector_Crystal]}}  \\
   \midrule

    \cpp{CDMS_HVeV_2020_LL}
    & \cpp{calc_CDMS_HVeV_2020_LL} (\cpp{double}) & {\cpp{
    LocalHalo_GeV [LocalMaxwellianHalo]} \newline \cpp {RD_fraction [double]} \newline \cpp{sigma_e [double]} \newline \cpp{CDMS_HVeV_2020 [obscura::DM_Detector_Crystal]}}  \\
   \midrule

    \cpp{DAMIC_M_2023_LL}
    & \cpp{calc_DAMIC_M_2023_LL} (\cpp{double}) & {\cpp{
    LocalHalo_GeV [LocalMaxwellianHalo]} \newline \cpp {RD_fraction [double]} \newline \cpp{sigma_e [double]} \newline \cpp{DAMIC_M_2023 [obscura::DM_Detector_Crystal]}}  \\
   \midrule

    \cpp{XENON1T_Migdal_LL}
    & \cpp{calc_XENON1T_Migdal_LL} (\cpp{double}) & {\cpp{
    LocalHalo_GeV [LocalMaxwellianHalo]} \newline \cpp {RD_fraction [double]} \newline \cpp{sigma_SI_p [double]}  \newline \cpp{sigma_SI_n [double]} \newline \cpp{XENON1T_Migdal [obscura::DM_Detector_Ionization_Migdal]}}  \\
   \midrule
   
    \cpp{DarkSide50_Migdal_LL}
    & \cpp{calc_DarkSide50_Migdal_LL} (\cpp{double}) & {\cpp{
    LocalHalo_GeV [LocalMaxwellianHalo]} \newline \cpp {RD_fraction [double]} \newline \cpp{sigma_SI_p [double]}  \newline \cpp{sigma_SI_n [double]} \newline \cpp{DarkSide50_Migdal [obscura::DM_Detector_Ionization_Migdal]}}  \\
   \midrule

    \cpp{DarkSide50_Migdal_2023_LL}
    & \cpp{calc_DarkSide50_Migdal_2023_LL} (\cpp{double}) & {\cpp{
    LocalHalo_GeV [LocalMaxwellianHalo]} \newline \cpp {RD_fraction [double]} \newline \cpp{sigma_SI_p [double]}  \newline \cpp{sigma_SI_n [double]} \newline \cpp{DarkSide50_Migdal_2023 [obscura::DM_Detector_Ionization_Migdal]}}  \\
   \midrule
   
    \cpp{PandaX_4T_Migdal_LL}
    & \cpp{calc_PandaX_4T_Migdal_LL} (\cpp{double}) & {\cpp{
    LocalHalo_GeV [LocalMaxwellianHalo]} \newline \cpp {RD_fraction [double]} \newline \cpp{sigma_SI_p [double]}  \newline \cpp{sigma_SI_n [double]} \newline \cpp{PandaX_4T_Migdal [obscura::DM_Detector_Ionization_Migdal]}}  \\
   \midrule

    \cpp{RD_oh2_underprediction} & \cpp{RD_oh2_underprediction_SubGeVDM} (\cpp{double}) & \\
    \midrule
    
   \end{tabular}
}
  \label{tab:capabilities_2}
\end{table*}

\begin{table*}[t]
\centering
  \caption{Capabilities added to \cosmobit associated with this work. \label{tab:capabilities_3}}
  \scriptsize
   \makebox[\linewidth]{
   \begin{tabular}{p{4.5cm} p{6.3cm} p{5.4cm} }
    \toprule
   \textbf{Capability}  & \textbf{Function} (\textbf{type}) & \textbf{Dependencies [type] / \newline Backend reqs [type (args)]} \\ \midrule
   \cpp{Neff_after_BBN} 
      &\cpp{extract_Neff_after_BBN} (\cpp{double}) & {\cpp {primordial_abundances} \newline \cpp{[BBN_container]} }  \\
   \midrule
   
   \cpp{N_eff_likelihood_Planck_BAO}
      &\cpp{compute_N_eff_likelihood_Planck_BAO} (\cpp{double}) & {\cpp {Neff_after_BBN [double]} }  \\
   \midrule

   \end{tabular}
}

\vspace{5mm}

  \caption{Capabilities added to \colliderbit associated with this work. \label{tab:capabilities_4}}
  \scriptsize
   \makebox[\linewidth]{
   \begin{tabular}{p{4.5cm} p{6.3cm} p{5.4cm} }
    \toprule
   \textbf{Capability}  & \textbf{Function} (\textbf{type}) & \textbf{Dependencies [type] / \newline Backend reqs [type (args)]} \\ \midrule
   
   \cpp{BaBar_single_photon_LogLike}
      &\cpp{BaBar_single_photon_LogLike_SubGeVDM} (\cpp{double}) & {\cpp {dark_photon_decay_rates [DecayTable::Entry]} }  \\
   \midrule
   
   \cpp{AllAnalysisNumbers}
   & \cpp{SubGeVDM_results} (\cpp{AnalysisDataPointers}) & \cpp{SubGeVDM_spectrum [Spectrum]}

   \end{tabular}

}
\end{table*}

In addition to the new models, this work has also expanded the likelihood computations of the \darkbit~\cite{GAMBITDarkMatterWorkgroup:2017fax}, \cosmobit~\cite{GAMBITCosmologyWorkgroup:2020htv} and \colliderbit~\cite{GAMBIT:2017qxg} modules. The full list of new capabilities, along with their module functions, dependencies and backend requirements, is given in tables \ref{tab:capabilities_1}-\ref{tab:capabilities_4}.

In \darkbit, many capabilities are supplemented with module functions for the sub-GeV models, including those for computing the process catalog, \cpp{TH_ProcessCatalog_SubGeVDM_fermion} and \cpp{TH_ProcessCatalog_SubGeVDM_scalar}, those that calculate the direct detection couplings, \cpp{DD_couplings_SubGeVDM_fermion} and \cpp{DD_couplings_SubGeVDM_scalar}, and those storing the DM properties, \cpp{DarkMatter_ID_SubGeVDM_fermion} and \cpp{DarkMatterConj_ID_SubGeVDM_fermion}, for fermion DM, and \cpp{DarkMatter_ID_SubGeVDM_scalar}, and \cpp{DarkMatterConj_ID_SubGeVDM_scalar} for scalar DM.

New capabilities added for this project include \cpp{RD_oh2_aDM}, which computes the relic abundance for an asymmetric DM model, \cpp{ID_suppression}, which calculates the suppression factor of ID signals due to under abundant DM, \cpp{DM_mass_loss} computes the loss in mass of subcluster due to self-interactions, \cpp{BulletCluster_lnL} is the likelihood function for the Bullet Cluster, \cpp{Xray_loglikelihoods} calculates the constraints from the annihilation of DM into X-rays, \cpp{set_gamLike_GC_halo} initialises the DM halo in \textsc{gamlike}, \cpp{LocalHalo_GeV} provides and alternative local halo parametrisation with all parameter in GeV and \cpp{sigma_e} computes the DM-electron cross section. Furthermore, many new experimental likelihoods are added for direct detection of DM using electron recoils and nuclear recoils via the Migdal effect. The capabilities for these likelihoods are of the form \cpp{<experiment>_LogLikelihood}, where the \cpp{<experiment>} label takes into account not only the experiment but also the type of interaction if there is more than one type. The likelihoods included are for the experiments:  \cpp{XENON1T_ER}, \cpp{DarkSide50_ER}, \cpp{DarkSide50_ER_2023}, \cpp{PandaX_4T_ER}, \cpp{SENSEI_at_MINOS}, \cpp{CDMS_HVeV_2020}, \cpp{DAMIC_M_2023}, \cpp{XENON1T_Migdal}, \cpp{DarkSide50_Migdal}, \cpp{DarkSide50_Migdal_2023} and \cpp{PandaX_4T_Migdal}. Lastly, the capability \cpp{RD_oh2_underprediction} calculates the expected underprediction of the relic abundance by not using the standard Boltzmann solution (see appendix \ref{app:RD_cBE} above) from the tabulated results from \cite{Binder:2021bmg}.

The only changes done to \cosmobit for this work are those to enable the usage of \alterbbn with the sub-GeV DM models, as well as new capabilities to extract the value of $N_{\rm eff}$ after BBN. These are \cpp{Neff_after_BBN}, which get the $N_{\rm eff}$ value after BBN from \alterbbn, and \cpp{N_eff_likelihood_Planck_BAO}, which computes the likelihood for $N_{\rm eff}$ given Planck + BAO data.

Finally, the modifications to \colliderbit include a new capability to calculate the likelihood to see dark photons at BaBar, \cpp{BaBar_single_photon_LogLike}, and a new module function \cpp{SubGeVDM_results} for the capability \cpp{AllAnalysisNumbers} that implements limits from the beam dump experiments LSND, MiniBoone and NA64 as interpolated yields.

\subsection{Backend interfaces}
\label{app:backends}

For the purpose of this study two new backend interfaces have been implemented, to \darkcast and to \obscura. Furthermore, important changes were performed to the backend interface to \darksusy in order to allow the computation of the relic abundance in asymmetric DM models (more about that in appendix~\ref{app:darksusy}), and that of \alterbbn, to allow the input of the DM parameters and the return of $N_{\rm eff}$.

A new backend interface was created to \darkcast v1.1 for the calculation of the decay widths and branching fractions of the dark photon. Convenience backend functions were created, \cpp{dark_photon_decay_width} and \cpp{dark_photon_branching_fraction} for single final state,  and \cpp{dark_photon_decay_width_multi} and \cpp{dark_photon_branching_fraction_multi} for multiple final states.

Lastly, the external tool \obscura, designed to compute likelihoods from direct detection experiments, was used to calculate the constraints from electron recoils and nuclear recoils using the Migdal effect. Since \obscura is written in \texttt{C++}, we use \textsc{BOSS} to create the hierarchy of abstract and wrapper classes required to use them inside \gambit (see section~4.5 of ref.~\cite{GAMBIT:2017yxo} for more details on how \textsc{BOSS} works). The classes used from \obscura are \cpp{Standard_Halo_Model} for setting the DM halo model, \cpp{DM_Particle_SI} the DM particle model and three models for detectors: crystal experiments, \cpp{DM_Detector_Crystal}, ionization experiments with electron recoils, \cpp{DM_Detector_Ionization_ER}, and ionization experiments with nuclear recoils via the Migdal effect, \cpp{DM_Detector_Ionization_Migdal}. Using those classes, the \obscura functions for which backend interfaces have been implemented are \cpp{XENON1T_S2_ER}, \cpp{DarkSide50_S2_ER,} \cpp{DarkSide50_S2_ER_2023}, \cpp{PandaX_S2_4T_ER}, \cpp{SENSEI_at_MINOS}, \cpp{CDMS_HVeV_2020}, \cpp{DAMIC_M_2023}, which compute the likelihood of DM-electron electron scatterings, and \cpp{XENON1T_Migdal}, \cpp{DarkSide50_Migdal}, \cpp{DarkSide50_Migdal_2023} and \cpp{PandaX_4T_Migdal}, that compute the likelihood of DM-nucleon scattering with Migdal effect.

\bibliographystyle{JHEP_improved}
\bibliography{references}

\end{document}